\documentclass[aps,prd,superscriptaddress,floatfix,nofootinbib,notitlepage,twocolumn,showkeys]{revtex4-1}

\usepackage{amsmath,amssymb}
\usepackage{graphicx}
\usepackage{natbib}
\usepackage{color}
\usepackage{url}
\usepackage{ulem}
\usepackage{rotating}
\usepackage{multirow}
\usepackage{dcolumn}
\usepackage{lineno}
\usepackage{breakurl}
\usepackage[usenames,dvipsnames]{xcolor}

\usepackage{hyperref}
\hypersetup{colorlinks=true,citecolor=blue}

\newcommand{\WISC}{WI$\times$SC}
\newcommand{\phz}{photo-$z$}
\newcommand{\phzs}{photo-$z$s}

\definecolor{Black}{named}{Black}
\definecolor{Red}{named}{Red}

\newcommand{\MP}{{\sc MontePython}}
\newcommand{\CLASS}{{\sc class}}

\newcommand\aapr{A\&ARv}             % Astronomy and Astrophysics Review (the)
\newcommand\mnras{MNRAS}             % Monthly Notices of the Royal Astronomical Society
\newcommand\aap{A\&A}                % Astronomy and Astrophysics
\newcommand\aj{AJ}                   % Astronomical Journal (the)
\newcommand\apjl{ApJ}                % Astrophysical Journal, Letters
\newcommand\apjs{ApJS}               % Astrophysical Journal, Supplement
\newcommand\jcap{J.~Cosmology Astropart. Phys.} % Journal of Cosmology and Astroparticle Physics
\newcommand\na{New~Astron.}          % New Astronomy

\begin{document}

\title[Updated Tomography of the ISW]{An Updated Tomographic Analysis of the Integrated Sachs-Wolfe Effect and Implications for Dark Energy}

\author{Benjamin St\"olzner}
\email{stoelzner@physik.rwth-aachen.de}

\author{Alessandro Cuoco}
\email{cuoco@physik.rwth-aachen.de}

\author{Julien Lesgourgues}
\email{lesgourg@physik.rwth-aachen.de}

\affiliation{Institute for Theoretical Particle Physics and Cosmology (TTK), RWTH Aachen University, Otto-Blumenthal-Strasse, 52057, Aachen, Germany.}

\author{Maciej Bilicki}
\email{bilicki@strw.leidenuniv.nl}
\affiliation{Leiden Observatory, Leiden University, P.O. Box 9513,  NL-2300 RA Leiden, The Netherlands}
\affiliation{National Centre for Nuclear Research, Astrophysics Division, P.O.Box 447, 90-950 \L{}\'{o}d\'{z}, Poland}
\affiliation{Janusz Gil Institute of Astronomy, University of Zielona G\'ora, ul. Lubuska 2, 65-265 Zielona G\'ora, Poland}

\label{firstpage}

%%%%%%%%%%%%%%%%%%%%%%%%%%%%%%%%%%%%%%%%%%%%%%%%%%%%%%%%%%%%%%%%%%%%%

\begin{abstract}

We derive updated constraints on the Integrated Sachs-Wolfe (ISW) effect through
cross-correlation of the cosmic microwave background 
with galaxy surveys. 
We improve with respect to similar previous analyses  in several ways.
First, we use the most recent versions of extragalactic object catalogs: 
SDSS DR12 photometric redshift (photo-$z$) 
and 2MASS Photo-$z$ datasets, 
as well as employed earlier for ISW, SDSS QSO photo-$z$ and NVSS samples.
Second, we use for the first time the WISE~$\times$~SuperCOSMOS catalog, which allows us
to perform an all-sky analysis of the ISW  up to $z\sim0.4$.
Third,   thanks to the  use of photo-$z$s,  
we separate each dataset
into different redshift bins, 
deriving the cross-correlation in each bin.
This last step leads to a significant improvement in
sensitivity.
We remove
cross-correlation between 
catalogs using masks which mutually exclude common regions of the sky.
We use two methods to quantify the significance of the ISW effect.
In the first one, we fix the 
cosmological model,
derive 
linear galaxy biases of the catalogs, and then
evaluate the significance of the ISW using a single parameter.
In the second approach we perform a global fit of the ISW and of the galaxy biases varying the cosmological model.
We find significances of the ISW in the range  4.7-5.0 $\sigma$  thus reaching, for the first time in such an analysis, the threshold
of 5 $\sigma$.
Without the redshift tomography we find a significance of $\sim$ 4.0 $\sigma$, which shows
the importance of the binning method.
Finally we use the ISW data to infer constraints on the Dark Energy redshift evolution and equation of state.
We find that the redshift range covered by the catalogs is still not optimal to derive strong constraints,
although this goal will be likely reached using future datasets such as from Euclid, LSST, and SKA.

\end{abstract}


\maketitle

%%%%%%%%%%%%%%%%%%%%%%%%%%%%%%%%%%%%%%%%%%%%%%%%%%%%%%%%%%%%%%%%

\renewcommand{\arraystretch}{1.2}

\section{Introduction}
\label{sec:intro}

We have, at present, strong evidence for Dark Energy (DE) from the 
large amount of available cosmological data \citep[e.g.,][]{Ade:2015xua}.
Nonetheless, this evidence is mostly based on  precise constraints
from the Cosmic Microwave Background (CMB) epoch extrapolated to the present time.
Local, or present-day, constraints on DE are, instead, mostly given by SuperNovae (SN) data,
which are not yet precise enough for accurately constraining the properties and
time evolution of DE \citep[e.g.,][]{Betoule14}.

Thus, it is important to look for alternative local DE probes.
In this respect such a DE-sensitive measurement is given by the
late-time Integrated Sachs-Wolfe effect (ISW) on the CMB \citep{sachs_perturbations_1967}.
{This effect is  imprinted in the angular pattern of the CMB 
in the presence of a time-varying cosmological gravitational potential,
which appears in the case of a non-flat universe \cite{Kamionkowski:1996ra,Kinkhabwala:1998zj},  as well as for a flat one in the presence of DE,
but also for various modified gravity theories \cite[e.g.,][]{Song:2006ej,Barreira12}.
Thus, for standard General Relativity (GR) and flat cosmology a non-zero ISW implies
the presence of DE.} 
The effect is very small and cannot be well measured 
using the CMB alone since it peaks at large angular scales (small multipoles, $\ell \lesssim 40$)
which
are cosmic-variance limited.
On the other hand, it was realized that this effect
can be more efficiently isolated by cross-correlating the CMB
with tracers of the Large-Scale Structure (LSS) of the Universe 
at low ($z\lesssim1$) redshift \citep{1998NewA....3..275B,boughn_cross_2001},
with most of the signal lying in the range $z\in [0.3,1.5]$ 
for a standard $\Lambda$CDM cosmological model \citep{Afshordi:2004kz}. 

%\vspace{10mm}

In the past, many ISW analyses were performed using
a large variety of tracers at different redshifts 
\citep{dupe_measuring_2011,2013arXiv1303.5079P,Planck15_ISW,2004PhRvD..69h3524A,2004Natur.427...45B,Fosalba:2003ge,
2007MNRAS.381.1347C,2006MNRAS.372L..23C,2006PhRvD..74f3520G,2008MNRAS.386.2161R,2007MNRAS.377.1085R,xia09,
Ferraro:2014msa,Hernandez-Monteagudo:2013vwa,Shajib:2016bes,Pietrobon:2006gh,McEwen:2006my,Vielva:2004zg}.
In a few cases, global analyses were performed combining
different LSS tracers, giving the most stringent constraints and evidence
for the ISW effect at the level of $\sim 4\ \sigma$ \citep{giannantonio_combined_2008,2012MNRAS.426.2581G,ho_correlation_2008}.
Related methodology, which has been explored more recently, consists in stacking CMB patches overlapping with locations of large-scale structures, such as superclusters or voids 
\cite{Granett:2008ju,Papai2011,Ilic:2013cn,Cai2014,Granett2015,Kovacs:2015bda,Kovacs2017}.
A further idea, which was sometimes exploited,
is to use the redshift information of a given catalog
to divide it into different redshift bins,  compute the cross-correlation in
each bin, and then  combine the information.
This tomographic approach was pursued, for example, in the study of 2MASS \citep{francis_integrated_2010} or
SDSS galaxies \citep{2003astro.ph.Scranton,sawangwit_cross-correlating_2010}. 
Typically, the use of tomography does not provide  strong
improvement over the no-binning case, either because the catalog does not contain
a large enough number of objects and splitting them increases the shot-noise,
or because the redshift range is not well suited for ISW studies.

Nonetheless, in the recent years, several catalogs with redshift information and 
with a very large number of objects have become
available  thanks to the use of 
photometric redshifts (\phzs) instead of spectroscopic ones.
Although \phzs\
are not as accurate
as their spectroscopic counterparts, the former
are sufficient for performing a tomographic
analysis of the ISW with coarse $z$ bins. Hence we can
exploit these large catalogs, which have the advantage of giving a
low shot noise even when divided into sub-samples.
In this work, we combine for the first time the two above approaches:
we use several datasets covering different redshifts ranges, and we bin them
into redshift sub-samples to perform a global tomography. 
We show that in this way we are able to improve the significance of
the ISW effect from $\sim 4\ \sigma$ without redshift binning
to $\sim 5\ \sigma$ exploiting the full tomography information. 
When combining the various catalogs, we take special care to minimize their overlap both in terms of common sources and the same LSS traced, 
in order not to use the same information many times. This is done by appropriate data cleaning and masking.
We then use these improved measurement of the ISW effect to
study deviations of DE from the simplest assumption of a cosmological constant.

Finally, the correlation data derived in this work and the associated likelihood will soon be made publicly available, 
in the next release of the \MP\footnote{See \url{http://baudren.github.io/montepython.html} and \\
\url{https://github.com/brinckmann/montepython_public} } package~\citep{Audren:2012wb}.

%%%%%%%%%%%%%%%%%%%%%%%%%%%%%%%%%%%%%%%%%%%%%%

\section{Theory}
\label{sec:theory}

The expression for the cross-correlation angular power spectrum  (CAPS)
between two fields $I$ and $J$  is given by:
\begin{equation}
C_l^{I,J} =\frac{2}{\pi}
\int k^2 P(k) [G_{\ell}^I(k)]
[G_{\ell}^J(k)]  dk,
\label{eq:angularspectrum}
\end{equation}
where $P(k)$ is the present-day power spectrum  of matter fluctuations. 
In the above expression we have  assumed an underlying cosmological model, like $\Lambda$CDM,
in which the evolution of density fluctuations is separable in wavenumber $k$ and redshift $z$ on linear scales.
A different expression applies, for example, in the presence of massive neutrinos \citep{Lesgourgues:2007ix},
where the $k$ and $z$ evolution is not separable.
{Moreover, in the following, we assume standard GR and a flat $\Lambda$CDM model.
For studies of the ISW effect for non-zero curvature or modified gravity see \cite{Kamionkowski:1996ra,Kinkhabwala:1998zj,Song:2006ej,Barreira12}.}

For the case $I=c$ of the fluctuation field of a catalog of
discrete objects, one has 
\begin{equation}
G_{\ell}^c(k)=\int \frac{dN(z)}{dz} b_c(z) D(z) j_{\ell}[k \chi(z)]dz,
\label{eq:crossocorr}
\end{equation}
where $dN(z)/dz$ and $ b_c(z)$ represent the redshift distribution
and the galaxy bias factor of the sources, respectively,
$j_{\ell}[k \chi(z)]$ are spherical Bessel functions, $D(z)=(P(k,z)/P(k))^{1/2}$ is the
linear growth factor of density fluctuations 
and $\chi(z)$ is the comoving distance to redshift $z$. 

For the case of cross-correlation with the
temperature fluctuation field obtained from the CMB maps
($J=T$), the ISW effect in real space is given by \citep[e.g.,][]{nishizawa_integrated_2014} 
\begin{equation}
\label{eqISW}
\Theta(\hat{n})=-2\int \frac{d\Phi(\hat{n}\chi,\chi)}{d\chi}d\chi,
\end{equation}
where $\Phi$ represents the gravitational potential.
%\footnote{\AC{ With respect to  \cite{nishizawa_integrated_2014} Eq.\ref{eqISW} 
%has a minus sign since it is expressed in terms of comoving distance instead of comoving time.}}.
In the expression, we neglect a factor of $\exp ( -\tau)$, which introduces
an error of the order of $10 \%$, smaller than the typical accuracy
achieved in the determination of the ISW itself.
Furthermore, using the Poisson and Friedmann equations,
\footnote{{Eqs.~\ref{eqISW}-\ref{eqPoisson}  are valid assuming GR. For modified gravity different appropriate expressions would apply (see, e.g., \cite{Song:2006ej,Barreira12}).} }
{and considering scales sufficiently within the horizon}
\begin{equation}
\label{eqPoisson}
\Phi(k,z)= - \frac{3}{2\, c^2}    \frac{\Omega_m}{a(z)}  \frac{H_0^2}{k^2} \ \delta(k,z) 
\end{equation}
where $c$ is the {speed} of light, $a(z)$ is the cosmological scale factor,
$H_0$ is the Hubble parameter today, $\Omega_m=\Omega_b+\Omega_c$ is the fractional density of matter today, and $\delta(k,z)$ is the matter fluctuation field in Fourier space, we can write
\begin{equation}
G_{\ell}^{T}(k)=   \frac{3\ \Omega_m}{c^2}   \frac{\mathrm{H}_0^2}{k^2}   \int \frac{d}{dz}\left(\frac{D(z)}{a(z)}\right) j_{\ell}[k\chi(z)]dz~.
\label{eq:isw}
\end{equation}
Finally, the equations above can be combined through Eq.~\eqref{eq:angularspectrum} to give the CAPS expected  for the
ISW effect resulting from the correlation between a catalog of extragalactic
objects,  tracing the underlying mass distribution, and the CMB.
Using the Limber approximation \citep{1953ApJ...117..134L} the correlation becomes \citep{ho_correlation_2008}
\begin{equation}
\begin{split}
C_{\ell}^{cT} = \frac{3\Omega_m\mathrm{H}_0^2}{c^3\left(l+\frac{1}{2}\right)^2}\int dz \, &b_c(z)\frac{dN}{dz}H(z)D(z)\frac{d}{dz}\left(\frac{D(z)}{a(z)}\right)\\
&\times P\left(k=\frac{l+\frac{1}{2}}{\chi(z)}\right).
\end{split}
\label{eq:isw2}
\end{equation}
The Limber approximation is very accurate at $\ell>10$ and accurate at the level
of $10\%$  at $\ell<10$ \citep{1953ApJ...117..134L}, which is sufficient 
for the present analysis.

In our study, we use the public
code \CLASS\footnote{See \url{http://class-code.net}}~\citep{blas_cosmic_2011} to compute the linear power spectrum of density fluctuations.
As an option, this code can compute internally the spectra $C_{\ell}^{cT}$ and $C_{\ell}^{cc}$, for arbitrary redshift distribution functions, using either the Limber approximation or a full integral in $(k,z)$ space.
We prefer, nonetheless, to use the Limber approximation since CAPS calculations are  significantly faster.
Also, to get better performances and more flexibility, we choose to perform these calculations directly inside our python likelihood, reading only $P(k,z)$ from the \CLASS{} output. We checked on a few examples that our spectra do agree with those computed internally by \CLASS. 

\section{CMB maps}
\label{sec:cmbmaps}

We use CMB maps from the Planck 2015 data release\footnote{See \url{http://pla.esac.esa.int/pla/\#maps}}  \citep{Ade:2015xua}
 which have been produced using four different methods of foreground subtraction: {\tt Commander}, {\tt NILC}, {\tt SEVEM}, and {\tt SMICA}.
Each method provides a confidence mask which defines the region of the sky in which the CMB 
 maps can be used. We construct a combined mask as the union of these four confidence masks. This mask is applied on the CMB maps before 
 calculating the cross-correlation.
We will use the   {\tt SEVEM} map as default for the analysis. Nonetheless, we have also
tested the other maps to check the robustness of the results.
The test is described in more detail in Sec.~\ref{sec:tests}.

As the ISW effect is achromatic, for 
further cross-checks we also use  CMB maps at different frequencies. In particular we use maps at 100 GHz, 143 GHz, and 217 GHz.
The results using these maps are also described in Sec.~\ref{sec:tests}.

\section{Additional cosmological datasets}
\label{sec:cosmodata}
In the following we will perform parameter fits using the ISW data obtained with the cross-correlation.
Beside this, in some setups, we will also use other cosmological datasets in conjunction. 
In particular, we will employ the Planck 2015 public likelihoods\footnote{See \url{http://pla.esac.esa.int/pla/\#cosmology}} \citep{Ade:2015xua} and the corresponding \MP\ interfaces {\tt Planck\_highl\_lite} (for high-$\ell$ temperature),  {\tt Planck\_lowl} (for low-$\ell$ temperature and polarization), and {\tt Planck\_lensing} (CMB lensing reconstruction). 
{The accuracy of the {\tt Planck\_highl\_lite}  likelihood (which performs an
internal marginalization over all the nuisance parameters except one)
with respect to the full Planck likelihood (where the nuisance parameters
are not marginalized) has been tested in \cite{Aghanim:2015xee,Ade:2015rim} where the authors find that
the difference in the inferred cosmological parameters is at the level of 0.1 $\sigma$.}
Finally we will use BAO data from 
6dF \citep{Beutler:2011hx}, 
SDSS DR7 \citep{Ross:2014qpa} and
BOSS DR10\&11 \citep{Anderson:2013zyy},
which are implemented as {\tt bao\_boss} and {\tt bao\_boss\_aniso} in \MP.

%%%%%%%%%%%%%%%%%%%%%%%%%%%%%%%%%%%%%%%%%%%%%%%%%%

\section{Catalogs of Discrete Sources}
\label{sec:catmaps}

For the cross-correlation with the CMB,  as tracers of matter distribution we use five catalogs of extragalactic sources. As the ISW is a wide-angle effect, they were chosen to cover as large angular scales as possible, and two of them are all-sky. Furthermore, our study does not require exact, i.e.\ spectroscopic, redshift information, thus photometric samples are sufficient. Except for one case, the datasets employed here include individual \phzs\ for each source, which allows us to perform a tomographic approach by splitting the datasets into redshift bins.

\begin{figure}
\includegraphics[width = 0.48\textwidth]{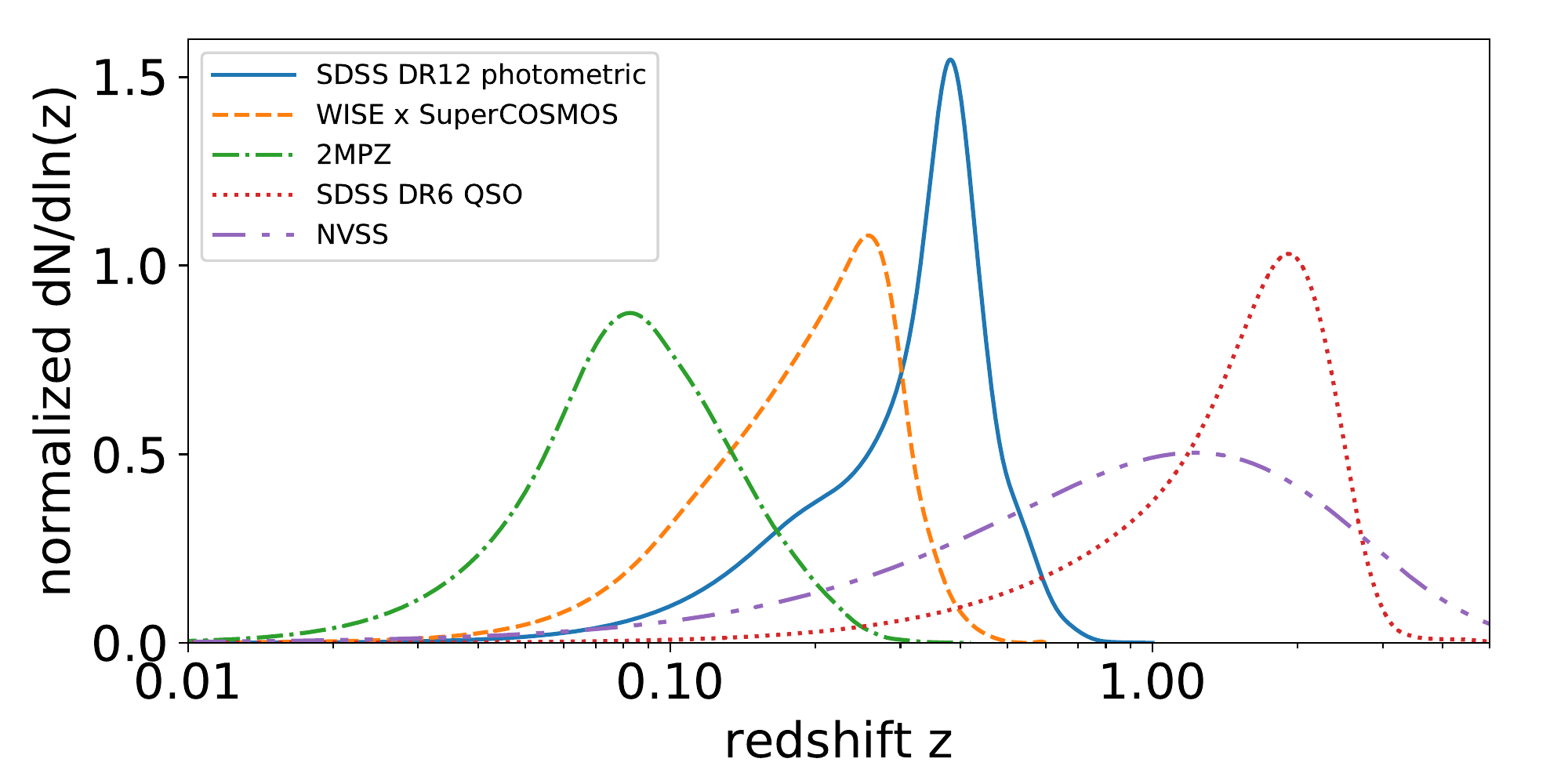}
\caption{Photometric redshift distributions for the five catalogs used for the cross-correlation. The  $dN/dz$ curves 
are normalized to a unit integral.
For the NVSS case the analytical approximation
described in the text  is used, since no redshifts information is available for the single catalog objects.}
\label{fig:dNdzs}
\end{figure}

The catalogs we use span a wide redshift range; see Fig.~\ref{fig:dNdzs} for their individual redshift distributions. Table  \ref{tab:catalogs} quantifies their properties (sky coverage, number of sources, mean projected density) as effectively used for the analysis, i.e., after applying  both the catalog and CMB masks. 

For a plot of the sky maps and masks of the catalogs described below, 
and for their 
detailed description,
see~\cite{Cuoco:2017bpv}.
Below we 
provide a short summary of the properties of the datasets.

\begin{table}
\begin{tabular}{lcrc}
\hline 
source & sky & number & mean surface \\
catalog & coverage & of sources & density [deg$^{-2}$] \\ 
\hline 
NVSS & 62.3\% & 431,724 & 67.2 \\ 
2MPZ & 64.2\% & 661,060 & 24.9 \\ 
WISE$\times$SCOS & 64.5\% & 17,695,635 & 665\\ 
SDSS DR12 & 18.7\% & 23,907,634 & 3095 \\ 
SDSS DR6 QSO  & 15.6\% & 461,093 & 71.8 \\ 
\hline 
\end{tabular} 
\caption{\label{tab:catalogs} Statistics of the catalogs used in the analysis. 
The numbers refer to the area of the sky effectively employed in the analysis,
i.e.,  applying both the catalog and CMB masks.}
\end{table}

\subsection{2MPZ}
\label{sec:2MPZ}
As a tracer of the most local LSS in this study we use the 2MASS Photometric Redshift catalog\footnote{Available from \url{http://ssa.roe.ac.uk/TWOMPZ.html}.} \citep[2MPZ,][]{bilicki14}. This dataset was built by merging three all-sky photometric datasets covering optical, near-infrared (IR), and mid-IR passbands: SuperCOSMOS  scans of UKST/POSS-II photographic plates \citep{peacock16}, 2MASS Extended Source Catalog \citep{jarrett2000}, and Wide-field Infrared Survey Explorer \citep[WISE,][]{wright10}. Photo-$z$s were subsequently estimated for all the included sources, by calibrating on overlapping spectroscopic datasets.

2MPZ includes $\sim 935,000$ galaxies over almost the full sky. Part of this area is however undersampled due to the Galactic foreground and instrumental artifacts, we thus applied a mask described in \cite{alonso15}. When combined with the CMB mask, this leaves over 660,000 2MPZ galaxies on $\sim64\%$ of the sky (Table~\ref{tab:catalogs}).

2MPZ provides the best-constrained \phzs\ among the catalogs used in this paper. They are practically unbiased ($\langle \delta z \rangle \sim 0$) and their random errors have RMS scatter $\sigma_{\delta z} \simeq 0.015$, to a good accuracy independent of redshift. We show the 2MPZ redshift distribution in Fig.~\ref{fig:dNdzs} with the dot-dashed green line; the peak is at $z\sim 0.06$ while the mean $\langle z \rangle \sim 0.08$. The overall surface density of 2MPZ is $\sim25$ sources per square degree.

For the tomographic analysis we split the catalog in three redshift bins: $z \in [0.00,0.105]$, $[0.105,0.195]$ and $[0.195, 0.30]$.
The first two include the bulk of the distribution, approximately divided into two comparable sub-samples,
while the third bin explores the tail of the $dN/dz$ where most of the ISW signal is expected.

A precursor of 2MPZ, based on 2MASS and SuperCOSMOS only, was used in a tomographic ISW analysis by \cite{francis_integrated_2010}, while an early application of 2MPZ itself to ISW tomography is presented in \cite{Steward14}. In both cases no significant ISW signal was found, consistent with expectations. Another ISW-related application of 2MPZ is presented in \cite{Planck15_ISW}, where it was applied to reconstruct ISW anisotropies caused by the LSS.

\subsection{WISE $\times$ SuperCOSMOS}
\label{sec:WIxSC}

The WISE~$\times$~SuperCOSMOS \phz\ catalog\footnote{Available from \url{http://ssa.roe.ac.uk/WISExSCOS.html}.} \citep[\WISC,][]{bilicki16} is an all-sky extension of 2MPZ obtained by cross-matching WISE and SuperCOSMOS samples. \WISC\ reaches roughly 3 times deeper than 2MPZ and has almost 30 times larger surface density. However, it suffers from more severe foreground contamination, and its useful area is $\sim70\%$ of the sky after applying its default mask. This is further reduced to $\sim65\%$ once the Planck mask is also used; the resulting \WISC\ sample includes about 17.5 million galaxies.

\WISC\ \phzs\ have overall mean error $\langle \delta z \rangle \sim 0$ and distance-dependent scatter of $\sigma_{\delta z} \simeq 0.033(1+z)$. The redshift distribution is shown in Fig.~\ref{fig:dNdzs} with the dashed orange curve. The peak is at $z\sim0.2$, and the majority of the sources are within $z<0.5$. In the tomographic approach, the \WISC\ sample is divided into four redshift bins:
$z \in [0.00,0.09]$, $[0.09,0.21]$, $[0.21, 0.30]$, and $[0.30, 0.60]$,
with approximately equal number of galaxies in each bin.

As far as we are aware, our study employs the \WISC\ dataset for an ISW analysis for the first time.
Various studies based using WISE have been performed in the past \citep{Goto:2012yc,Kovacs13,Ferraro:2014msa,Shajib:2016bes}.
However, the samples used there differed significantly from \WISC, and none included individual redshift estimates which would allow for redshift binning.

\subsection{SDSS DR12 photometric}
\label{sec:DR12}

Currently there are no all-sky \phz\ catalogs available reaching beyond \WISC. Therefore, in order to look for the ISW signal at $z>0.5$, we used datasets of smaller sky coverage. The first of them, with the largest number density of all employed in this paper, is based on the Sloan Digital Sky Survey Data Release 12 (SDSS-DR12) \phz\ sample compiled by  \cite{beck16}; to our knowledge, our study is its first application to an ISW analysis,
although earlier versions (DR 6 and DR 8) were used in \cite{giannantonio_combined_2008,2012MNRAS.426.2581G} (but without $z$ binning).

The parent SDSS-DR12 \phz\ dataset includes over 200 million galaxies. Here we however use a subsample described in detail in \cite{Cuoco:2017bpv}, which was obtained via appropriate cleaning as recommended by \cite{beck16}, together with our own subsequent purification of problematic sky areas. In particular, as the SDSS galaxies are distributed in two disconnected regions in the Galactic  south and north, with most of the area in the northern part, and uneven sampling in the south, we have excluded the latter region from the analysis. After additionally employing the Planck CMB mask, we were left with about 24 million SDSS DR12 sources with mean $\langle z \rangle = 0.34$ and mostly within $z < 0.6$. The resulting sky coverage is $\sim 19\%$  and the mean surface density is $\sim 3100$ deg$^{-2}$. The redshift distribution is  shown in Fig.~\ref{fig:dNdzs} with the solid blue line.

Thanks to the very large projected density of objects, we were able to split the SDSS-DR12 sample into several redshift bins, keeping low shot-noise in each shell.
For the tomographic analysis we  divided the dataset  into six bins: $z \in [0.0,0.1]$, $[0.1,0.3]$, $[0.3, 0.4]$, $[0.4, 0.5]$, $[0.5, 0.7]$ and $[0.7, 1.0]$.
The range  $z \in [0.1,0.3]$ is not subdivided further since this redshift range
is best covered by \WISC, 
where we already have sub-bins. The \phz\ accuracy of SDSS-DR12 depends on the `\phz\ class' defined by \cite{beck16}, and each class has an associated error estimate.
Our specific preselection detailed in \cite{Cuoco:2017bpv} leads to an effective \phz\ scatter of $\sigma_{\delta z} =0.022(1+z)$ based on the overall error estimates from \cite{beck16}.

\subsection{SDSS DR6 QSO}
\label{sec:QSO}

As a tracer of high-$z$ LSS, we use a catalog of photometric quasars (QSOs) compiled by \cite{richards09} from the SDSS DR6 dataset (DR6-QSO in the following), used previously in ISW studies by e.g. \cite{giannantonio_combined_2008,xia09} and \cite{2012MNRAS.426.2581G}. We apply the same preselections as in \cite{xia09}, and the resulting sample includes $6\times10^5$ QSOs on $\sim25\%$ of the sky. We exclude from the analysis three narrow stripes present in the south Galactic sky and use only the northern region.

The DR6-QSO sources are provided with \phzs\ spanning formally $0<z<5.75$ but with a relatively peaked $dN/dz$ and mean $\langle z \rangle \simeq 1.5$ (dotted red line in Fig.~\ref{fig:dNdzs}). For tomographic analysis, this QSO dataset will be split into three bins of $z\in [0.5,1.0]$, $[1.0,2.0]$, and $[2.0, 3.0]$,  selected in a way to have similar number of objects in each bin. We excluded the QSOs in the range $z\in [0.0,0.5]$ in order to minimize the overlap with the other catalogs in this redshift range. Nonetheless, there are very few DR6-QSO catalog objects at these redshifts, thus this choice has only a very minor impact on the results. 
The typical \phz\ accuracy of this dataset is $\sigma_{\delta z} \sim 0.24$ as reported by \cite{richards09}, and we will use this number for the extended modeling of underlying $dN/dz$s per redshift bin in Sec.~\ref{sec:tests}. 

\subsection{NVSS}
\label{sec:nvss}

The NRAO VLA Sky Survey \citep[NVSS,][]{1998AJ....115.1693C} is a catalog of radio sources, most of which are extragalactic. This sample has already been used for multiple ISW studies \citep[e.g.][]{boughn_cross_2001,Pietrobon:2006gh,Vielva:2004zg,McEwen:2006my,2008MNRAS.386.2161R}.
The dataset covers the whole sky available for the VLA instrument; after appropriate cleanup of likely Galactic entries and artifacts, the NVSS sample includes $\sim5.7\times10^5$ objects flux-limited to $> 10$ mJy, located at declinations $\delta\gtrsim -40^\circ$ and Galactic latitudes $|b| > 5^{\circ}$. This is the only of the datasets considered in this work which does not provide even crude redshift information for the individual sources. We thus use it without tomographic binning and, where relevant, assume its $dN/dz$ to follow the model of \cite{dezotti10} (purple short-long-dashed line in Fig.~\ref{fig:dNdzs}). This sample spans the broadest redshift range of all the considered catalogs, namely $0<z<5$.

%%%%%%%%%%%%%%%%%%%%%%%%%%%%%%%%%%%%%%%%%%%%%%%%%

\subsection{Masks}

In the correlation of the CMB with each catalog we use the 
CMB mask, described in Sec.~\ref{sec:cmbmaps}, combined with 
the specific mask of the given catalog. 
Beside this, we define specific masks which we use when combining the signal 
 from the different catalogs in order to circumvent including the same
information twice, 
and to avoid the need to take into account the cross-correlations between various tracers of the same LSS.
We proceeded in the following way.
\begin{itemize}
\item  SDSS catalogs (i.e. SDSS DR6 QSOs and SDSS DR12 galaxies) are used  without additional masks.
When combining the information with other catalogs we, however, exclude the
first SDSS DR12 bin, since the region  $z \in [0.0,0.1]$ is best covered by 2MPZ.

\item  To avoid correlations with the SDSS catalogs, when using all the remaining ones (i.e. NVSS, 2MPZ, \WISC)
we apply a mask which is a 
complementary of
the joint mask of SDSS DR12 galaxies and SDSS DR6 QSOs
(in short, SDSS mask in the following).

\item For 2MPZ and \WISC\ it is not possible to define mutually exclusive masks since
both these datasets cover practically the same
part of the sky.  Nonetheless, we use them together, since  \WISC\ was 
built excluding most of 
the objects already
contained in 2MPZ \citep{bilicki16}. The two catalogs, thus, have practically no common sources. 
In this way the correlation among the two datasets
is significantly suppressed, although
not totally, since both trace the same underlying LSS in the overlapping redshift ranges.
We will, however, not consider the first bin, $z \in [0.0,0.1]$, of \WISC\ in the combined analysis since
in this redshift range 2MPZ has better redshift determination and basically no
stellar contamination.
Nonetheless, as we will show in Sec.~\ref{sec:method}, the evidence for ISW in the range $z \in [0.0,0.2]$,
where 2MPZ and \WISC\ have most of the overlap, is very small, so, in practice,
this has only a marginal effect on the final ISW significance.

\item Similarly, also for NVSS, 
2MPZ and \WISC\ it is not possible to define a
mutually exclusive mask due to the large common area of the sky.
In this
case, we note that
2MPZ and \WISC\ cover
only the low redshift tail
of NVSS. Thus, the overlap and correlation among them is minimal.

\end{itemize}

We will thus use the above setup when reporting combined significances of the ISW
from the different catalogs.
For simplicity, we will use the same setup also to derive auto-correlations of
the single catalogs. In this case the significances could be increased
slightly for NVSS, 2MPZ, and \WISC\ if their proper masks were used, but we
checked that the improvement is only marginal.

\begin{figure*}
\includegraphics[width=0.45\textwidth]{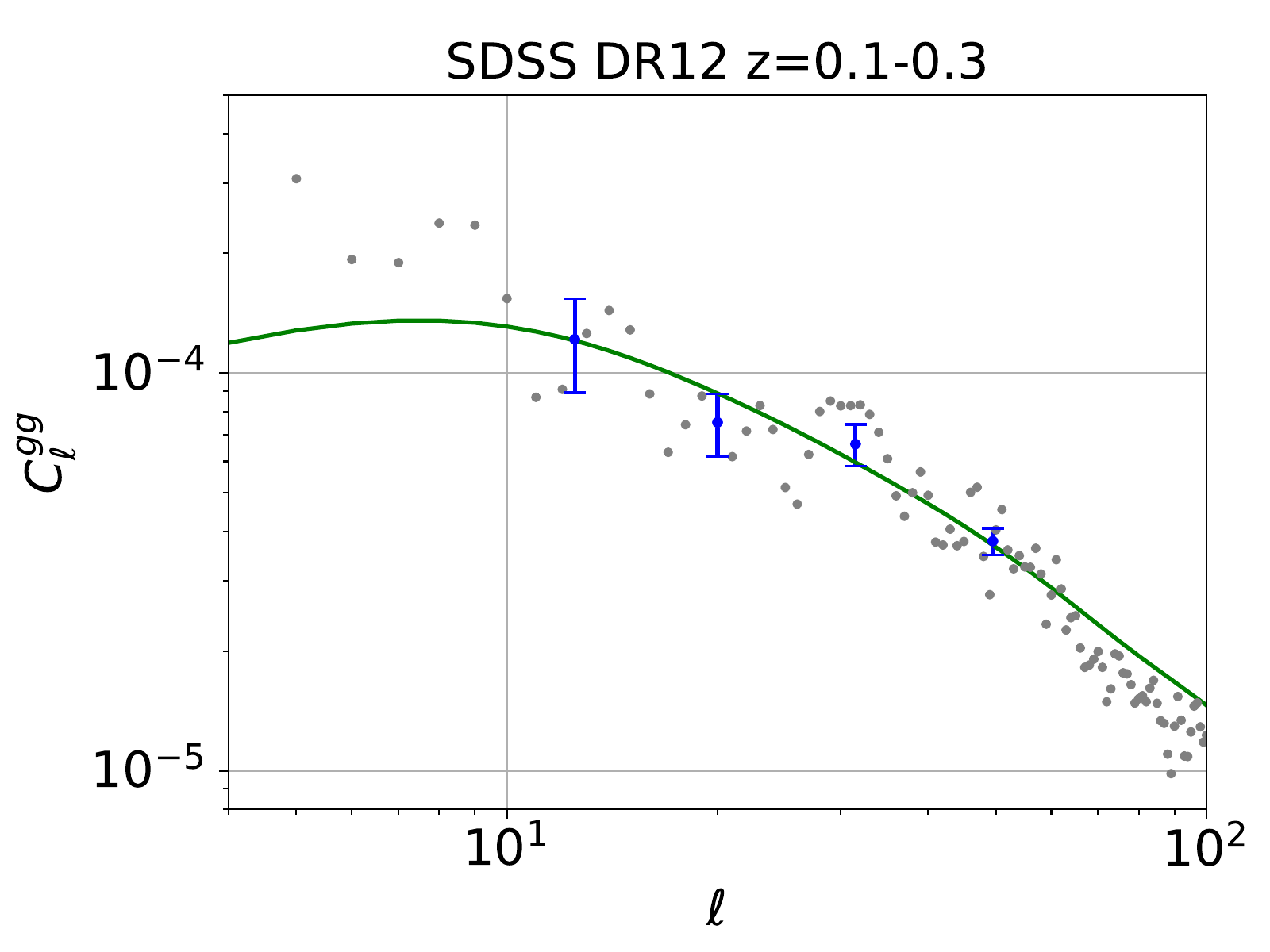}
\includegraphics[width=0.45\textwidth]{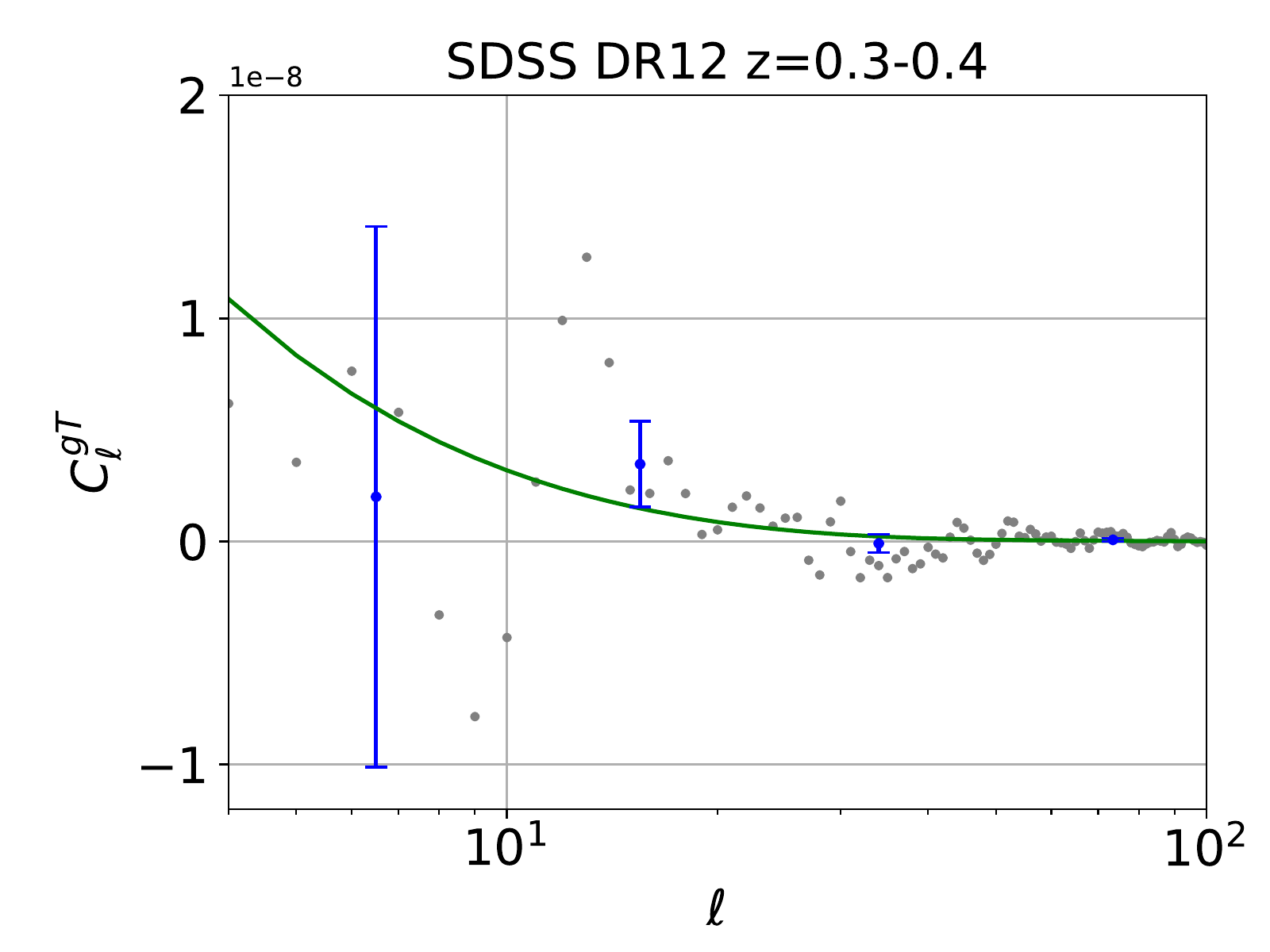}
\caption{Left: Example of measured source catalog auto-correlation and best-fit model with free galaxy bias, referring to the case of SDSS-DR12 in the labeled $z$ bin. Right: example of measured cross-correlation between sources and CMB temperature and best-fit model, referring to the case of SDSS-DR12 in the labeled $z$ bin.   Dots refer to the measured single multipoles, while data points with error bars refer to binned measurements. }
\label{fig:biasexample}
\end{figure*}

\section{Cross-Correlation Analysis}
\label{sec:corranalysis}

In the previous section we have presented the catalogs of extragalactic objects that
we use in the analysis. Their input format is that of a 2D pixelized map of object counts
$n(\hat{\Omega}_i)$, where $\hat{\Omega}_i$ specifies the angular coordinate of the $i$-th pixel.
For the cross-correlation analysis we consider  maps of normalized counts $n(\hat{\Omega}_i)/\bar{n}$, where $\bar{n}$
is the mean object density in the unmasked area, and CMB temperature maps, also pixelized with a matching angular resolution.

In our  analysis we compute both the angular 2-point cross-correlation function, CCF,
$w^{(cT)}(\theta)$, and its harmonic transform, the
angular power spectrum $\bar C_\ell^{(cT)}$, CAPS.
However, we restrict the quantitative analysis to the CAPS only.
The reason for this 
choice is that the CAPS has the advantage that  
different multipoles are almost uncorrelated, especially after binning.
Their covariance matrix is therefore close to diagonal, which simplifies the comparison  between models and data.
Similarly, we compute also the auto-correlation power spectrum of the catalogs (APS)
and the related auto-correlation function (ACF).

\begin{table*}
\begin{tabular}{cccc|cc|cc}
\hline
 catalog   & z           & b                 &   $\chi^2_{min}$ & $b_{\rm Halofit}$                 &   $\chi^2_{min}$& $b_{\rm Halofit+\sigma_{\delta z}}$                 &   $\chi^2_{min}$\\
\hline
  SDSS        & 0-0.1       & $ 0.70 \pm 0.02 $ &             3.59& $ 0.69 \pm 0.02 $ &             3.76& $ 0.71 \pm 0.02 $ &             4.11 \\
             & 0.1-0.3     & $ 1.03 \pm 0.03 $ &             1.71 & $ 1.03 \pm 0.03 $ &             1.68& $ 1.02 \pm 0.03 $ &             1.63 \\
             & 0.3-0.4     & $ 0.88 \pm 0.03 $ &             0.64 & $ 0.88 \pm 0.03 $ &             0.63& $ 0.87 \pm 0.03 $ &             0.61 \\
             & 0.4-0.5     & $ 0.79 \pm 0.02 $ &             4.84 & $ 0.80 \pm 0.02 $ &             4.65& $ 0.84 \pm 0.03 $ &             4.99 \\
             & 0.5-0.7     & $ 1.14 \pm 0.04 $ &             6.16 & $ 1.13 \pm 0.04 $ &             5.86& $ 1.23 \pm 0.04 $ &             6.35 \\
             & 0.7-1       & $ 1.02 \pm 0.11 $ &            15.04 & $ 1.03 \pm 0.11 $ &            14.99& $ 1.23 \pm 0.13 $ &            15.16  \\
 WIxSC       & 0-0.09      & $ 0.62 \pm 0.03 $ &             0.46 & $ 0.60 \pm 0.03 $ &             0.38 & $ 0.57 \pm 0.03 $ &             0.28 \\
             & 0.09-0.21   & $ 0.89 \pm 0.03 $ &             2.38 & $ 0.87 \pm 0.03 $ &             2.67 & $ 0.88 \pm 0.03 $ &             2.62 \\
             & 0.21-0.3    & $ 0.80 \pm 0.02 $ &            10.07 & $ 0.81 \pm 0.02 $ &            10.14 & $ 0.80 \pm 0.02 $ &            10.09 \\
             & 0.3-0.6     & $ 0.96 \pm 0.03 $ &             5.62 & $ 1.03 \pm 0.04 $ &             5.88 & $ 1.24 \pm 0.04 $ &             5.55 \\
 QSO         & 0-1         & $ 1.55 \pm 0.16 $ &             5.9  & $ 1.56 \pm 0.16 $ &             5.93 & $ 1.45 \pm 0.15 $ &             4.97 \\
 		         & 0.5-1       & $ 1.54 \pm 0.26 $ &             3.09 & $ 1.55 \pm 0.26 $ &             3.07 & $ 1.52 \pm 0.26 $ &             3.07 \\
             & 1-2         & $ 2.64 \pm 0.27 $ &             3.61 & $ 2.66 \pm 0.27 $ &             3.59 & $ 2.61 \pm 0.27 $ &             3.6 \\
             & 2-3         & $ 3.19 \pm 0.50 $ &             7.08 & $ 3.21 \pm 0.51 $ &             7.05 & $ 3.51 \pm 0.55 $ &             7.08 \\
 2MPZ        & 0-0.105     & $ 1.09 \pm 0.03 $ &            4.41 & $ 1.03 \pm 0.03 $ &             1.30 & $ 1.03 \pm 0.03 $ &             1.26 \\
             & 0.105-0.195 & $ 1.12 \pm 0.04 $ &             2.00  & $ 1.12 \pm 0.04 $ &             2.07 & $ 1.19 \pm 0.04 $ &             2.17 \\
             & 0.195-0.3   & $ 1.84 \pm 0.09 $ &            6.54  & $ 1.86 \pm 0.09 $ &            6.67& $ 2.03 \pm 0.09 $ &             6.34  \\
 NVSS        & 0-6         & $ 2.18 \pm 0.08 $ &             3.02 & $ 2.04 \pm 0.08 $ &             0.64  & ---&---\\
\hline
\hline
 catalog   & z     & b                 &   $\chi^2_{min}$&$b_{\rm Halofit}$                 &   $\chi^2_{min}$ \\
\hline
  SDSS        & 0-1   & $ 1.34 \pm 0.04 $ &             1.25& $ 1.35 \pm 0.04 $ &             1.27 & $ 1.39 \pm 0.04 $ &             1.59  \\
 WIxSC       & 0-0.6 & $ 1.08 \pm 0.03 $ &             3.15 & $ 1.07 \pm 0.03 $ &             3.74& $ 1.12 \pm 0.03 $ &             3.94  \\
 QSO         & 0-3   & $ 2.67 \pm 0.23 $ &             2.77 & $ 2.68 \pm 0.23 $ &             2.76& $ 2.66 \pm 0.23 $ &             2.4   \\
 2MPZ        & 0-0.3 & $ 1.23 \pm 0.04 $ &             5.19& $ 1.17 \pm 0.04 $ &             2.04  & $ 1.20 \pm 0.04 $ &             1.98 \\
\hline
\end{tabular}

%\begin{tabular}{lrrr}
%\hline
% catalogue   &   $l_{min}$ &   $l_{max}$ &   $n_{bins}$ \\
%\hline
% SDSS        &         10 &        60 &            4 \\
% WIxSC       &         10 &        50 &            4 \\
% QSO         &         10 &        60 &            4 \\
% 2MPZ        &         10 &        50 &            4 \\
% NVSS        &         10 &       100 &            4 \\
%\hline
%\end{tabular}
\caption{\label{tab:biases}Linear biases for the different redshift bins of the various catalogs fitted for a fixed cosmological model. 
The reported errors on the bias are derived from the fit of Eq.\ \eqref{eq:chi21}; goodness of fit is quantified in the relevant $\chi^2$ columns. The $\chi^2$ refers to the case of a fit with 4 bins in the multipole range 10-60.}
\end{table*}

We use the  {\it PolSpice}\footnote{See \url{http://www2.iap.fr/users/hivon/software/PolSpice/}}
statistical toolkit \citep{szapudi01,chon04,efstathiou04,challinor05} to estimate the correlation functions and power spectra. 
{\it PolSpice} automatically corrects for the effect of the mask.
In this respect, we point out that the effective geometry of the mask used for the correlation analysis is
obtained by combining that of the CMB maps with those of each catalog of astrophysical objects.
The accuracy of the {\it PolSpice} estimator has been assessed in \cite{xia15} by comparing the measured CCF
with the one computed using the popular Landy-Szalay  method \citep{ls93}. The two were found to be in very good agreement.
 {\it PolSpice} also provides the  covariance matrix for the angular power spectrum,
 $\bar V_{\ell\ell'}$ \citep{2004MNRAS.349..603E}.

For the case of source catalog APS a further step is required. 
Contrary to the CAPS, the APS contains shot noise due to the discrete nature  
of the objects in the map. The shot noise is constant in multipole
and can be expressed as $C_{\rm N}=4 \pi f_{\rm sky}/N_{\rm gal}$,
where  $f_{\rm sky}$ is the fraction of sky covered by the catalog in the unmasked area
and $N_{\rm gal}$ is the number of catalog objects, again in the unmasked area.
The above shot-noise has been  subtracted  from our final estimated APS.

The Planck  Point Spread Function and the map pixelization affect in principle the
estimate of the CAPS. However, the CAPS contains information
on the ISW only up to $\ell \sim100$ where these effects are negligible.
We will thus not consider them further.

Finally, to reduce the correlation in nearby multipoles induced by the angular mask, we use 
an $\ell-$binned version of  the measured CAPS.
The number of bins and the maximum  and minimum $\ell$ used in the analysis will be varied
to assess the robustness of the results.
We indicate the  binned CAPS with the same symbol as the unbinned one, $C_\ell^{(cT)}$.
It should be clear from the context  which one is used.
The $C_\ell^{(cT)}$ in each bin is given by the simple unweighted average
of the $C_\ell^{(cT)}$ within the bin.
For the binned $C_\ell^{(cT)}$ we build
the corresponding covariance matrix as a block average of the unbinned covariance matrix $V_{\ell\ell'}$, i.e., $\sum_{\ell\ell'} V_{\ell\ell'}/\Delta\ell/\Delta\ell'$,
where $\Delta\ell, \Delta\ell'$ are the widths of the two multipole bins, and $\ell, \ell'$ run over
the multipoles of the first and the second bin.
The binning procedure is very efficient in removing correlation among nearby multipoles,
resulting in a block covariance matrix that is, to a good approximation, diagonal. 
We will use nonetheless the full block covariance matrix in the following,
although we have checked  that using  the diagonal only gives minor differences.
When showing CAPS plots, however, we use the diagonal terms to plot
the errors on the $C_\ell$,   $\left(\Delta C_{\ell} \right)^2=\sum_{\ell \ell'} V_{\ell\ell'}/\Delta\ell^2$,
where the sum runs over the multipoles of the bin contributing to $C_\ell$.

\begin{table*}
\begin{tabular}{lllrrrr}
\hline
 catalog   & z           & $A_{\rm ISW}$                  &   $\frac{A}{\sigma_A}$ &   $\chi^2_0$ &   $\chi^2_{min}$ &   $\Delta{\chi^2}$ \\
\hline
SDSS        & 0-0.1       & $ 0.23 \pm 3.35 $ &              0.07 &    1.224 &        1.219 &        0.005\\
	        & 0.1-0.3     & $ 0.90 \pm 1.03 $ &               0.87 &    3.89 &        3.12 &          0.76 \\
             & 0.3-0.4     & $ 1.94 \pm 1.24 $ &               1.57  &    4.47 &        2.01 &          2.45  \\
             & 0.4-0.5     & $ 2.77 \pm 1.36 $ &               2.03  &    6.57 &        2.45 &          4.12  \\
             & 0.5-0.7     & $ 2.59 \pm 1.13 $ &               2.28  &    9.28 &        4.06 &          5.22  \\
             & 0.7-1       & $ 1.00 \pm 2.72 $ &               0.37 &    6.76 &        6.62 &          0.13 \\
 WIxSC       & 0-0.09      & $ 5.24 \pm 4.86 $ &              1.08      &    2.84    &        1.68    &        1.16       \\             
 	          & 0.09-0.21   & $ 0.34 \pm 1.01 $ &               0.33     &    4.63    &        4.52    &          0.11     \\
             & 0.21-0.3    & $ 1.04 \pm 0.94 $ &               1.1      &    3.62    &        2.4     &          1.21     \\
             & 0.3-0.6     & $ 1.33 \pm 0.94 $ &               1.41     &    4.91    &        2.92    &          1.99     \\
 QSO         & 0-1         & $ 2.50 \pm 1.64 $ &              1.52      &    5.95    &        3.64    &        2.31       \\
             & 0.5-1       & $ 2.39 \pm 1.65 $ &               1.45     &    7.46    &        5.34    &          2.11     \\
             & 1-2         & $ 2.49 \pm 1.64 $ &               1.52     &    3.99    &        1.68    &          2.31     \\
             & 2-3         & $ 1.83 \pm 4.80 $ &               0.38     &    3.11    &        2.96    &          0.14     \\
 2MPZ        & 0-0.105     & $ 1.25 \pm 3.43 $ &               0.36     &    1.26    &        1.13    &          0.13     \\
             & 0.105-0.195 & $ 0.53 \pm 1.77 $ &               0.3      &    1.12    &        1.03    &          0.09     \\
             & 0.195-0.3   & $ 1.04 \pm 1.47 $ &               0.71     &    1.66    &        1.16    &          0.5      \\
 NVSS        & 0-6         & $ 1.70 \pm 0.57 $ &               2.97     &   14.9     &        6.11    &          8.79     \\
\hline
\end{tabular}

\vspace{0.5cm}

\begin{tabular}{llrrrr}
\hline
 catalog                    & $A_{\rm ISW}$                 &   $\frac{A}{\sigma_A}$ &   $\chi^2_0$ &   $\chi^2_{min}$ &   $\Delta{\chi^2}$ \\
\hline
 SDSS                         & $ 1.89 \pm 0.57 $ &                   3.29 &      30.96 &          20.11 &              8.46 \\
 WIxSC                        & $ 0.93 \pm 0.56 $ &                   1.67 &      13.16 &          10.39 &              2.76 \\
 Quasars                      & $ 2.41 \pm 1.13 $ &                   2.13 &      14.55 &          10.01 &              2.99 \\
 2MPZ                         & $ 0.87 \pm 1.07 $ &                   0.81 &       4.04 &           3.38 &              0.65 \\
 SDSS+WIxSC                   & $ 1.39 \pm 0.40 $ &                   3.49 &      44.12 &          31.94 &             11.21 \\
 SDSS+Quasars                 & $ 1.99 \pm 0.51 $ &                   3.9  &      45.51 &          30.28 &             11.45 \\
 SDSS+WIxSC+Quasars           & $ 1.51 \pm 0.38 $ &                   4    &      58.67 &          42.66 &             14.2  \\
 SDSS+WIxSC+Quasars+NVSS+2MPZ & $ 1.51 \pm 0.30 $ &                   5    &      77.61 &          52.61 &             22.16 \\
\hline
 SDSS+WIxSC+Quasars+NVSS      & $ 1.56 \pm 0.31 $ &                   4.97 &      73.57 &          48.85 &             21.52 \\
 SDSS+WIxSC+NVSS+2MPZ         & $ 1.44 \pm 0.31 $ &                   4.6  &      63.06 &          41.92 &             19.17 \\
 SDSS+Quasars+NVSS+2MPZ       & $ 1.75 \pm 0.36 $ &                   4.88 &      64.45 &          40.67 &             19.41 \\
 SDSS+WIxSC+Quasars+2MPZ      & $ 1.44 \pm 0.36 $ &                   4.04 &      62.71 &          46.35 &             14.85 \\
 WIxSC+Quasars+NVSS+2MPZ      & $ 1.36 \pm 0.35 $ &                   3.84 &      46.65 &          31.9  &             13.71 \\
\hline
\end{tabular}
%\begin{tabular}{lrrr}
%\hline
% catalogue   &   $l_{min}$ &   $l_{max}$ &   $n_{bins}$ \\
%\hline
% SDSS        &          4 &       100 &            4 \\
% WIxSC       &          4 &       100 &            4 \\
% QSO         &          4 &       100 &            4 \\
% 2MPZ        &          4 &       100 &            4 \\
% NVSS        &          4 &       100 &            4 \\
%\hline
%\end{tabular}
\caption{\label{tab:AISW}Summary of the measured ISW and related significances for the single redshift bins of each catalogs (top table) and for various combinations of the catalogs, where, in the latter case, also the individual redshift bins of each catalog were combined (bottom table). 
{The last five rows give the cases in which a single catalog is excluded from the fit each time.}
The $\chi^2$ refers to the case of a fit with 4 bins in the multipole range 4-100.}
\end{table*}

\begin{table*}
\begin{tabular}{llrrrr}
\hline
 catalog                    & $A_{\rm ISW}$                 &   $\frac{A}{\sigma_A}$ &   $\chi^2_0$ &   $\chi^2_{min}$ &   $\Delta{\chi^2}$ \\
\hline
 SDSS                         & $ 0.96 \pm 0.65 $ &                   1.49 &         5.3  &             3.09 &                2.21 \\
 WIxSC                        & $ 0.62 \pm 0.61 $ &                   1.02 &         5.28 &             4.24 &                0.65 \\
 Quasars                      & $ 1.28 \pm 0.63 $ &                   2.03 &         5.55 &             1.41 &                3.94 \\
 2MPZ                         & $ 0.90 \pm 2.32 $ &                   0.39 &         0.87 &             0.72 &                0.15 \\
 NVSS                         & $ 1.70 \pm 0.57 $ &                2.97    &     14.9     &          6.11    &             8.79    \\
 SDSS+WIxSC                   & $ 0.94 \pm 0.42 $ &                   2.23 &        18.47 &            13.48 &                4.96 \\
 SDSS+Quasars                 & $ 1.32 \pm 0.56 $ &                   2.35 &        19.85 &            14.33 &                5.2  \\
 SDSS+WIxSC+Quasars           & $ 1.12 \pm 0.40 $ &                   2.84 &        33.02 &            24.97 &                7.95 \\
 SDSS+WIxSC+Quasars+NVSS      & $ 1.31 \pm 0.33 $ &                   4.02 &        47.91 &            31.76 &               15.27 \\
 SDSS+WIxSC+Quasars+NVSS+2MPZ & $ 1.27 \pm 0.31 $ &                   4.08 &        51.95 &            35.28 &               15.92 \\
\hline
\end{tabular}
\caption{\label{tab:AISWnocatbin}Summary of the measured ISW and related significances for the the case of no redshift binning of the catalogs. Various combinations of the catalogs are shown. }
\end{table*}

\begin{table}
\begin{center}
\begin{tabular} { l  c}

 Parameter &  68\% limits\\
\hline
{\boldmath$10^{-2}\omega_{b }$} & $2.226\pm 0.019            $\\
{\boldmath$\omega_{cdm }  $} & $0.1187\pm 0.0012          $\\
{\boldmath$n_{s }         $} & $0.9674\pm 0.0043          $\\
{\boldmath$10^{-9}A_{s }  $} & $2.152\pm 0.052            $\\
{\boldmath$h              $} & $0.6780\pm 0.0053          $\\
{\boldmath$\tau_{\rm reio }   $} & $0.068\pm 0.013            $\\
{\boldmath$10^{-2}A_{\rm Planck }$} & $100.01\pm 0.25  $\\
\hline
{\boldmath$\Omega_{\Lambda }$} & $0.6916\pm 0.0071          $\\
%\hline
\end{tabular}
\end{center}
\caption{\label{tab:PlanckBAO}Results of the \MP{} fit to Planck + BAO data only. 
Here $\Omega_{\Lambda }$ is a derived parameter and $A_{\rm Planck }$ a Planck nuisance parameter.}
\end{table}

\section{Derivation of the ISW significance}
\label{sec:method}

In this section we illustrate the two methods we use to
quantify the significance of the ISW. We will assume for the first method a flat $\Lambda$CDM
model with  cosmological parameters
$\Omega_{\rm b} h^2 = 0.022161$,
$\Omega_{\rm c} h^2 = 0.11889$, $\tau= 0.0952$, $h = 0.6777$, $\ln{10^{10}A_{\rm s}} = 3.0973$ at $k_0=0.05$ Mpc$^{-1}$,
and $n_{\rm s} =0.9611$, in accordance with the most recent Planck results \citep{Ade:2015xua}.

\subsection{Method 1}
This is the usual method employed in previous publications  to study the significance of the ISW.
In this case we fix the cosmological model to the best-fit  one measured by Planck, and we derive with \CLASS{} the matter
power spectrum $P(k,z)$, which is used to calculate the expected auto-correlation $C_{\ell}$ for each catalog for the appropriate redshift bin.
The measured auto-correlation is then used to fit the linear bias, as a proportionality constant in the predicted  $C_{\ell}$.
An example of this fit is shown in the left panel of Fig.~\ref{fig:biasexample}. 
A simple $\chi^2$ over the bins of the auto-correlation is used for the fit:
\begin{equation}
  \chi^2_{AC} \equiv \chi^2(b^2)=\sum_{\rm \ell~bins} {\frac{(\hat{C}^{\rm c}_{\ell}(b^2) - C^{\rm  c}_{\ell})^2}{(\Delta C^{\rm  c}_{\ell})^2} } \, ,
\label{eq:chi21}
\end{equation}
where $\hat{C}^{\rm  c}_{\ell}$ and $C^{\rm  c}_{\ell}$  represent the model and the measured CAPS, and the sum is over all $\ell$ bins.

As mentioned in Sec.~\ref{sec:corranalysis} we tested that the use of the full covariance matrix
with respect to the diagonal expression for the $\chi^2$ above does not give appreciable differences.
Table~\ref{tab:biases} summarizes the various measured biases, and the default binning used for the auto-correlations.
We tested the robustness of the fitted biases changing the number of bins from 4 to 6 and 
the maximum $\ell$ from 40-80, and we found stable results, with variations of the order of 10\%.
A maximum $\ell$ of 40-80 is chosen since above this range typically non-linear effects
become significant. 
As the default case, we use 4 bins in the range 10-60.

As a further test we checked the impact of using non-linear corrections to the matter power
spectrum to model the auto-correlation of the catalogs.
The non-linear corrections were implemented through the version of  Halofit~\citep{Takahashi:2012em} implemented in \CLASS{} v2.6.1. 
The last 2 columns of Table~\ref{tab:biases}  show the bias and the best-fit $\chi^2$ obtained
using the non-linear model. It can be seen that the biases obtained with and without non-linear corrections
are fully compatible. The only exception is the first redshift bin of 2MPZ where the best-fit bias changes
at the $2\ \sigma$ level. More importantly,  the fit shows a visible improvement from $\chi^2\sim4.4$
to $\chi^2\sim1.3$.
This is expected, since at these low redshifts  even the small $\ell$s correspond
mostly to small, non-linear, physical scales.
As we show below, however, 2MPZ presents little or no imprint of the ISW effect, so we conclude
that the use of the linear  $P(k,z)$ has a negligible impact on the study of the ISW effect
in this analysis.

As an additional comment about the galaxy biases reported in Table~\ref{tab:biases},  we note that the $\sim$10\% variation quoted above is
typically larger than the statistical errors given in that Table,
the latter being sometimes only a few \%; this means that the bias errors are systematics- rather statistics-limited. 
Also, in some cases, for example most notably in the $z\in [0.7, 1.0]$ bin of SDSS DR12 galaxies,
the minimum $\chi^2$ is quite large, indicating a poor quality of the fit. This is also visible in some of the AC plots provided in Appendix \ref{acplots}. This is likely related to non-uniformities of the catalogs,
which are more severe in the tails of the redshift distribution, which in particular leads to excessively large measured low-$\ell$ AC power in some cases.
Therefore, in such instances, the small statistical errors on $b$ should be taken with care.
In general, we stress that the precise determination of the bias error is not
crucial in this analysis, which is, instead, focused on the determination
of the significance of the ISW effect. To this aim, the error, and even the value of the bias,
have  only a limited impact. See further discussion below.

In the second step, all the galaxy biases are fixed to best-fit values previously derived, and only the
measured cross-correlations are used. At this point only a single parameter $A_{\rm ISW}$
is fitted using as data either a single measured cross-correlation or a combination of them,
with the   $\chi^2$ statistics:
\begin{equation}
  \chi^2_{CC} \equiv \chi^2(A_{\rm ISW})=\sum_{z-\rm{bins} }  \sum_{\rm cat.} \sum_{\rm \ell~bins} {\frac{(A_{\rm ISW} \hat{C}^{\rm T c}_{\ell} - C^{\rm T c}_{\ell})^2}{(\Delta C^{\rm T c}_{\ell})^2} } \, ,
\label{eq:chi22}
\end{equation}
where $\hat{C}^{\rm T c}_{\ell}$ and $C^{\rm T c}_{\ell}$  represent the model (for the standard $\Lambda$CDM cosmological model considered)
and the measured catalog -- CMB temperature
cross-correlation for a given redshift bin, respectively, the sum is over all the $\ell$ bins, and over different catalogs and different redshift bins.
The linear parameter $A_{\rm ISW}$ quantifies the agreement with the above standard model expectation.
In the denominator we use the error provided by \textit{Polspice} discussed
in the previous section. In principle, however, one should use an error where the  model is taken into account.
For the case of binned data, however, this is a small effect
(see for example discussion in \cite{Fornasa:2016ohl}).

An example of measured cross-correlation and  fit to
the model is shown in the right panel of Fig.~\ref{fig:biasexample}. 
Table~\ref{tab:AISW} summarizes the results of the fit for each single $z$-bin of each catalog, 
for each catalog combining the different  $z$-bins, and for different combinations of the catalogs,
where, again, for each catalog $z$-binning has been used. 
For the default case we use four multipole 
bins between 
$\ell$ of 4 and 100,
but, again, we have verified that the results are stable when changing the number of bins from 4 to 6 and
the maximum $\ell$ from 60 to 100, which is expected,
since the ISW effect is rapidly decreasing as a function of $\ell$, and not much signal is 
expected beyond  $\ell \sim 60$.

To quantify the significance of the measurement we use as test-statistic
the quantity
\begin{equation}
\rm{TS} = \chi^2(0)-\chi_{min}^2 \; ,
\end{equation}
 where $\chi_{min}^2$ is the minimum $\chi^2$,
and $\chi^2(0)$  is the $\chi^2$  of the null hypothesis of no ISW effect, i.e.\ of the case $A_{\rm ISW} = 0$.
TS is expected to behave asymptotically as a $\chi^2$ distribution with a number of degrees of freedom
equal to the number of fitted parameters, allowing us to derive the significance level
of a measurement based on the measured TS.
In this case, since there is only one fitted parameter, the significance in sigma is
just given by $\sqrt{TS}$.
From Table~\ref{tab:AISW} one can see that the maximum significance achieved with Method 1 when
using all the catalogs in combination is $\sqrt{22.16}=4.7 \sigma$. 
From the different results it can also be seen that the main contribution
is given by NVSS and SDSS DR12 galaxies. We remind that the cross-correlation with NVSS
is calculated masking the area of the sky used to calculate the correlation with SDSS.
The two are, thus, completely independent. 
A smaller, and comparable, contribution, is given by \WISC\ and SDSS-QSO.
2MPZ instead show basically no sign of ISW, which is expected given the very
low $z$ range.
In the Table we also include   a column with the signal to noise (S/N=$A/\sigma_A$) of the ISW measurement
for comparison with other works since this quantity is often reported in the literature.
We can see that the global fit reaches a S/N of 5.

We also show in Table~\ref{tab:AISWnocatbin} the result of the fit when no
redshift binning is used. It is clear that without 
such binning the significance 
of the ISW  is significantly reduced, especially for SDSS-DR12 and  \WISC\ ,
while the significance of SDSS-QSO is almost unchanged.
Overall, when no redshift binning is used, the significance of the ISW effect
combining all the catalog is 4.0 $\sigma$, which is significantly reduced
with respect to  the 4.7 $\sigma$  achieved with the redshift binning.

As mentioned above, the derived significance is very weakly dependent on the exact values
of the biases used. For the case of a single catalog redshift bin, this is clear looking at 
Eqs.~\eqref{eq:isw2} \& \eqref{eq:chi22}, which show that the ISW signal is linear in $b$.
The fit to the cross-correlation thus constraints the quantity $b A_{\rm ISW}$ and the value
of $b$ is not important for the determination of the significance, although, clearly,
is relevant in determining the value of $A_{\rm ISW}$. When several redshift bins and catalogs are used,
the above argument is not exact anymore, but remains approximately valid. 
We checked, indeed, that using different biases derived from the autocorrelation fits using different $\ell_{max}$ 
and different number of $\ell$ bins, gives unchanged significances.

{We can see that the preferred $A_{\rm ISW}$ value from the combined fit is slightly larger than 1 at a bit more 
than 1 $\sigma$.    In the single catalog fits, both NVSS, QSOs and SDSS seem to drive the   $A_{\rm ISW}$ value
above 1.  This is confirmed in the last 5 rows of Table~\ref{tab:AISW} where different fits are performed
each time excluding only one catalog and  combining the remaining four. All the fits
give compatible results with  $A_{\rm ISW}$  above 1 at around  1 $\sigma$ or a bit more.
This result is further scrutinized in Section~\ref{sec:DE} where we investigate
if this slight difference of $A_{\rm ISW}$ from 1 can be interpreted as an indication of 
departure of DE from the simple case of a cosmological constant. }

\begin{table*}
\begin{center}
\begin{tabular}{l ||ccc}
\hline
Parameter &  AC+CC&CC&PL+AC+CC\\
\hline
{\boldmath$10^{-2}\omega_{b }$} & $2.230\pm 0.014$		& $2.229\pm 0.013            			$& $2.228\pm 0.020$				\\
{\boldmath$\omega_{cdm }  $} 	  & $0.1060\pm 0.0062$   & $0.1045^{+0.0093}_{-0.023} 	$& $0.1185\pm 0.0012$			\\
{\boldmath$n_{s }         $} 			  & $0.9670\pm 0.0039$   & $0.9667\pm 0.0036          		$& $0.9678\pm 0.0043$ 		\\
{\boldmath$10^{-9}A_{s }  $}  		  & $2.132\pm 0.049$		& $2.142\pm 0.044            			$& $2.149\pm 0.051$ 			\\
{\boldmath$h              $} 				  & $0.6770\pm 0.0044$   & $0.6775\pm 0.0044            			$& $0.6790\pm 0.0053$	\\
{\boldmath$\tau_{\rm reio }   $} 			  &--- 									&---													  & $0.068\pm 0.013$	\\
{\boldmath$10^{-2}A_{\rm Planck }$}   &---									&---													  & $100.01\pm 0.25     $\\
{\boldmath$A_{\rm ISW}       $} 			  & $1.53\pm 0.29          $ & $1.57\pm 0.29              				$& $1.62\pm 0.30        $\\
{\boldmath$b_{0,{\rm 2MPZ} }      $} 		  & $1.276\pm 0.059       $& $1.37^{+0.29}_{-0.24}              $& $1.194\pm 0.028    $\\
{\boldmath$b_{1,{\rm 2MPZ} }      $} 		  & $1.243\pm 0.049      $ & $1.31^{+0.22}_{-0.31}      		$& $1.188\pm 0.030            $\\
{\boldmath$b_{2,{\rm 2MPZ} }      $} 		  & $1.795\pm 0.080      $ & $1.89\pm 0.27              				$& $1.743\pm 0.070            $\\
{\boldmath$b_{0,{\rm SDSS} }     $} 		  & $1.104\pm 0.043      $ & $1.11^{+0.15}_{-0.24}      		$& $1.060\pm 0.030            $\\
{\boldmath$b_{1,{\rm SDSS} }     $} 		  & $0.904\pm 0.030      $ & $0.887^{+0.11}_{-0.089}    		$& $0.883\pm 0.025            $\\
{\boldmath$b_{2,{\rm SDSS} }     $} 		  & $0.820\pm 0.027       $& $0.84^{+0.21}_{-0.13}      		$& $0.800\pm 0.023            $\\
{\boldmath$b_{3,{\rm SDSS} }     $} 		  & $1.178\pm 0.038       $& $1.11^{+0.22}_{-0.14}      		$& $1.160\pm 0.034            $\\
{\boldmath$b_{4,{\rm SDSS} }     $} 		  & $1.12^{+0.12}_{-0.11}$& $0.99^{+0.22}_{-0.25}           	$& $1.11^{+0.12}_{-0.10}  $\\
{\boldmath$b_{0,{\rm WISC} }     $} 		  & $0.951\pm 0.040        $& $1.01^{+0.15}_{-0.23}      		$& $0.914\pm 0.030           $\\
{\boldmath$b_{1,{\rm WISC} }     $} 		  & $0.851\pm 0.032        $& $0.85^{+0.16}_{-0.21}      		$& $0.828\pm 0.026           $\\
{\boldmath$b_{2,{\rm WISC} }     $} 		  & $1.005\pm 0.038        $& $0.99^{+0.16}_{-0.20}     			$& $0.988\pm 0.034           $\\
{\boldmath$b_{0,{\rm QSO} }  $} 		  & $1.44^{+0.25}_{-0.22}$& $1.26^{+0.45}_{-0.32}      				$& $1.40^{+0.27}_{-0.22}  $\\
{\boldmath$b_{1,{\rm QSO} }  $} 		  & $2.46^{+0.26}_{-0.22}$& $1.90^{+0.60}_{-0.41}      				$& $2.47^{+0.27}_{-0.24}  $\\
{\boldmath$b_{2,{\rm QSO} }  $} 		  & $3.35^{+0.41}_{-0.33}$& $2.68^{0.60}_{-0.52}             			$& $3.34^{+0.46}_{-0.39}  $\\
{\boldmath$b_{\rm NVSS }      $} 			  & $2.54\pm 0.11            $& $2.31\pm 0.39				      		$& $2.479\pm 0.097           $\\
\hline
{\boldmath$\Omega_{\Lambda }$} & $0.720\pm 0.014      $ & $0.722^{+0.050}_{-0.022}   		$&$0.694\pm 0.005$	\\
{\boldmath TS }&$22.0$&$26.5$&$24.9$\\
{\boldmath $\sigma$ }&$4.7$&$5.1$&$5.0$\\
\hline
{\boldmath $\Delta \log({\rm ev})$ }&$ 11.9$&$ 11.5$&$12.7$\\
\hline
\end{tabular}
\end{center}
\caption{Result of the \MP{} fits in the $\Lambda$CDM model with using several combinations of Planck data, AC data and CC data. When the Planck data is not used, Gaussian priors on the cosmological parameters except $\omega_{cdm}$ are assumed. Here $\Omega_{\Lambda }$ is a derived parameter.
The third to last row gives the Test Statistics (TS) which is equal to $\Delta\chi^2$ for the fit in the first two column and $- 2\, \Delta \log {\cal L}$ for the fit in the third column. The second to last row gives the significance $\sigma=\sqrt{TS}$. 
{Finally, the last row gives the logarithm of the Bayes factor, representing the evidence for non-zero  $A_{\rm ISW}$ in Bayesian terms.} }
\label{tab:lcdm_fit}
\end{table*}

\subsection{Method 2}
The first method is, in principle, not fully self-consistent, because
the auto-correlations, and hence the biases, are sensitive to the underlying matter power spectrum.
We fixed the matter power spectrum to the Planck $\Lambda$CDM best-fitting model, but this
may not be the best fit to the auto-correlation data.
The induced error should be negligible when CMB and BAO are also used, since they 
impose $P(k)$ to be very close to the fiducial model. But more importantly,
the cross-correlation determines a given amount of ISW, and this has in principle
an effect on cosmology, since a different ISW means a different Dark Energy model and
thus also a different $P(k)$.
For these reasons it is more consistent to fit to the data at the same time as the bias parameters, the cosmological parameters,
 and the $A_{\rm ISW}$ parameter used to assess the detection significance.

We perform such a fit using the \MP{} environment.
The fit typically involves many parameters ($>15$) which can present degeneracies
which are not known in advance. To scan efficiently this parameter space we
run \MP{} in the Multinest mode~\citep{Feroz:2008xx}. In this way we can robustly 
explore the posterior with typically $\sim 10^6$ likelihood evaluations,
and efficiencies of the order of 10\%.
We consider two cases. 

In the first case, we only use cross-correlation and auto-correlation measurements.
We call this dataset AC+CC, and we fit a total of 22 parameters, i.e, 15 biases, $A_{\rm ISW}$,
and the six $\Lambda$CDM parameters ($\omega_b$, $\omega_{cdm}$, $n_s$, $h$, $A_s$, $\tau_{\rm reio}$). 
When Planck data are
used, we also include the nuisance parameter $A_{\rm Planck }$ \citep{Ade:2015xua}.
For all cosmological parameters except $\omega_{cdm}$,
we use Gaussian priors derived from a fit of Planck+BAO summarized in Table \ref{tab:PlanckBAO},
which are consistent with those published in \cite{Ade:2015xua}. 
The error bars from Planck+BAO are so small that we find essentially the same result for $A_{\rm ISW}$
when fixing these five parameters to their best fit values instead of marginalizing over them with Gaussian priors. 
Our results for this 
fit are shown in the first column of Table~\ref{tab:lcdm_fit}.
As expected, the constraint on  $\omega_{cdm}$ coming from the AC+CC data
is weaker than that from Planck+BAO data, by about a factor 6.
Also,  the $\omega_{cdm}$ best-fit of the AC+CC analysis is lower than the Planck+BAO fit, by about 2$\sigma$.
The fitted galaxy biases are typically compatible with those of Method 1, although
in several cases they are 10-20\% larger, which can be understood as  a consequence of the lower $\omega_{cdm}$,
resulting in a lower $P(k)$ normalization. Indeed, the measured auto-correlations basically fix the product
of the squared biases and of the overall $P(k)$ amplitude.
Comparing the case with free $A_{\rm ISW}$ to the one with $A_{\rm ISW}=0$,
we find TS=$\Delta \chi^2$= 22, giving a significance of 4.7 $\sigma$,
identical  to the one found in \mbox{Method 1}.
With the same setup we also perform a fit using CC data only. The results are shown in the second
column of Table~\ref{tab:lcdm_fit}.  In this case the biases are determined from the
cross-correlation only, without relying on the autocorrelation. It is interesting to see that, 
even in this case, good constraints on the biases can be achieved,
although, clearly, the errors  are much larger (by a factor of $\sim$ 4-5) than  when including the AC data.
We find for this case TS=26.5 corresponding to a significance of 5.1 $\sigma$,
thus reaching the 5 $\sigma$ threshold. The increase in significance seems to be due to the larger
freedom  in the fit of the biases which allows to reach an overall better best-fit of the CC data
with respect to the case in which the biases are constrained by the AC data.

In the second case, we fit the same parameters to the data, but we now include the full Planck+BAO likelihoods instead of Gaussian priors on five parameters. 
Formally, we use the Planck and BAO likelihoods combined with
the $\chi^2$ from the AC+CC data:
\begin{equation}
    \log L = \log L_{\rm PL} + \log L_{\rm BAO} - \chi^2_{AC}/2 - \chi^2_{CC}/2.
\end{equation}
It should be noted that the use of other data besides AC+CC does not affect the ability
to derive the significance of the ISW detection, which is only encoded in the parameter $A_{\rm ISW}$
entering the AC+CC likelihood.
Results of this fit are shown in the third column of Table~\ref{tab:lcdm_fit}.
The main difference with respect to the previous fit is the value of $\omega_{cdm }$,
now driven back to the Planck best-fit. 
This upward shift in $\omega_{cdm }$ results, again, in a global
downward shift of the biases, by about 10-20\%,  giving now
a better compatibility with the results of Method 1.

In general, apart from the small degeneracy with $\omega_{cdm }$ resolved by the inclusion of Planck+BAO data, the biases are well constrained by the fit.
This means that the sub-space of biases 
is approximately orthogonal to the rest of the global parameter space,
which simplifies the fit and speeds up its convergence.
To measure the significance, in this case we define the test statistic as TS= $- 2\, \Delta \log {\cal L}$,
which shares the same properties of the TS defined in terms of the $\chi^2$.
Comparing the case with free $A_{\rm ISW}$ to the one with $A_{\rm ISW}=0$, we now get TS= $- 2\, \Delta \log {\cal L}$ = 24.9,
which gives a significance of 5.0 $\sigma$.
Since the cosmology is basically fixed by the Planck+BAO data 
to a point in parameter space very close to the fiducial model of Method 1,
this improvement in significance comes, apparently, from fitting jointly the biases and $A_{\rm ISW}$
 (while in Method 1 the biases were kept fixed using the results of the first step of the method).
The joint fit explores the correlations which exist between the biases and $A_{\rm ISW}$.
This results in a better global fit, and also in a slightly enhanced $A_{\rm ISW}$ significance,
reaching the 5 $\sigma$ threshold. 

{Finally, since the fit performed with Multinest automatically provides
also the {\it evidence} of the Posterior, in the last row of Table \ref{tab:lcdm_fit}  we additionally report the logarithm of the Bayes factor, 
i.e., the logarithm of the ratio of the evidences for the two fits where $A_{\rm ISW}$ is free and where it is fixed to $A_{\rm ISW}=0$.
We find in all cases values around $\sim$12.
Logarithm of the Bayes factors larger than 5 represents {\it strong evidence} according to Jeffreys' scale \cite{Trotta:2017wnx}.}

\section{Robustness tests}
\label{sec:tests}

In this section we describe some further tests performed to verify the robustness of the results.

As mentioned in Sec.~\ref{sec:cmbmaps}, several CMB maps are available from Planck, resulting from different foreground cleaning methods.
In Fig.~\ref{fig:2mpz_cmb} we show the results of the cross-correlation using
four CMB maps cleaned with four different methods.
We pick up as an example the cross-correlation with the full 2MPZ catalog,
without subdivision in redshift bins.
It clearly appears that the use of different maps has no appreciable impact on the result.

Another important aspect is the possible frequency dependence of the correlation.
In particular, while the ISW effect is expected to be {achromatic,
some secondary effects, like a correlation due to a Sunyaev-Zel'dovich \citep{SZeffect}
or Rees-Sciama \citep{RSeffect} 
imprint
in the CMB map, are expected to be frequency dependent.
To test this possibility, we use available Planck CMB maps at 100 GHz, 143 GHz and 217 GHz.
Again the the full 2MPZ catalog is used as example, since these effects are expected
to peak at low redshift.
Fig.~\ref{fig:2mpz_freq}  shows the result of the correlation at different frequencies.
We observe a very small trend of the CAPS with  frequency, especially for the first $\ell$  bin,
but this effect is negligible with respect to the error bars of the data points.
Results are similar for the other catalogs, showing no frequency dependence.

\begin{figure}
\includegraphics[width = 0.48\textwidth]{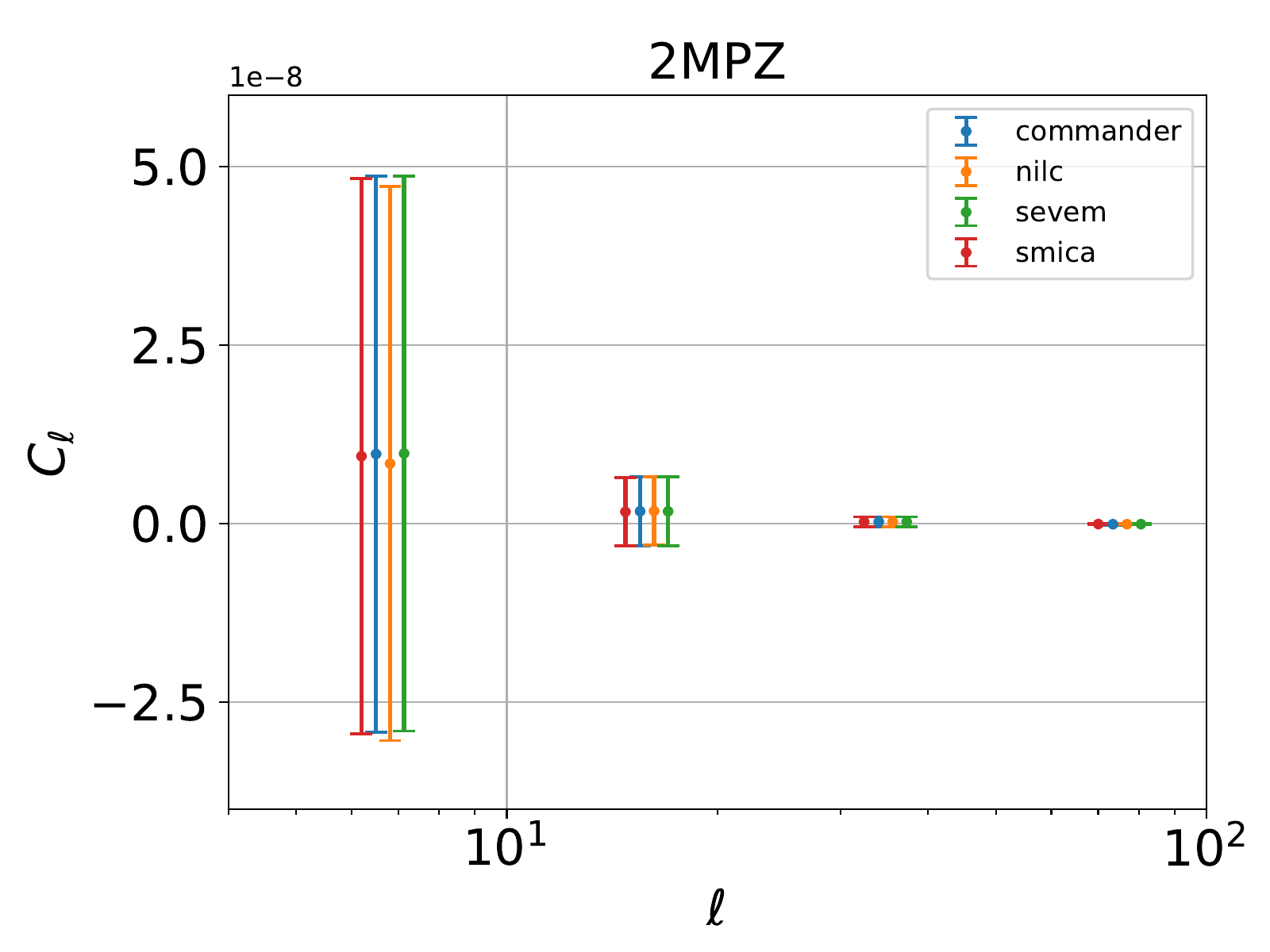}
\caption{Measured cross-correlation of 2MPZ in one single redshift bin with CMB maps from {\tt Commander}, {\tt NILC}, {\tt SEVEM}, and {\tt SMICA}.}
\label{fig:2mpz_cmb}
\end{figure}
\begin{figure}
\includegraphics[width = 0.48\textwidth]{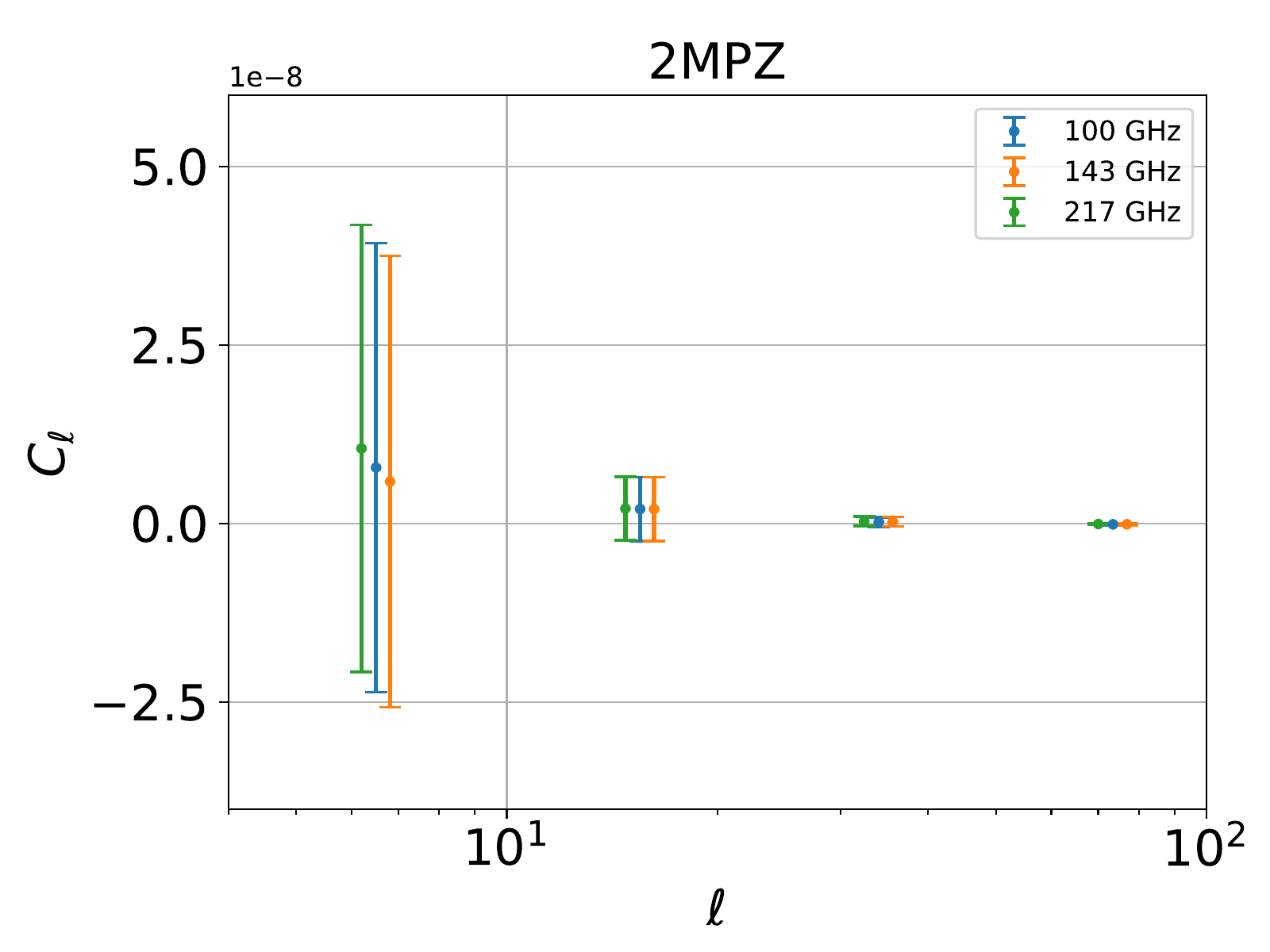}
\caption{Measured cross-correlation of 2MPZ in one single redshift bin with CMB at 100 GHz, 143 GHz and 217 GHz.}
\label{fig:2mpz_freq}
\end{figure}

Finally, we tested the effect of \phz\ errors.
In the basic setup, the theoretical predictions for the auto- and cross-correlation functions per redshift bin are modeled by assuming that the true redshift distribution is well approximated by the \phz\ one, i.e.\ $dN/dz_\mathrm{true} \simeq dN/dz_\mathrm{phot}$. In reality, sharp cuts in $dN/dz_\mathrm{phot}$ will correspond to more extended tails in $dN/dz_\mathrm{true}$ because the \phzs\ are smeared out in the radial direction. However, we 
can easily take \phz\ errors into account if we know their statistical properties.
In the case of 2MPZ, the \phz\ error is basically constant in $z$ and has roughly Gaussian scatter of $\sigma_{\delta z} \simeq 0.015$ centered at $\langle \delta z \rangle =0$, while for \WISC\ the scatter is $\sigma_{\delta z}(z)=0.033(1+z)$ with also approximately zero mean in $\delta z$. 
For SDSS QSOs it is also approximately constant in $z$ and equal to 0.24.
Finally, for SDSS DR12 the error is $\sigma_{\delta z}(z)=0.022(1+z)$
(see Sec.~\ref{sec:catmaps}).
We thus derive the effective true redshift distribution of a given bin by
convolving the measured photo-$z$ selection function in that bin with a $z$-dependent Gaussian of width $\sigma_{\delta z}(z)$.
The resulting true-$z$ distribution
is a smoothed version of the photo-$z$ distribution, presenting tails outside the 
edges of the bin.
We then use this distribution to fit again the auto- and cross-correlations data.
The results are shown in the last column of Table~\ref{tab:biases}.
We find that the effect of \phz\ errors has
some impact on the determination of the biases.
The effect is most important in the high-$z$ tails of various catalogs, and, in particular,
\WISC\ and SDSS DR12. This is not surprising since, in these cases, the \phz\ errors
increase with redshift and are largest at high-$z$.
The effect is at the level of 10-20\%.
This corresponds to a decrease in $A_{\rm ISW}$ of the same amount in these bins.
Nonetheless, since the above bins only have a limited weight on the combined fit, 
the impact on the final $A_{\rm ISW}$ determined from the global fit
of all bins and catalogs is basically negligible.

\begin{figure}
\includegraphics[width=0.45\textwidth]{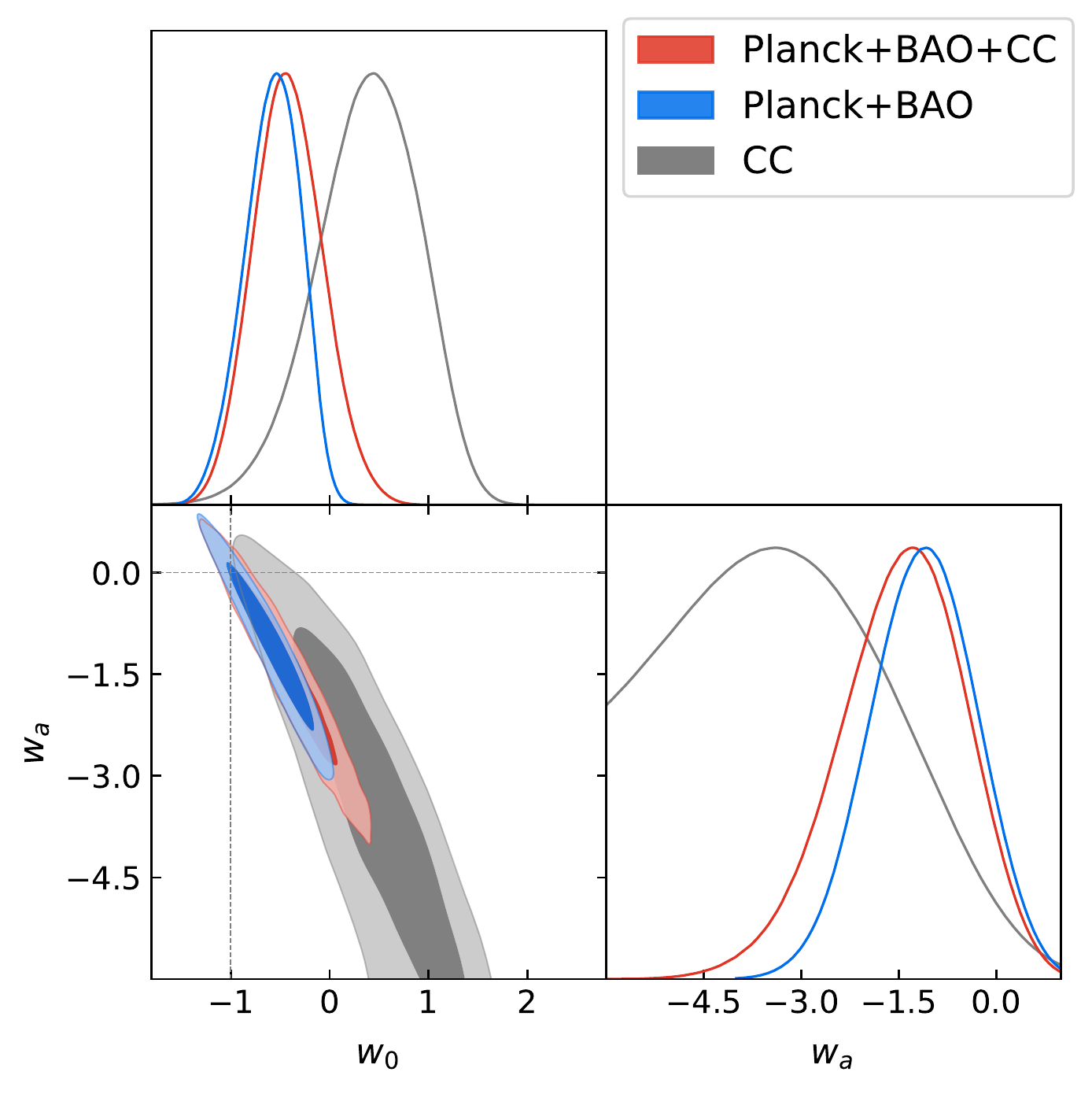}
\caption{Marginalized posterior in the $w_0-w_a$ plane for the three different fits,   Planck+BAO, Planck+BAO+CC and CC only.}
\label{fig:wowazoom}
\end{figure}

\begin{figure}
\includegraphics[width=0.45\textwidth]{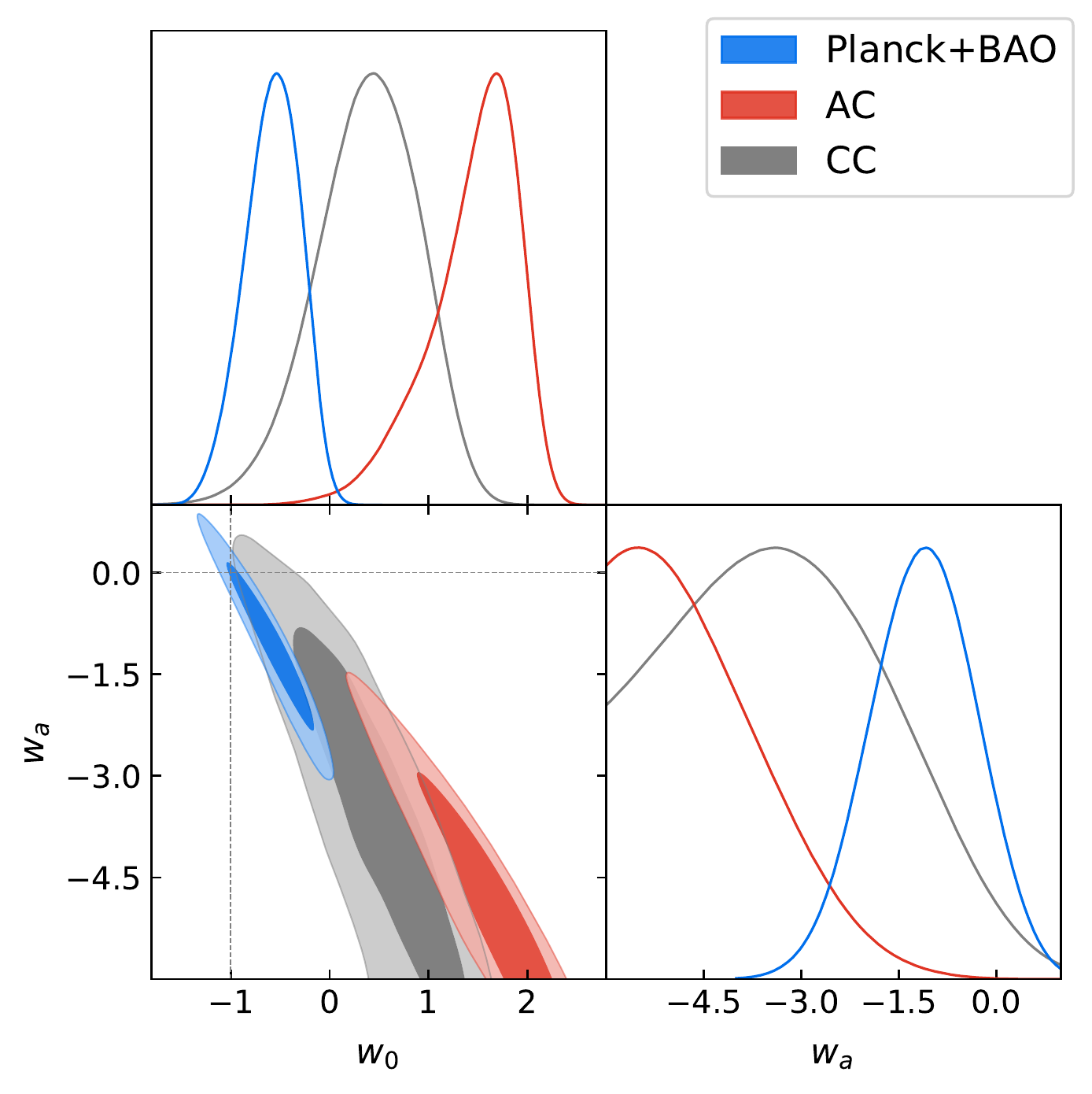}
\caption{Marginalized posterior in the $w_0-w_a$ plane for the three different fits,   Planck+BAO, CC only, and AC only.}
\label{fig:wowazoomAC}
\end{figure}

\begin{table}
\begin{center}
\begin{tabular} { l  c}
 Parameter &  68\% limits\\
\hline
{\boldmath$10^{-2}\omega_{b }$} & $2.224\pm 0.021            $\\

{\boldmath$\omega_{cdm }  $} & $0.1190\pm 0.0017          $\\

{\boldmath$n_{s }         $} & $0.9668\pm 0.0051          $\\

{\boldmath$10^{-9}A_{s }  $} & $2.137\pm 0.063            $\\

{\boldmath$h              $} & $0.639^{+0.018}_{-0.029}   $\\

{\boldmath$w_0            $} & $-0.58^{+0.30}_{-0.25}     $\\

{\boldmath$w_a            $} & $-1.10\pm 0.76             $\\

{\boldmath$\Omega_{0,fld }$} & $0.650^{+0.024}_{-0.029}   $\\
\hline
\end{tabular}
\end{center}
\caption{\label{tab:wowa_planck}Results of the \MP{} fit with using Planck + BAO data.}
\end{table}

\section{Dark energy fit}
\label{sec:DE}

In this section we investigate the power of the cross-correlation data to constrain DE, in a similar framework as presented in \cite{Corasaniti:2005pq,Pogosian:2005ez} and \cite{Pogosian:2004wa}. 
For this purpose, we do not use the $A_{\rm ISW}$ parameter employed in 
Sec.~\ref{sec:method},
since it is only an artificial quantity necessary to evaluate the ISW significance from the cross-correlation data. 
However, as shown in Sec.~\ref{sec:method}, there is indication that 
the best-fit value of $A_{\rm ISW}$
is
above 1 at slightly more than 1$\sigma$. This suggests (although with low statistical significance) that DE could differ from a simple cosmological constant.
To investigate this more in detail, we perform a fit with Method 2 of Sec.~\ref{sec:method}, 
but with $A_{\rm ISW}=1$,
and with extra parameters accounting for dynamical Dark Energy. For simplicity, we use the $w_0-w_a$ 
empirical parametrization \citep{Linder:2002et,Chevallier:2000qy}
and the parameterized post-Friedmann framework of \cite{hu_parameterized_2007} and \cite{fang_crossing_2008}, which are implemented in \CLASS, to study models with $w<-1$.
We test several different fit setups. In particular, since the AC dataset is
a cosmological probe with  its own sensitivity to the cosmological parameters,
we test various combinations in which the AC and CC data are used separately.
A further reason to study the AC data separately from the CC ones
is that
the APS of extragalactic objects
are typically difficult to model accurately, even at small $\ell$,
due to the non-linearity and possible stochasticity of the galaxy bias with respect to matter. 
Separate fits to the AC and CC data could then reveal inconsistencies 
that might be associated to our minimal assumption that the bias is linear and scale-independent.
A further reason to study separately the AC and CC data is the fact that the  AC ones are more prone
to possible systematic effects present in the catalogs like, for example,  non-uniform calibration across the sky. 
These systematics would more severely bias the AC-based cosmological inference, while the CC measurements
are more robust in this respect, since systematic offsets or mis-calibrations across the sky do not generally correlate with the LSS nor the CMB.

We perform the following fits:  (a) Planck+BAO,
(b) Planck+BAO+CC+AC, (c) Planck+BAO+CC, (d) CC only, (e) AC only,  (f) AC+CC.
Case (a) has the standard 6 $\Lambda$CDM parameters, plus $w_a$, $w_0$, and one Planck nuisance parameter, $A_{\rm Planck }$,
required for the evaluation of the Planck likelihood \citep{Ade:2015xua}, thus
 9 parameters in total. The results of this baseline fit are shown in Table~\ref{tab:wowa_planck}.
Case (b) includes CC and AC datasets, and uses additionally 15 bias parameters (24 parameters in total).
Case (c) is similar to (b) but without AC data. Since the biases are still needed for the CC fitting,
they are still included in the fit, but with a Gaussian prior coming from fit  (b). We verified that
just fixing the biases to the best fit (b), instead of including them in the fit with Gaussian priors,
does not actually change the results. Similarly, the result does not change if the biases
are taken from another fit than (b), like (e) or (f). 
For fit (d), featuring only CC data, all cosmological parameters except ($w_a$, $w_0$, $\omega_{cdm }$) and all bias parameters are either fixed or marginalized with Gaussian priors.
For fit (e), featuring only AC data, all cosmological parameters except ($w_a$, $w_0$, $\omega_{cdm }$) are fixed to best-fit values, while the biases are left free, since they are constrained by the AC data.
Finally fit (f) combines AC and CC data, and uses the same setup as fit (e).

\begin{table*}
\begin{center}
\begin{tabular} {l |ccccc }
 Parameter &  AC+CC&CC+bias priors&AC& PL+AC+CC&PL+CC+bias priors\\
\hline
{\boldmath$10^{-2}\omega_{b }$} & $2.222\pm 0.021$ 						& $2.222\pm 0.022            			$& $2.222\pm 0.021            	$	& $2.232\pm 0.022            $	&$2.227\pm 0.022            $\\
{\boldmath$\omega_{cdm }  $} 	  & $0.1134\pm 0.0075          $		& $0.111^{+0.016}_{-0.029}		$& $0.114\pm 0.011            $	  	& $0.1179\pm 0.0018          $& $0.1185\pm 0.0018          $ \\
{\boldmath$n_{s }         $} 			  & $0.9652\pm 0.0055          $		& $0.9642\pm 0.0057          			$& $0.9647\pm 0.0055          $	& $0.9691\pm 0.0056          $& $0.9681\pm 0.0054          $\\
{\boldmath$10^{-9}A_{s }  $}  		  & $2.162\pm 0.076            		$ 		& $2.187\pm 0.080            			$& $2.183\pm 0.077            $	  	& $2.151\pm 0.065            $& $2.152\pm 0.064            $ \\
{\boldmath$h              $} 				  & $0.624^{+0.023}_{-0.029}   	$	& $0.641\pm 0.031            			$& $0.592\pm 0.058            $		& $0.625^{+0.026}_{-0.030}   $& $0.625^{+0.028}_{-0.031}   $\\
{\boldmath$\tau_{\rm reio }   $} 			  &---													&---											  		  &---												& $0.069\pm 0.017            $& $0.068\pm 0.016            $\\
{\boldmath$\Omega_{\Lambda }$} & $0.650\pm 0.029            $			& $0.672^{+0.068}_{-0.048}		$& $0.605^{+0.069}_{-0.049}$	&$0.639\pm 0.038				$& $0.635^{+0.037}_{-0.032}   $	\\
{\boldmath$w_0$} 				  & $0.97^{+0.57}_{-0.44}      $		& $0.39^{+0.57}_{-0.46}      		$& $1.46^{+0.55}_{-0.27}      $	& $-0.37\pm 0.33             $& $-0.43^{+0.32}_{-0.36}     $\\
{\boldmath$w_a$} 				  & $-3.6^{+1.2}_{-1.5}        $			& $-3.2^{+1.4}_{-1.9}        			$& $-4.47^{+0.59}_{-1.4}      $	& $-1.63^{+1.0}_{-0.86}      $& $-1.44^{+1.0}_{-0.81}      $\\
{\boldmath$10^{-2}A_{\rm Planck }$}   &---													&---											  		  &---												& $100.02\pm 0.25            $	& $100.02\pm 0.25            $\\
{\boldmath$b_{0,{\rm 2MPZ} }      $} 		  & $1.56^{+0.13}_{-0.12}      $ 		& $1.2220^{+0.0073}_{-0.021}	$& $1.68^{+0.11}_{-0.042}     $	& $1.240\pm 0.040            $& $1.2220^{+0.0076}_{-0.021} $\\
{\boldmath$b_{1,{\rm 2MPZ } }      $} 		  & $1.46\pm 0.11              $ 			& $1.188\pm 0.030            	$		  & $1.56^{+0.10}_{-0.056}     $	& $1.228\pm 0.041            $& $1.188\pm 0.030            $\\
{\boldmath$b_{2,{\rm 2MPZ } }      $} 		  & $1.94\pm 0.15             	 $			& $1.743\pm 0.070            		    $& $2.04^{+0.14}_{-0.11}      $	& $1.773\pm 0.076            $& $1.744\pm 0.069            $\\
{\boldmath$b_{0,{\rm SDSS } }     $} 		  & $1.195\pm 0.090            $			& $1.060\pm 0.030            		  	$& $1.254^{+0.081}_{-0.056}  $& $1.078\pm 0.033            $& $1.060\pm 0.030            $\\
{\boldmath$b_{1,{\rm SDSS } }     $} 		  & $0.861^{+0.065}_{-0.086}  	 $	& $0.882\pm 0.030            			$& $0.879^{+0.057}_{-0.065}  $& $0.880\pm 0.027            $& $0.884\pm 0.030            $\\
{\boldmath$b_{2,{\rm SDSS } }     $} 		  & $0.743^{+0.057}_{-0.082}   	$	& $0.800\pm 0.025            			$& $0.747^{+0.052}_{-0.065}  $& $0.792\pm 0.024            $& $0.801\pm 0.025            $\\
{\boldmath$b_{3,{\rm SDSS } }     $} 		  & $1.016^{+0.074}_{-0.12}   		 $	& $1.161\pm 0.035            			$& $1.004^{+0.069}_{-0.11}    $& $1.141\pm 0.036            $& $1.161\pm 0.035            $\\
{\boldmath$b_{4,{\rm SDSS } }     $} 		  & $0.935^{+0.098}_{-0.13}   		 $	& $1.110\pm 0.020            			$& $0.902^{+0.089}_{-0.13}    $& $1.09^{+0.11}_{-0.10}      $& $1.110\pm 0.020            $\\
{\boldmath$b_{0,{\rm WISC } }     $} 		  & $1.085\pm 0.083            $			& $0.913\pm 0.030            			$& $1.155^{+0.078}_{-0.053}  $& $0.940\pm 0.035            $& $0.913\pm 0.030            $\\
{\boldmath$b_{1,{\rm WISC } }     $} 		  & $0.884^{+0.068}_{-0.077}   	$	& $0.828\pm 0.030            			$& $0.924^{+0.062}_{-0.055}  $& $0.840\pm 0.029            $& $0.828\pm 0.031            $\\
{\boldmath$b_{2,{\rm WISC } }     $} 		  & $0.981^{+0.078}_{-0.097}  	 $	& $0.987\pm 0.041            			$& $1.008\pm 0.070            $		& $0.990\pm 0.036            $& $0.988\pm 0.040            $\\
{\boldmath$b_{0,{\rm QSO } }  $} 		  		  & $1.14\pm 0.22              $				& $1.401\pm 0.030            $& $1.10\pm 0.20              $		& $1.40\pm 0.22              $& $1.401\pm 0.030            $\\
{\boldmath$b_{1,{\rm QSO } }  $} 		  		  & $1.77\pm 0.26              $				& $2.470\pm 0.030            $& $1.67^{+0.23}_{-0.32}      $	& $2.44^{+0.26}_{-0.23}      $& $2.470\pm 0.030            $\\
{\boldmath$b_{2,{\rm QSO } }  $} 		  		  & $2.47^{+0.35}_{-0.40}      $		& $3.341\pm 0.050            $& $2.34^{+0.27}_{-0.49}      $	& $3.34^{+0.41}_{-0.35}      $& $3.339\pm 0.049            $\\
{\boldmath$b_{\rm NVSS }      $} 			  & $2.36^{+0.17}_{-0.24}      $		& $2.47\pm 0.10              $& $2.38^{+0.17}_{-0.22}      $	& $2.484\pm 0.099            $& $2.487\pm 0.098            $\\
\hline
\end{tabular}
\end{center}
\caption{Result of the \MP{} fits in the $\Lambda$CDM +$w_0$ +$w_a$ model with using several combinations of Planck + BAO (PL) data, AC data and CC data. When the Planck data is not used, Gaussian priors on all cosmological parameters except ($\omega_{cdm}$, $w_0$, $w_a$)  are assumed.}
\label{tab:lcdm_w0wa_fit}
\end{table*}

Figs.~\ref{fig:wowazoom}-\ref{fig:wowazoomAC} show the results for $w_0$ and $w_a$ (marginalized over all the remaining parameters) for some of these fits.
Table~\ref{tab:lcdm_w0wa_fit} gives the confidence intervals on all the parameters for all our fits.
The most evident result is that the AC-only fit selects a region of parameter space
significantly in tension with the Planck+BAO constraints, basically excluding 
the standard case $(w_0, w_a) = (-1, 0)$ at more than 3 $\sigma$.
This is either a consequence of the linear bias model not being accurate enough
to provide reliable cosmological constraints, or an indication of some systematic effects in some of the catalogs.
Problems in the modeling of the bias might be particularly relevant for the auto-correlation of the catalogs in the highest redshift bins,
which are the most sensitive to deviations from a standard cosmological constant,
but also the ones lying  in the tail of the redshift distribution of the catalog,
where different population of galaxies are probably selected, which requires  more
accurate modeling.
More sophisticated approach to the modeling of the catalog auto-correlations might be thus required to address properly this issue.
Various bias models have been proposed beyond linear bias, like for instance
models based on the halo occupation distribution of the catalog objects (see for instance \cite{Ando:2017wff}).
We leave a systematic study of this subject for future work.
Intrinsic artifacts in the catalog, 
like non-uniformity in the sky coverage, or large errors in the \phz\ determination,
are also a likely issue.
These problems can become more evident especially in the tails of the redshift distribution.
Indeed, the largest $\chi^2$ for AC fits from Table~\ref{tab:biases} are for the $z$-bins in the tail
of the distribution, especially for SDSS DR12 and QSOs, indicating a poor match
between the model and the data.
This can be seen more explicitly also in the related plots in Appendix \ref{acplots}.

Hence, in deriving DE constraints it is more conservative to discard information from AC and focus on CC only.
We see that the constraints from the CC data are compatible with Planck+BAO results.
However, given the relatively low significance of the ISW effect, the former
are about three times weaker than the latter for 
each parameter.  
The direction of the degeneracy between $w_0$ and $w_a$ is
approximately the same in the two fits, which was not obvious a priori, since  the two data sets are sensitive to Dark Energy through different physical effects
{(the ISW effect in CMB temperature angular spectrum for the CC fit, and the constraint on the BAO scale for the Planck + BAO fit)}. 
It appears that the valley  of well-fitting models with $w_0>-1$ always corresponds 
to $w(z)$ crossing $-1$ in the range $0.0<z<1.5$, but with very different derivatives $w'(z)$. 
Even when $w_0$ is very large, all models in this valley  do feature accelerated expansion of the Universe 
in the recent past, but not necessarily today. 
In fact, when $w_0$ increases while $w_a$ decreases simultaneously, the stage of  accelerated expansion is preserved 
but translated backward in time.

Since the CC data are less sensitive  than Planck and do not feature a different direction of degeneracy,
the joint constraints from Planck+BAO+CC are basically
unchanged with respect to Planck+BAO only.

\section{Discussion and Conclusions}
\label{sec:discussion}

We derived an updated measurement of the ISW effect through
cross-correlations of the cosmic microwave background 
with several galaxy surveys, namely, 
2MASS Photometric Redshift catalog (2MPZ),
NVSS, SDSS QSOs, SDSS DR12 photometric redshift dataset, and WISE~$\times$~SuperCOSMOS; 
the two latter are here used for the first time for an ISW analysis.
We also improved with respect to previous analyses performing
tomography within each catalog, i.e., exploiting the photometric
redshifts and dividing each catalog into redshift bins.
We found that the current 
cross-correlation data provide  strong evidence
for the ISW effect and thus for Dark Energy, at the 5 $\sigma$ level. 

However, current catalogs are still not optimal
to derive cosmological constraints from the ISW, for two main reasons. First, the clustering of
objects requires complicated modeling, probably beyond the simple linear bias assumption. On this last point, improvements are
possible using more sophisticated  modeling,  but at a price of
introducing more nuisance parameters. Also, the tails of the redshift distributions of the objects might
be more strongly affected by catalog systematics such as uneven sampling or large \phz\ errors.

Second, the data used in this paper are  sensitive mostly to the redshift range $0<z< 0.6$,
while the ISW effect is expected to be important for $0.3<z< 1.5$. 
Several planned or forthcoming wide-angle galaxy surveys will cover this redshift range and should thus bring (major) improvement for ISW detection via cross-correlation with CMB. For the Euclid satellite, the predicted significance of such a signal is $\sim 8 \sigma$ \citep{Euclid_cosmol}, and one should expect similar figures from the Large Synoptic Survey Telescope \citep{LSST}, and the Square-Kilometer Array \citep{SKA}. The very high S/N of ISW from these deep and wide future catalogs will not only allow for much stronger constraints on dark energy than we obtained here, but even on some modified gravity models which often predict very different ISW signatures than $\Lambda$CDM \citep[e.g.][]{Renk17}.

\section*{acknowledgments}
Simulations were performed with computing resources granted by RWTH Aachen University under project thes0263.

MB is supported by the Netherlands Organization for Scientific Research, NWO, through grant number 614.001.451, and by the Polish National Science Center under contract \#UMO-2012/07/D/ST9/02785.

Some of the results in this paper have been derived using the HEALPix package\footnote{\url{http://healpix.sourceforge.net/}} \citep{2005ApJ...622..759G}.

This research has made use of data obtained from the SuperCOSMOS Science Archive, prepared and hosted by the Wide Field Astronomy Unit, Institute for Astronomy, University of Edinburgh, which is funded by the UK Science and Technology Facilities Council.

Funding for SDSS-III has been provided by the Alfred P. Sloan Foundation, the Participating Institutions, the National Science Foundation, and the U.S. Department of Energy Office of Science. The SDSS-III web site is \url{http://www.sdss3.org/}. SDSS-III is managed by the Astrophysical Research Consortium for the Participating Institutions of the SDSS-III Collaboration including the University of Arizona, the Brazilian Participation Group, Brookhaven National Laboratory, Carnegie Mellon University, University of Florida, the French Participation Group, the German Participation Group, Harvard University, the Instituto de Astrofisica de Canarias, the Michigan State/Notre Dame/JINA Participation Group, Johns Hopkins University, Lawrence Berkeley National Laboratory, Max Planck Institute for Astrophysics, Max Planck Institute for Extraterrestrial Physics, New Mexico State University, New York University, Ohio State University, Pennsylvania State University, University of Portsmouth, Princeton University, the Spanish Participation Group, University of Tokyo, University of Utah, Vanderbilt University, University of Virginia, University of Washington, and Yale University.

Some of the results in this paper have been derived using the GetDist package\footnote{\url{https://github.com/cmbant/getdist}}.

\bibliography{BibISW}
\appendix
\section{Auto- and cross-correlation results}
\label{acplots}

In this appendix we show the measured APS and CAPS and the related best-fit model
for all the catalogs and $z$-bins considered in the analysis. 
Dots refer to the measured single multipoles, while data points with error bars refer to binned measurements.

\begin{figure*}
\includegraphics[width = 0.45\textwidth]{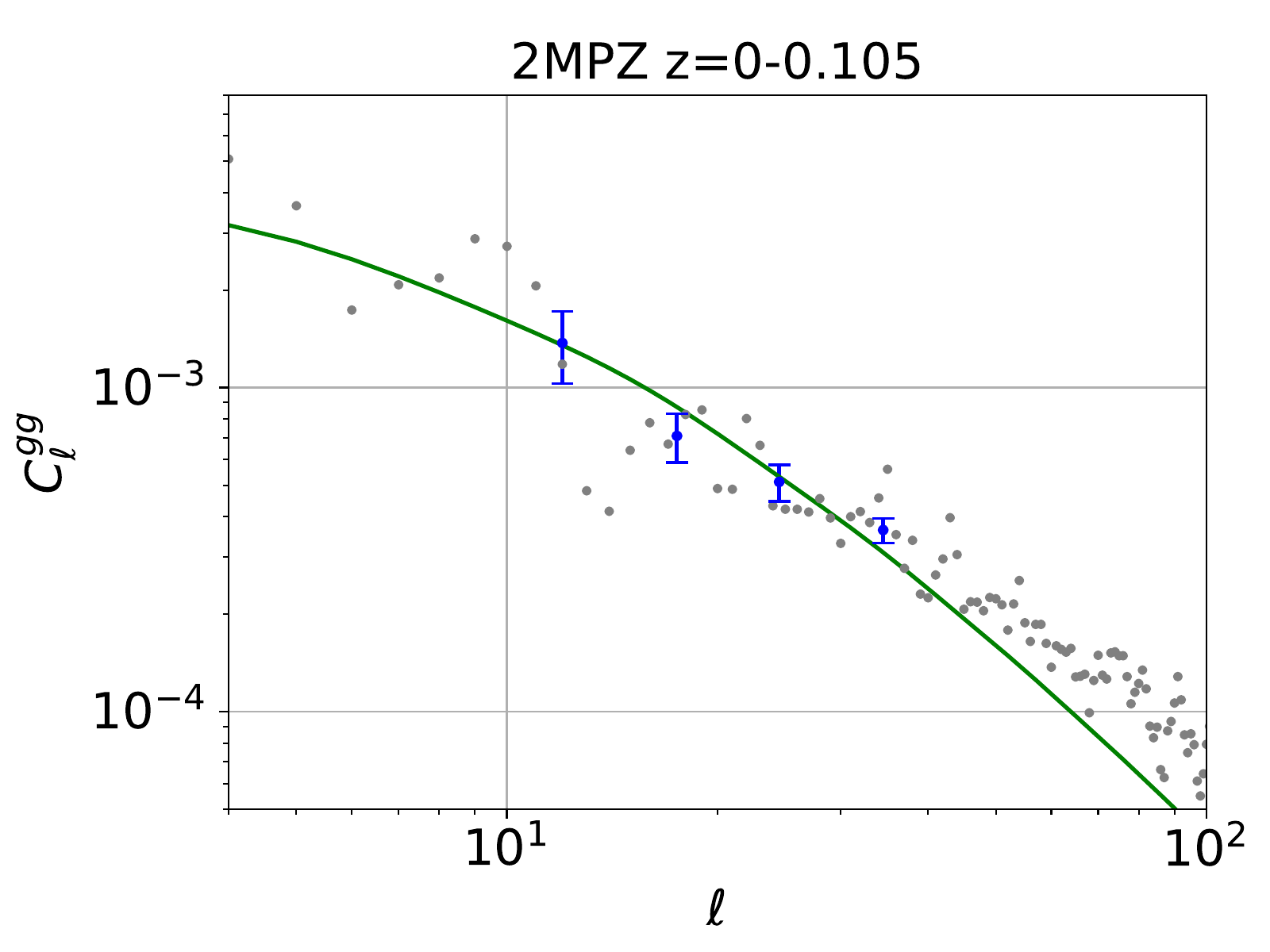}
\includegraphics[width = 0.45\textwidth]{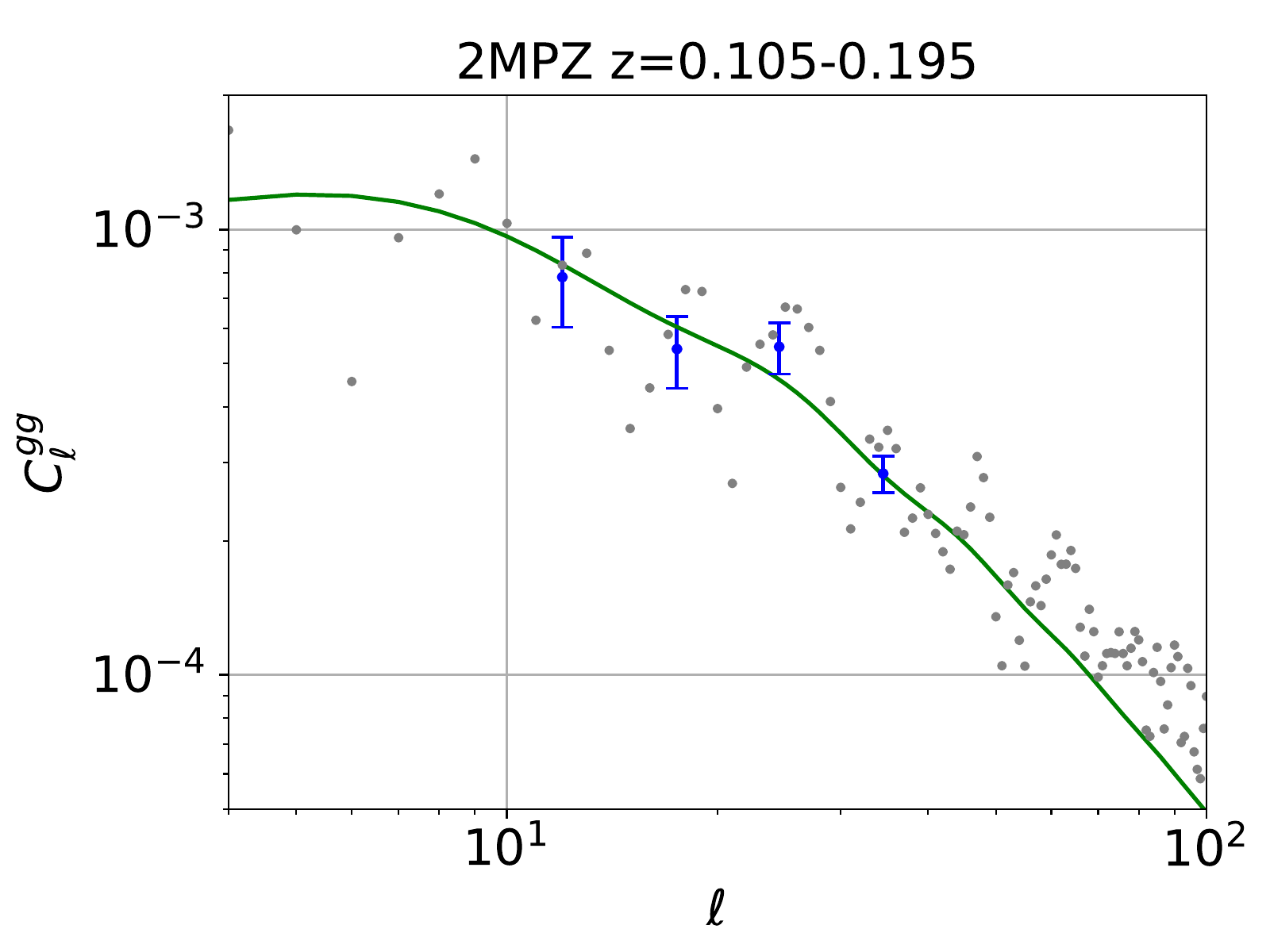}
\includegraphics[width = 0.45\textwidth]{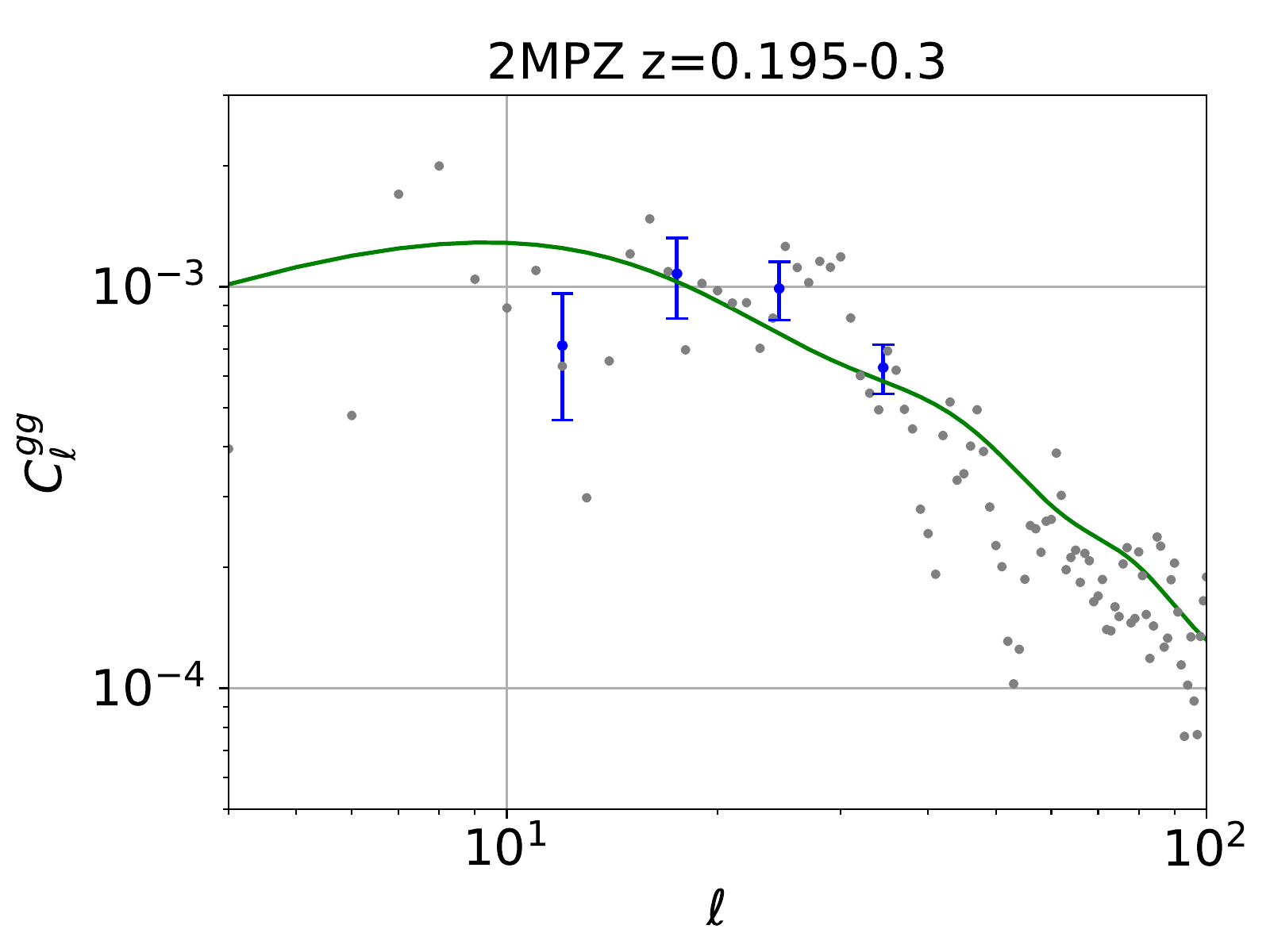}
\includegraphics[width = 0.45\textwidth]{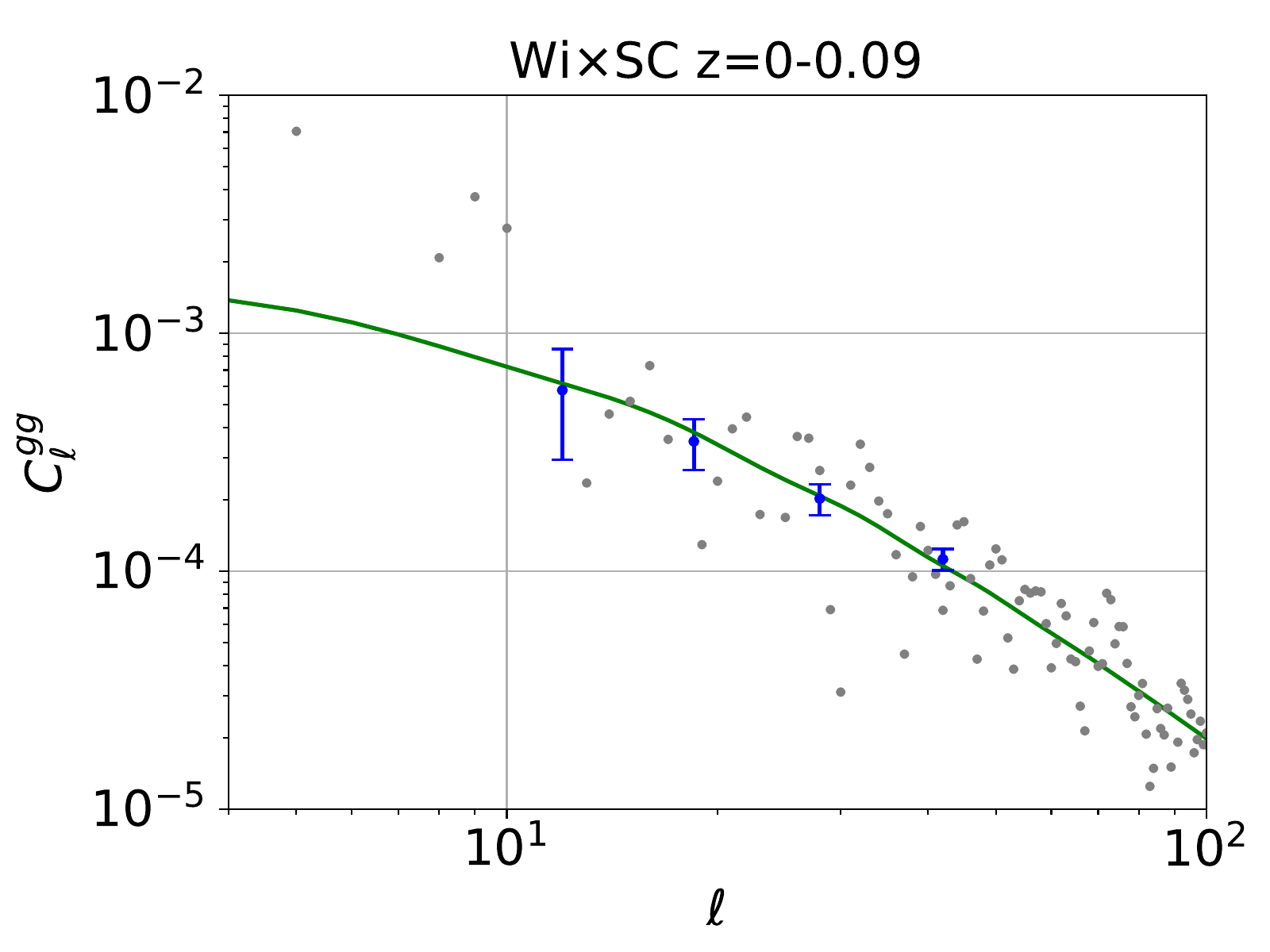}
\includegraphics[width = 0.45\textwidth]{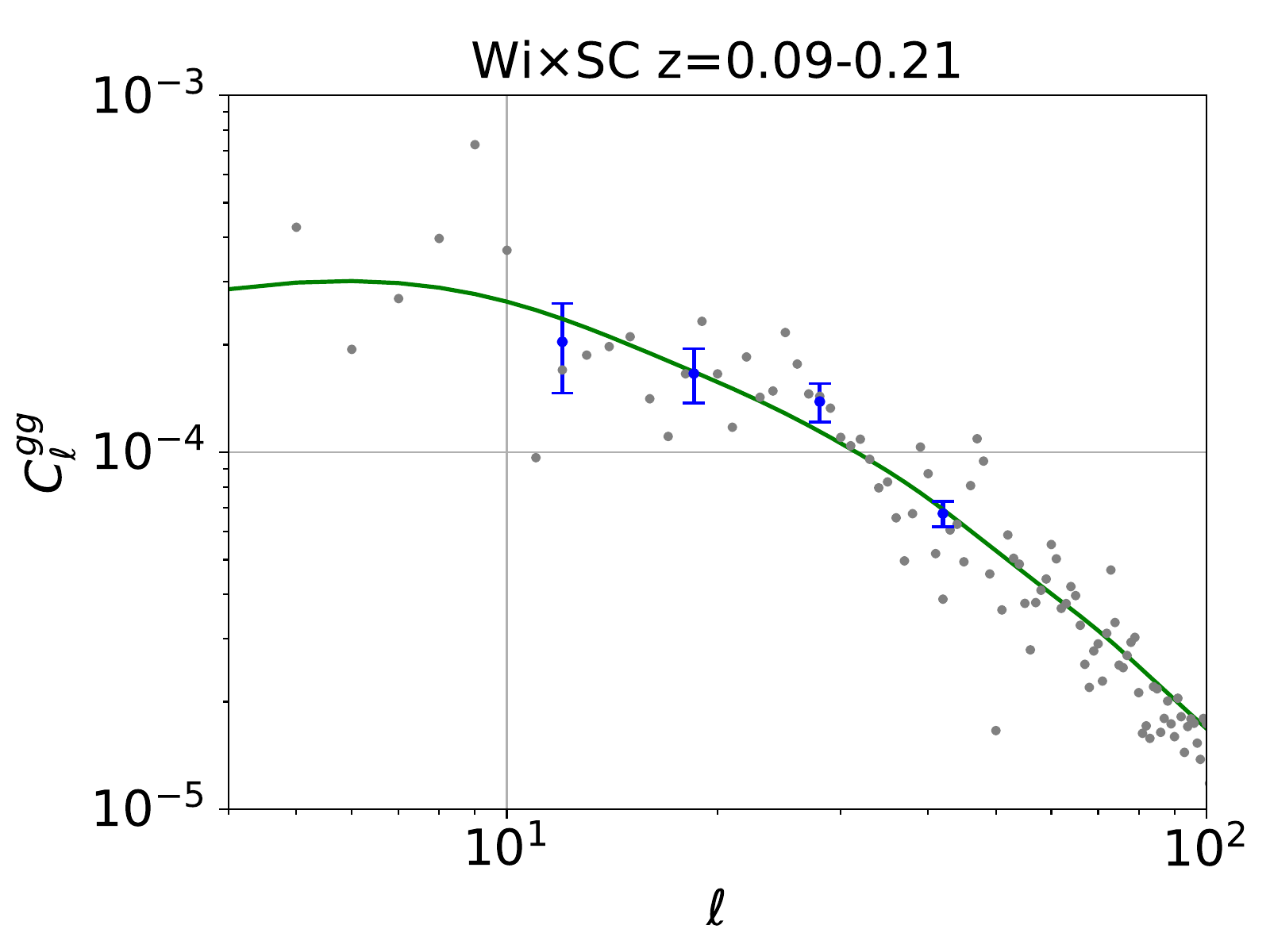}
\includegraphics[width = 0.45\textwidth]{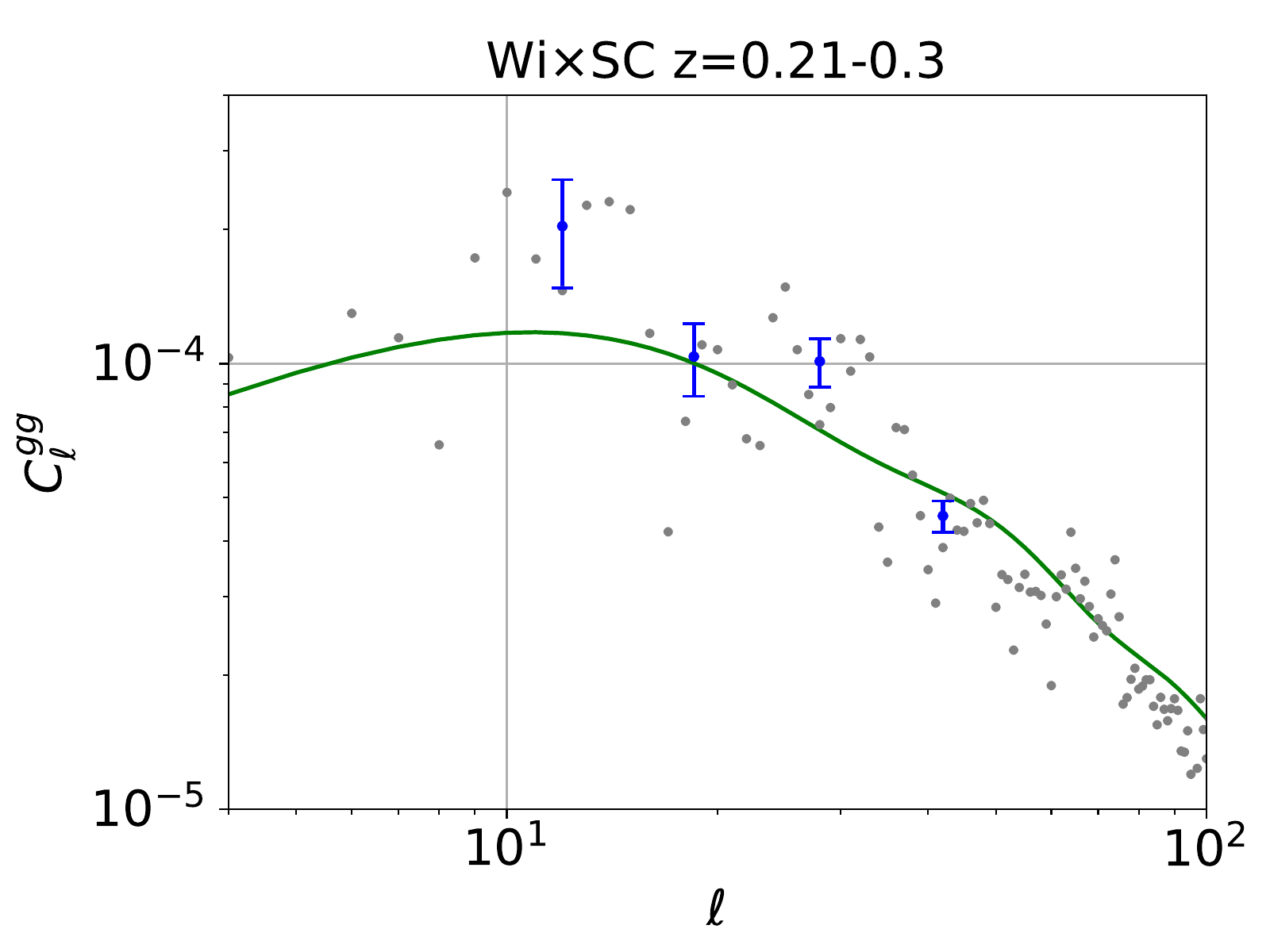}
\caption{Measured auto-correlation for different catalogs and redshift bins.}
\end{figure*}

\begin{figure*}
\includegraphics[width = 0.45\textwidth]{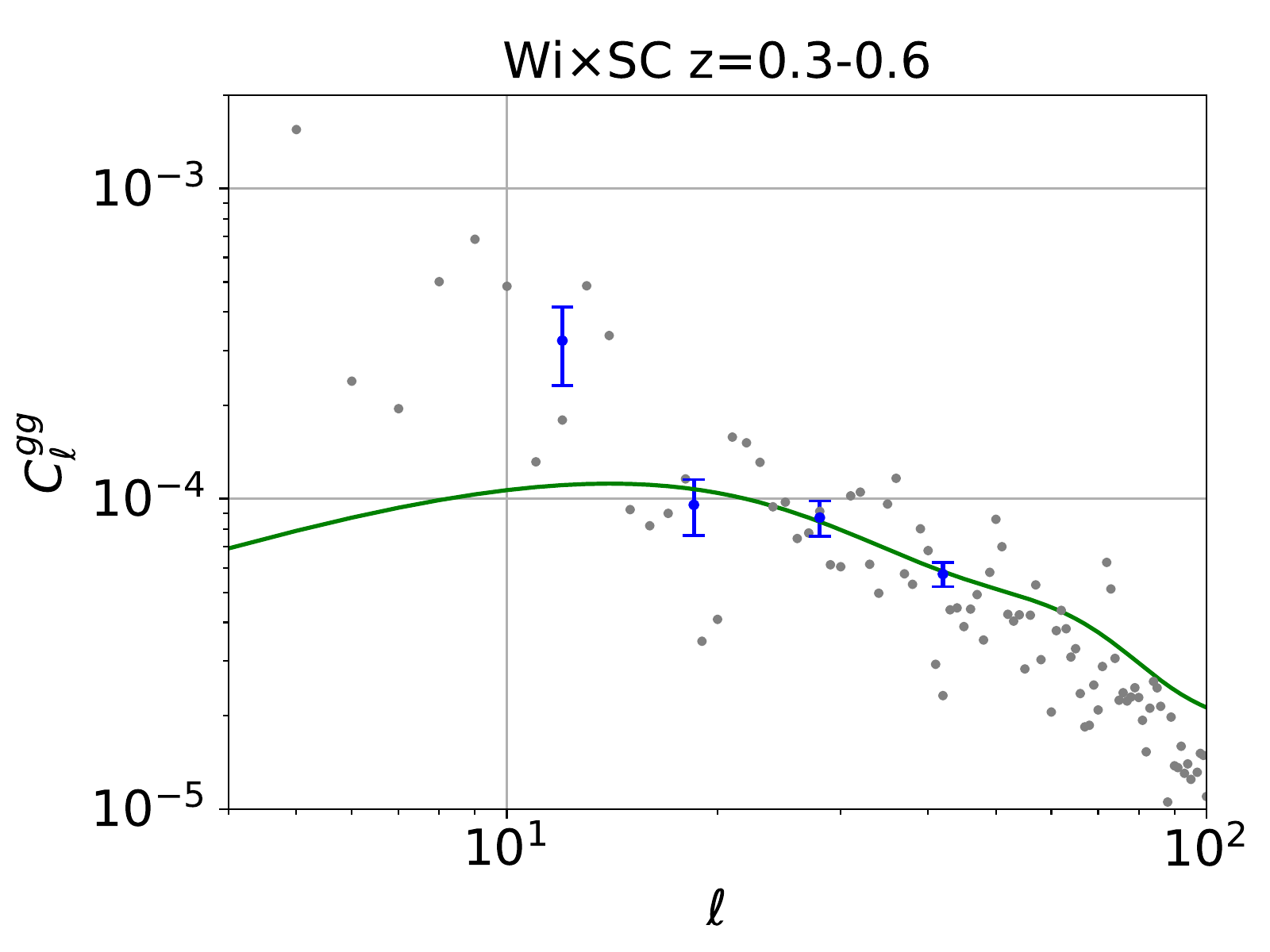}
\includegraphics[width = 0.45\textwidth]{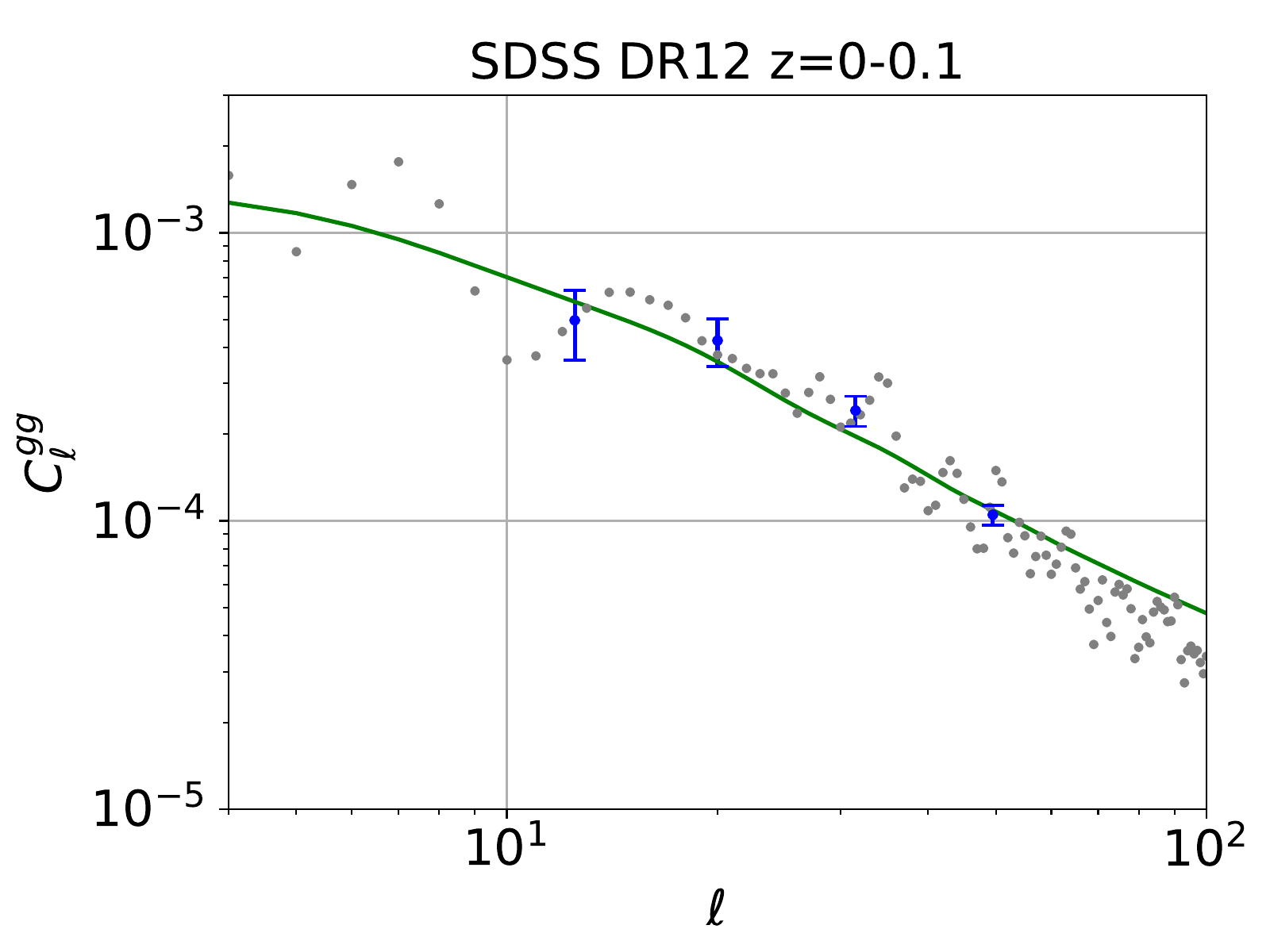}
\includegraphics[width = 0.45\textwidth]{autocorrelation/sdss_bin_1.pdf}
\includegraphics[width = 0.45\textwidth]{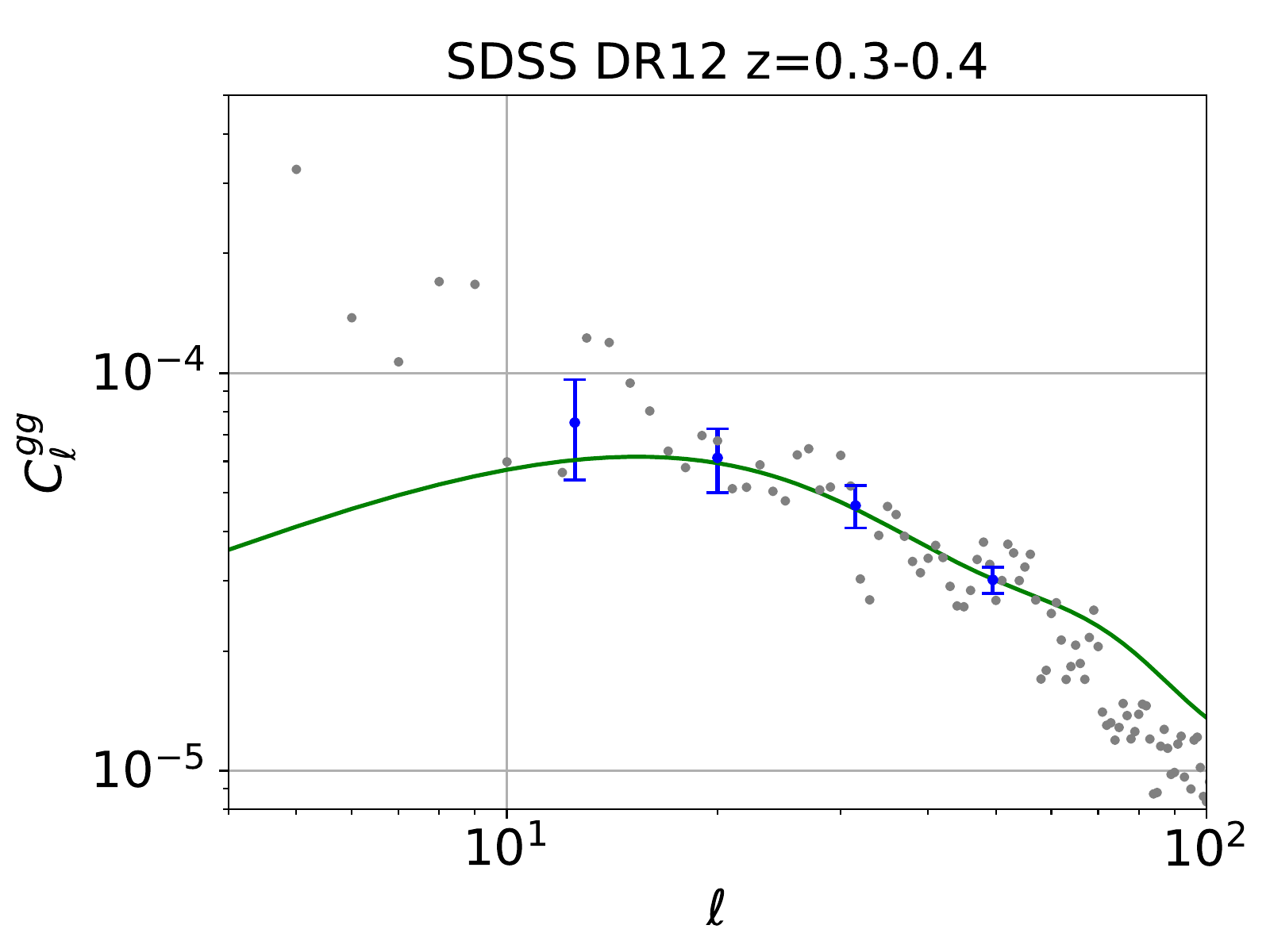}
\includegraphics[width = 0.45\textwidth]{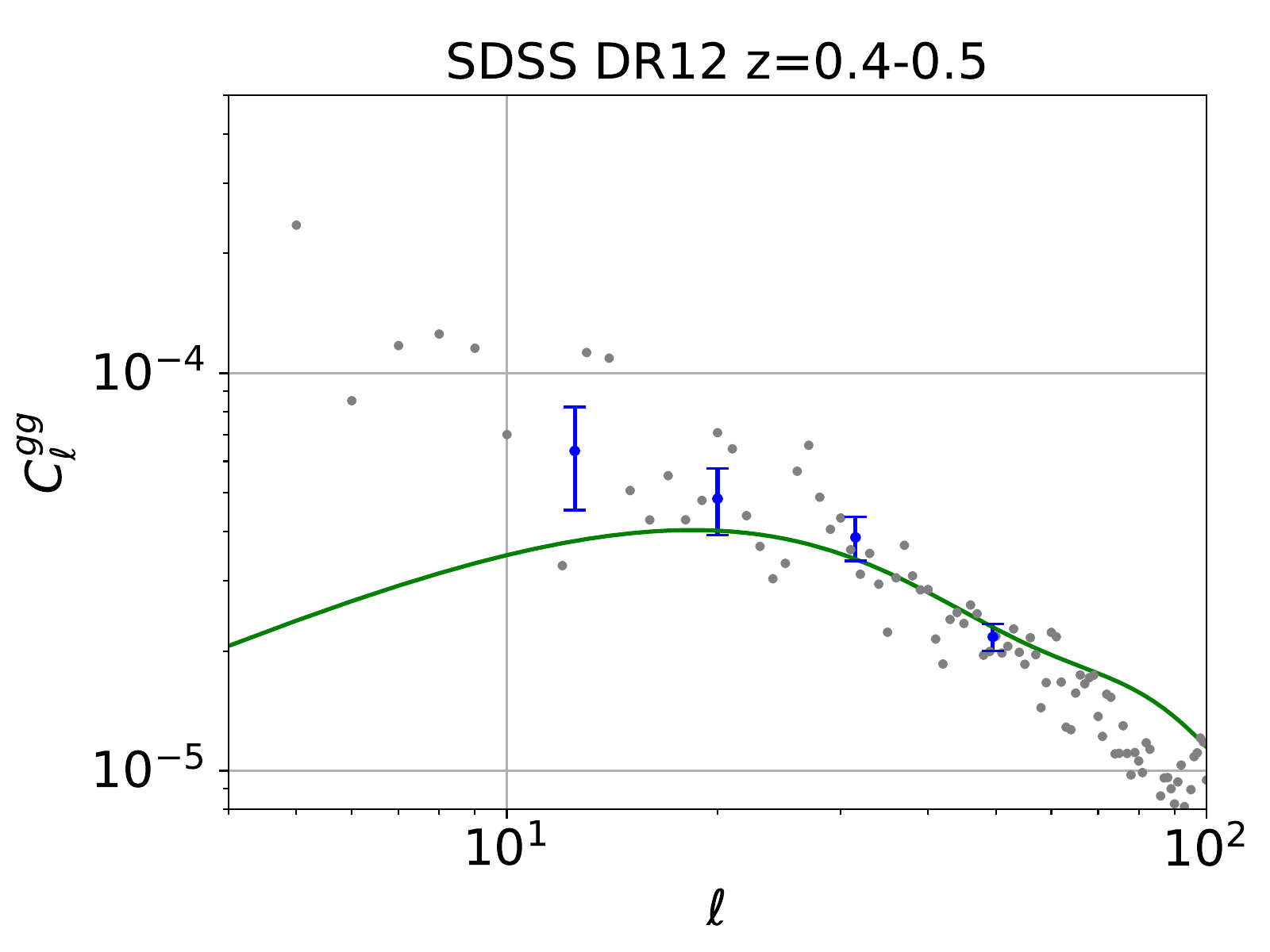}
\includegraphics[width = 0.45\textwidth]{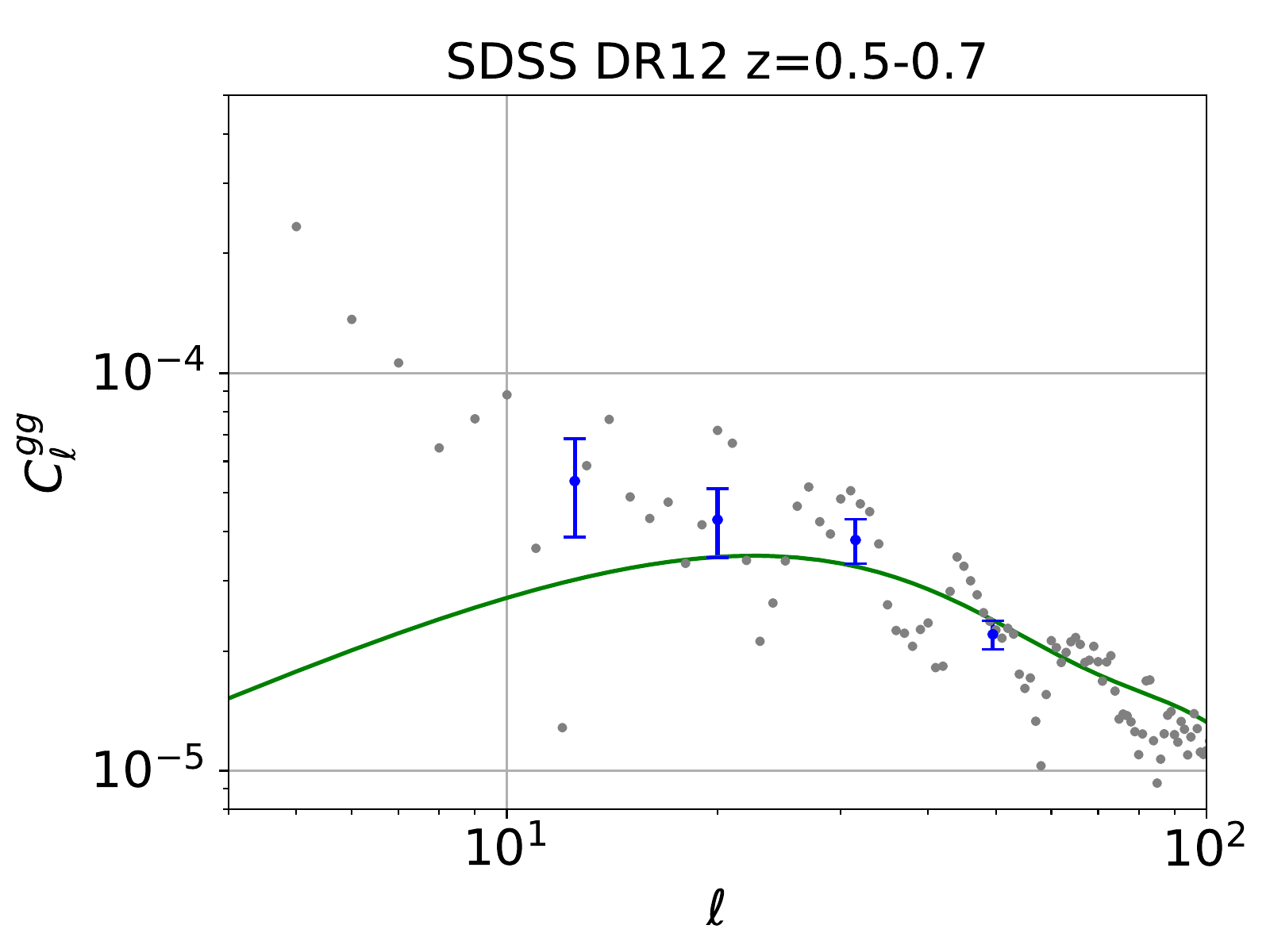}
\caption{Measured auto-correlation for different catalogs and redshift bins.}
\end{figure*}

\begin{figure*}
\includegraphics[width = 0.45\textwidth]{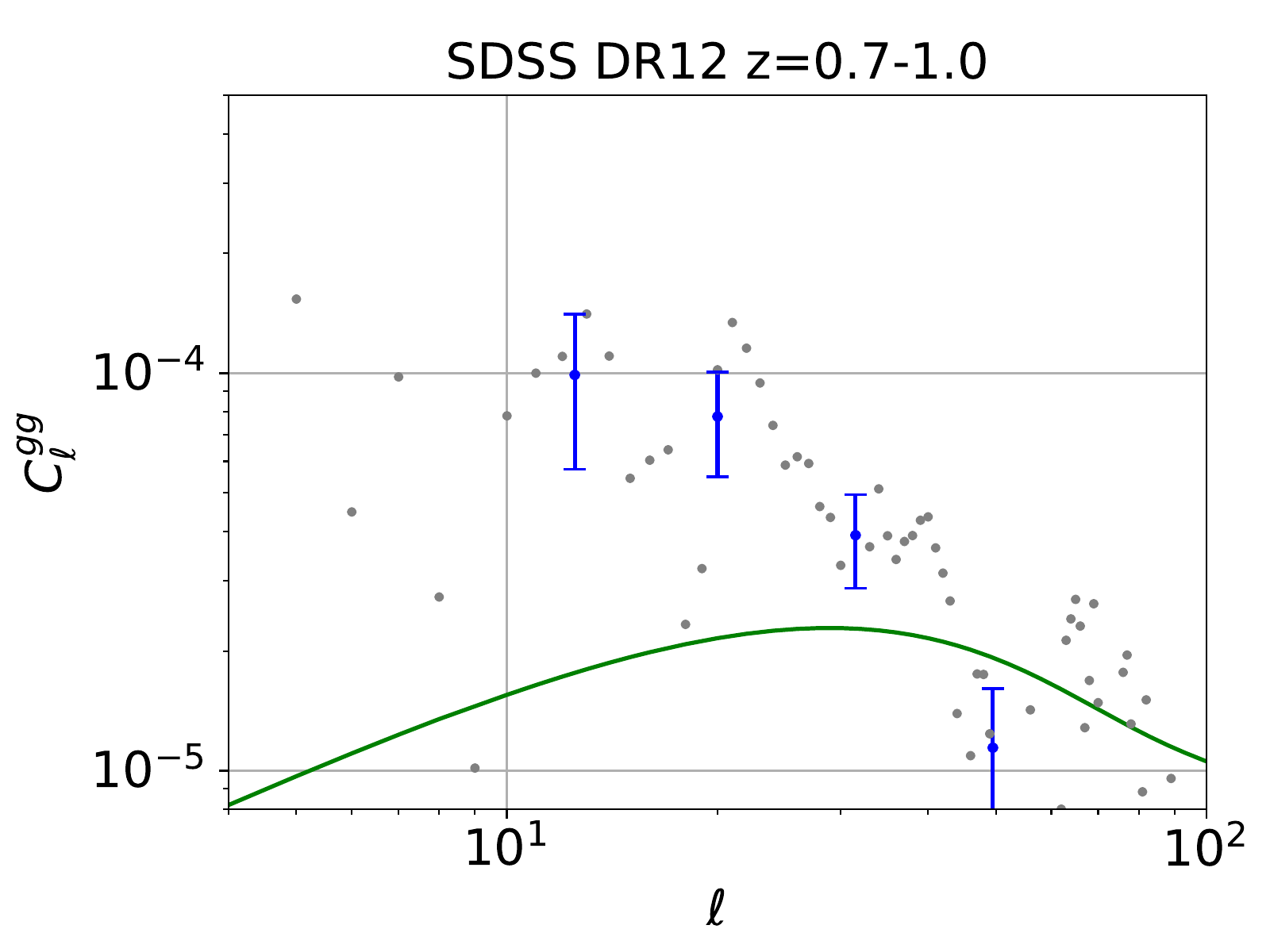}
\includegraphics[width = 0.45\textwidth]{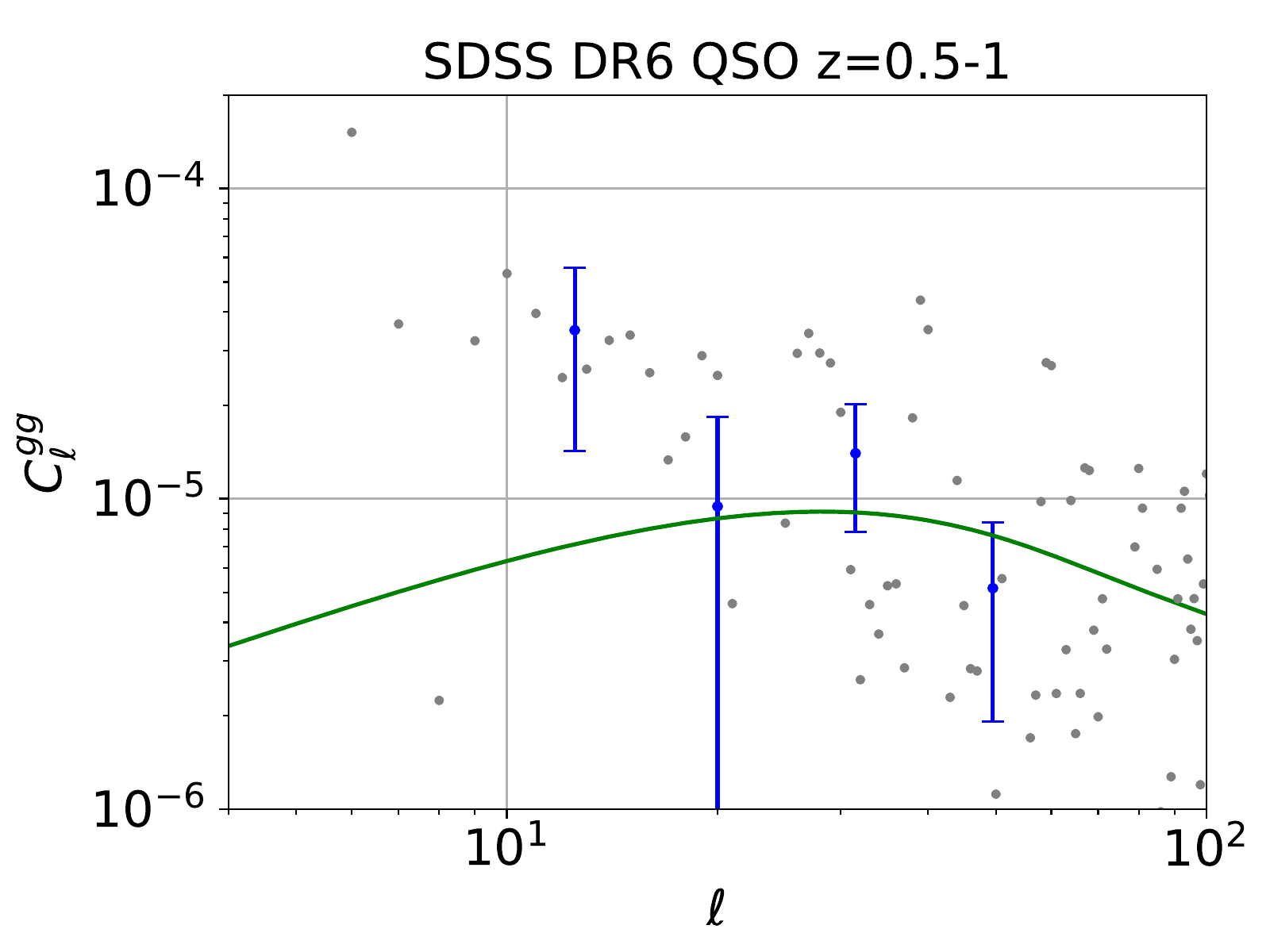}
\includegraphics[width = 0.45\textwidth]{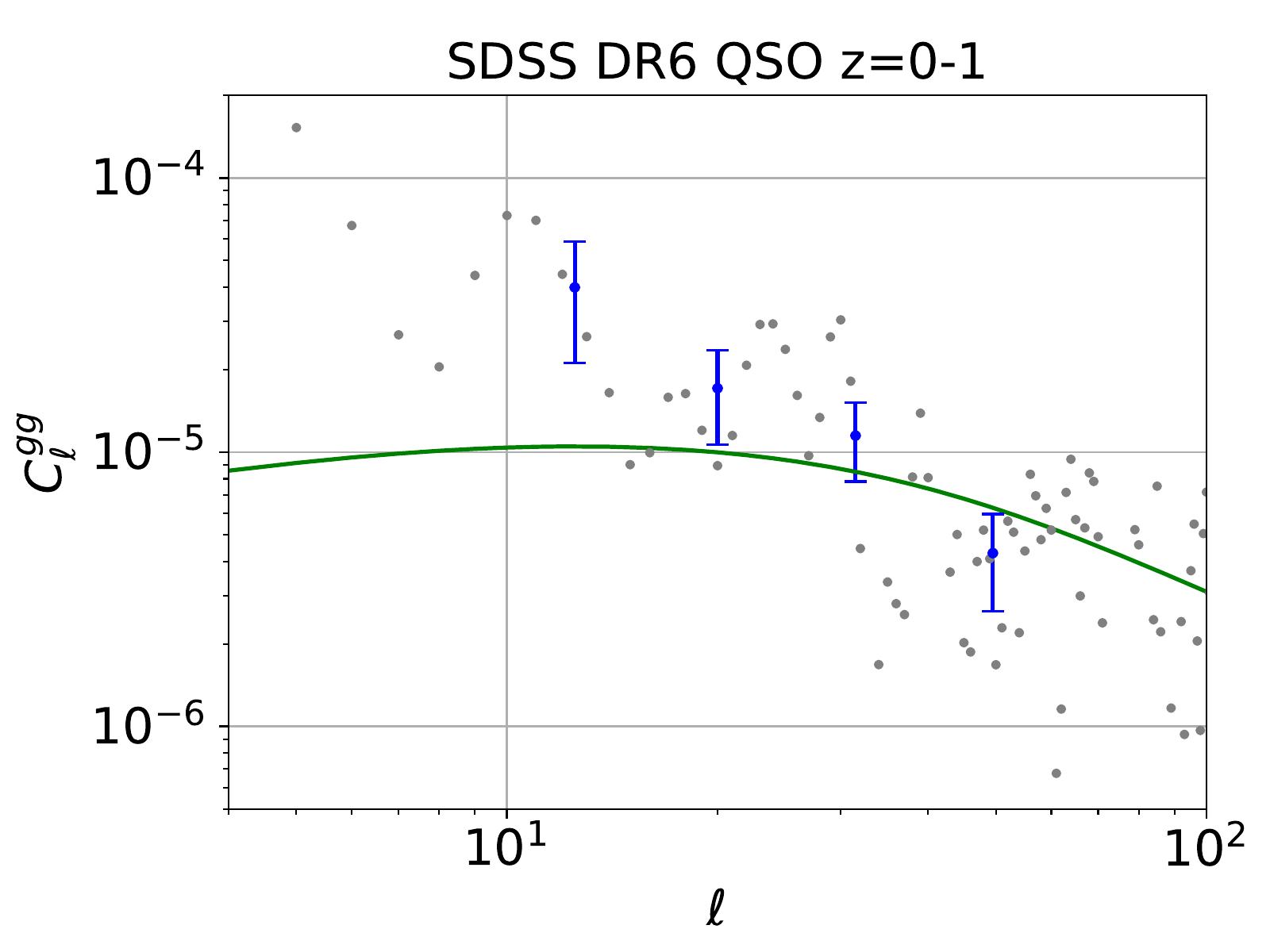}
\includegraphics[width = 0.45\textwidth]{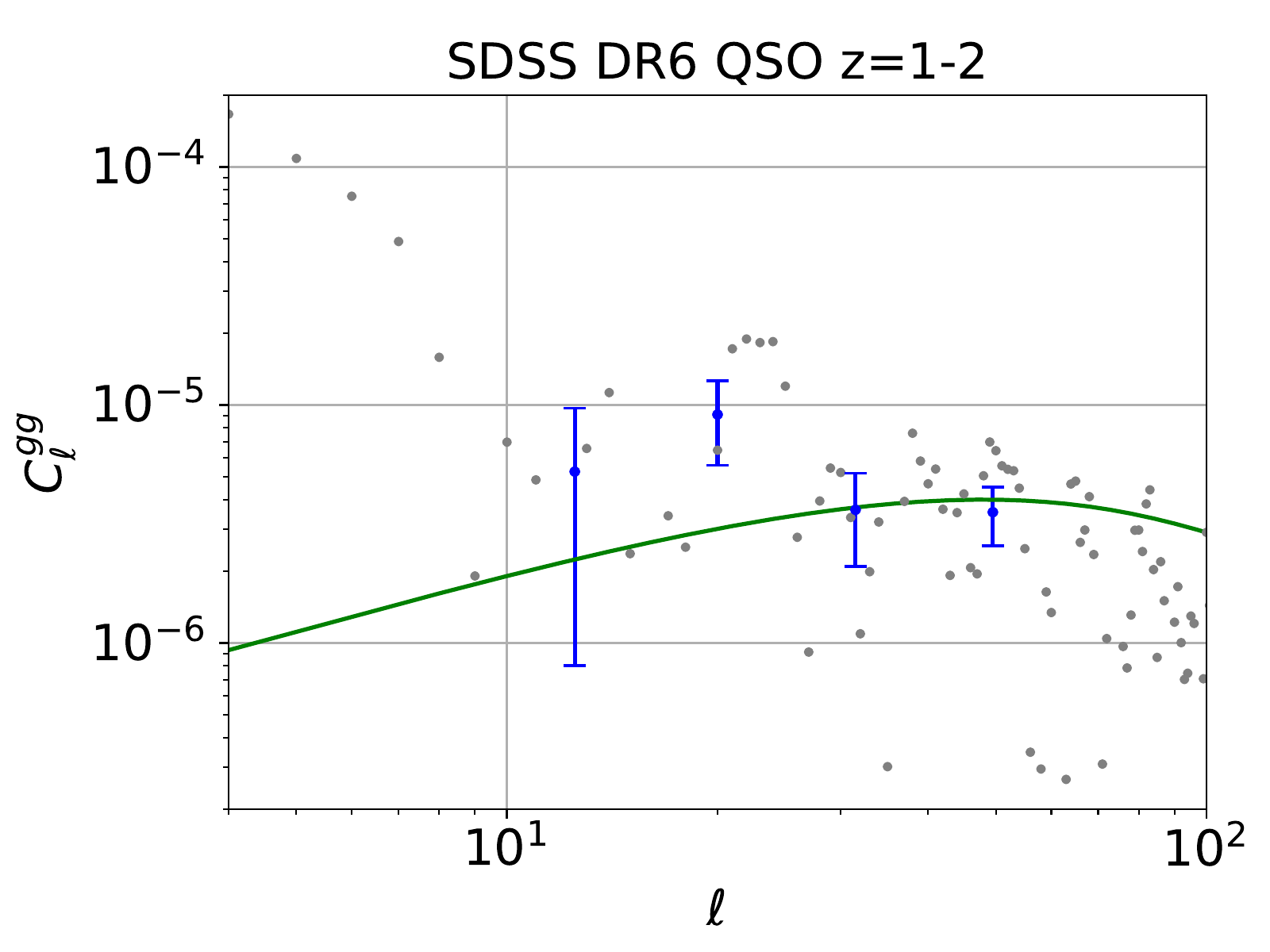}
\includegraphics[width = 0.45\textwidth]{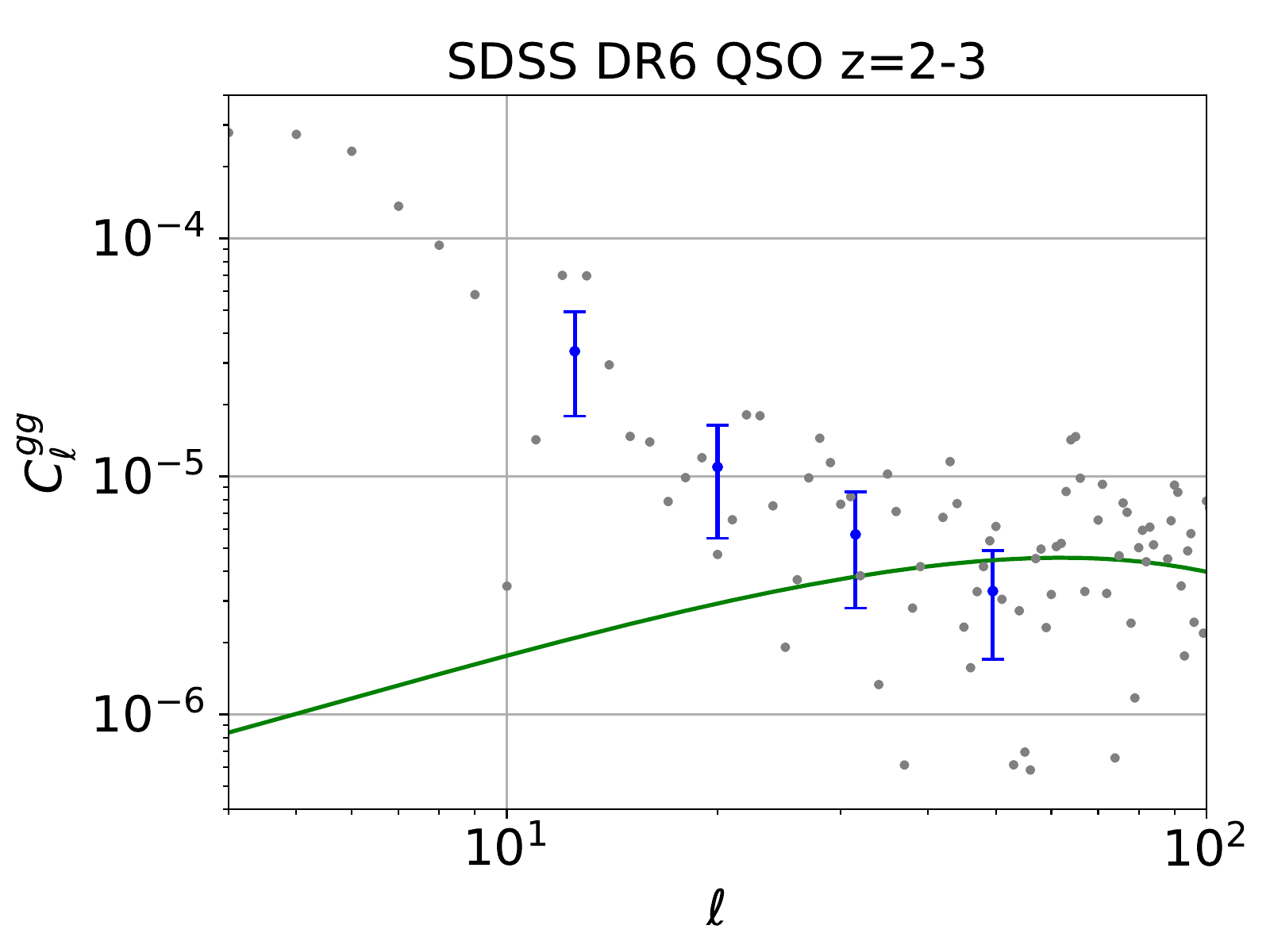}
\includegraphics[width = 0.45\textwidth]{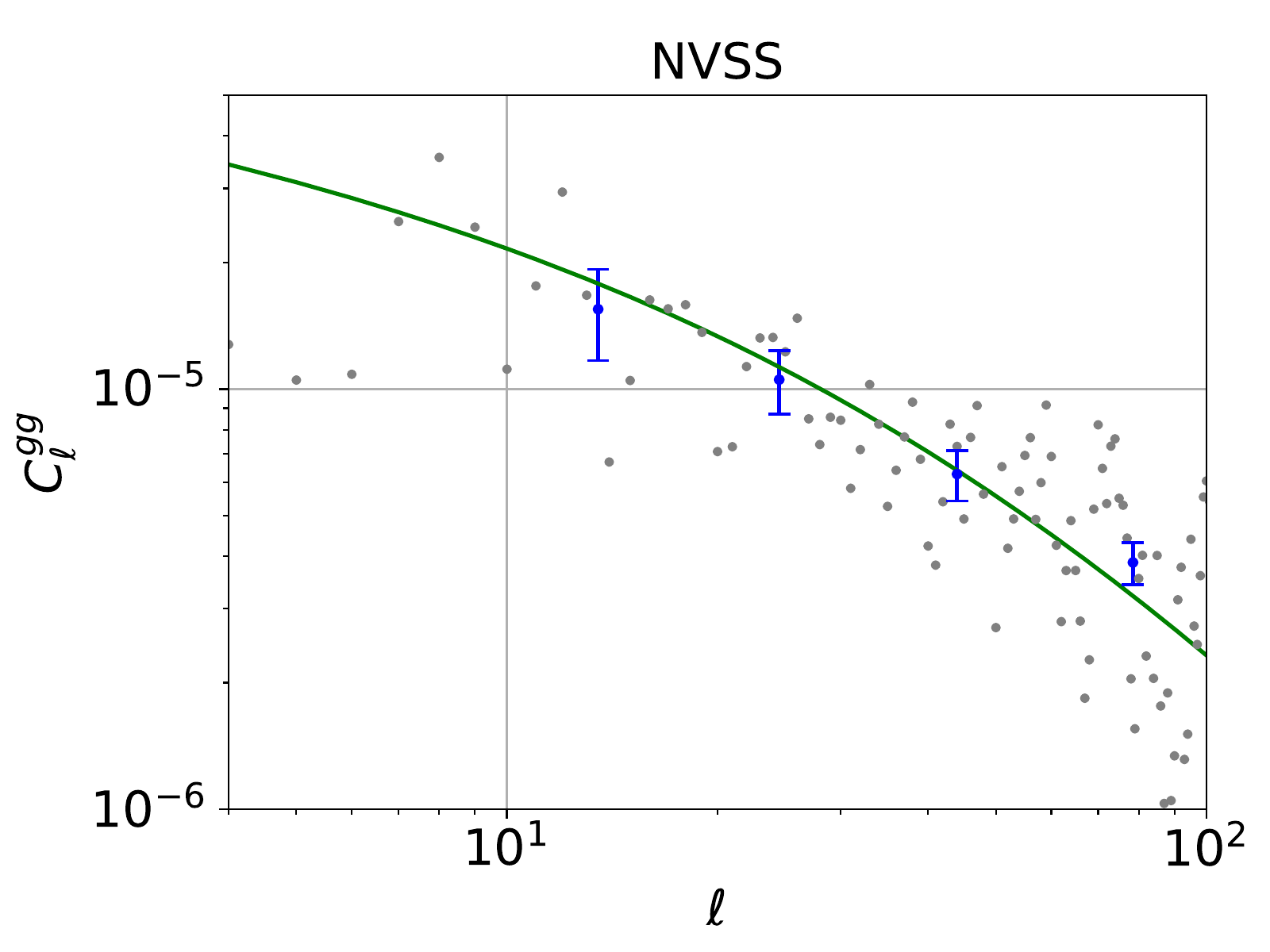}
\caption{Measured auto-correlation for different catalogs and redshift bins.}
\end{figure*}

%\section{Cross-correlation results}

%In this appendix we show the measured CAPS and the related best-fit model
%for all the catalogs and $z$-bins considered in the analysis. 
%Dots refer to the measured single multipoles, while data points with error bars refer to binned measurements. 

\begin{figure*}
\includegraphics[width = 0.45\textwidth]{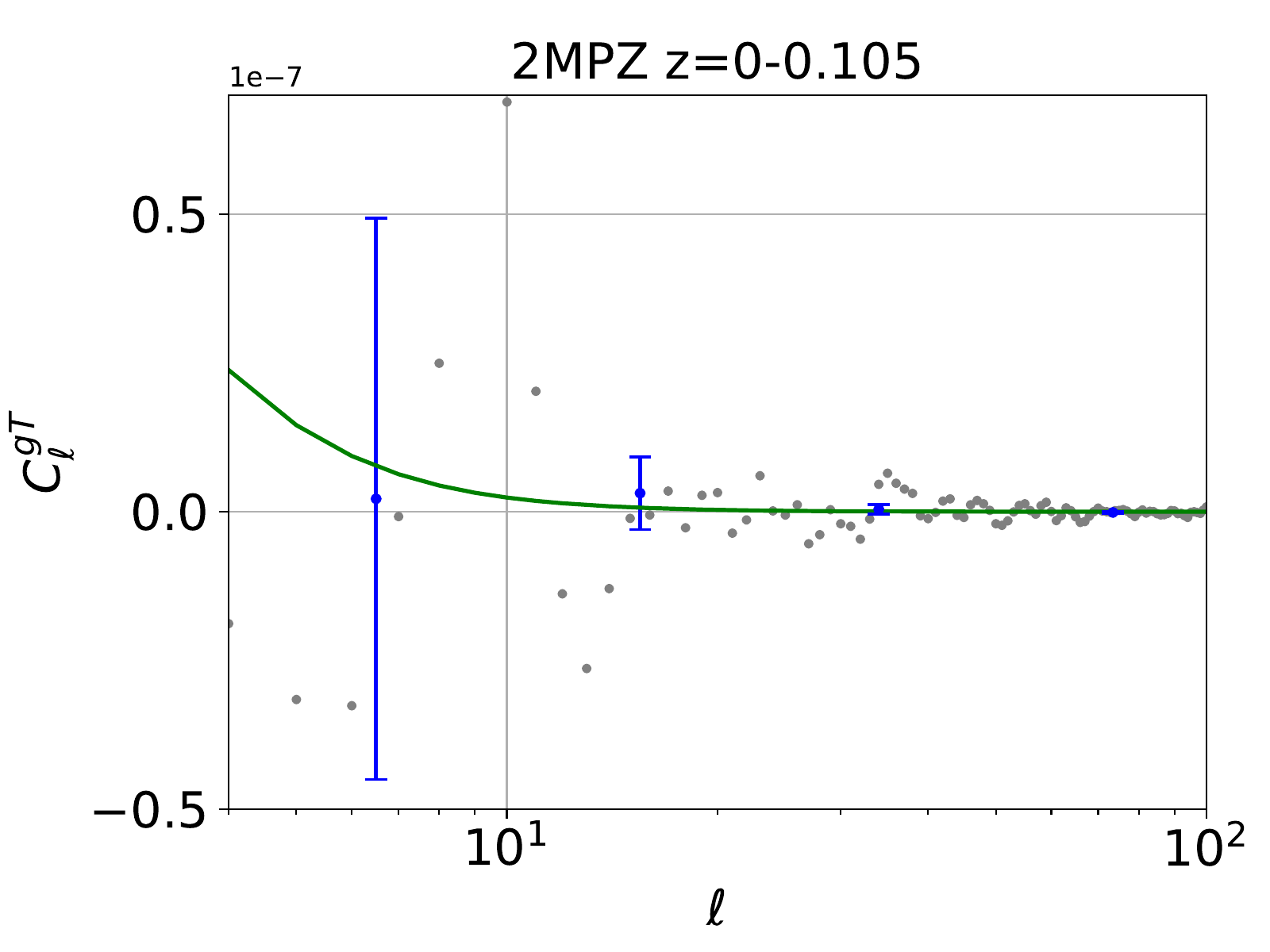}
\includegraphics[width = 0.45\textwidth]{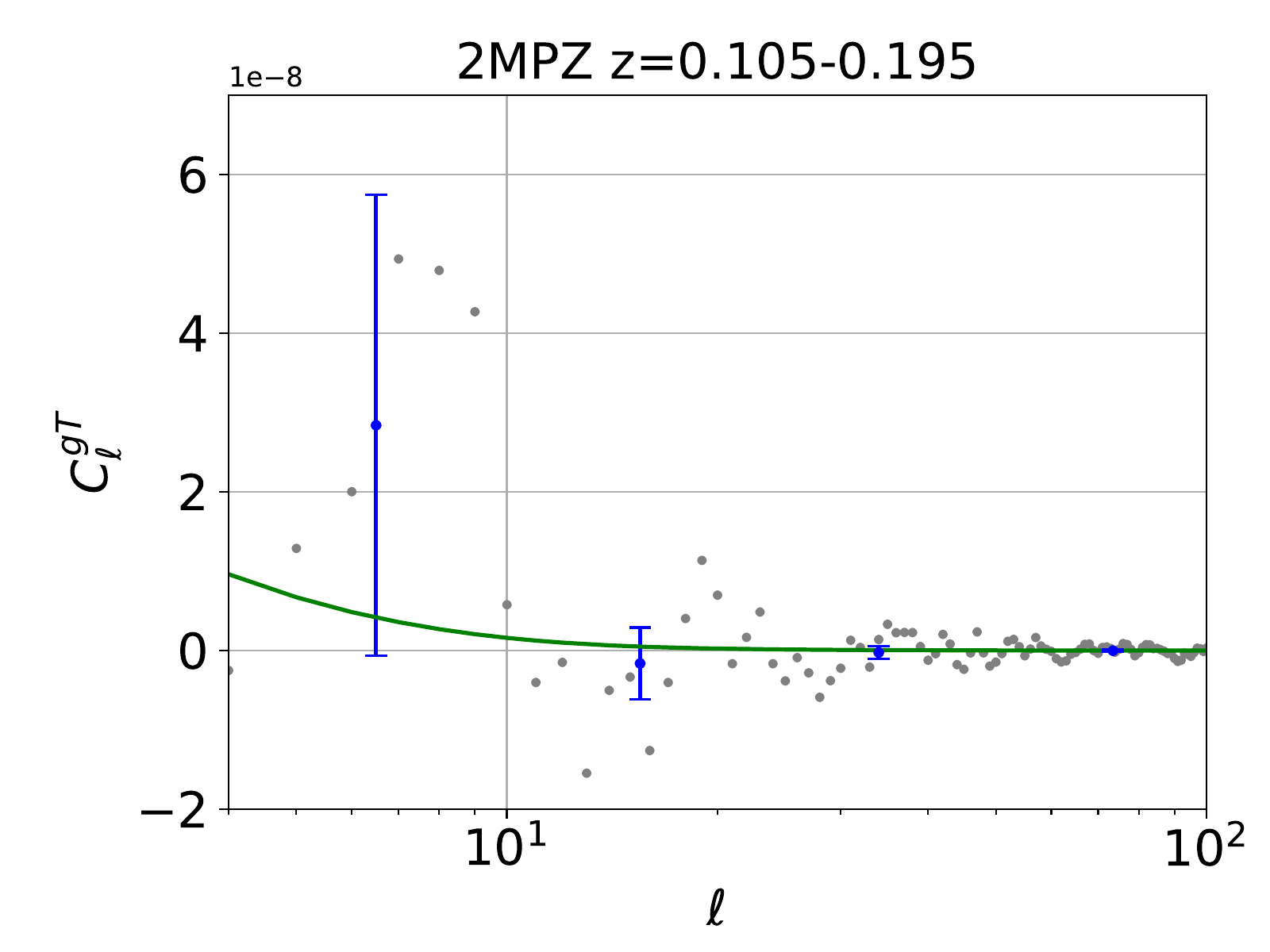}
\includegraphics[width = 0.45\textwidth]{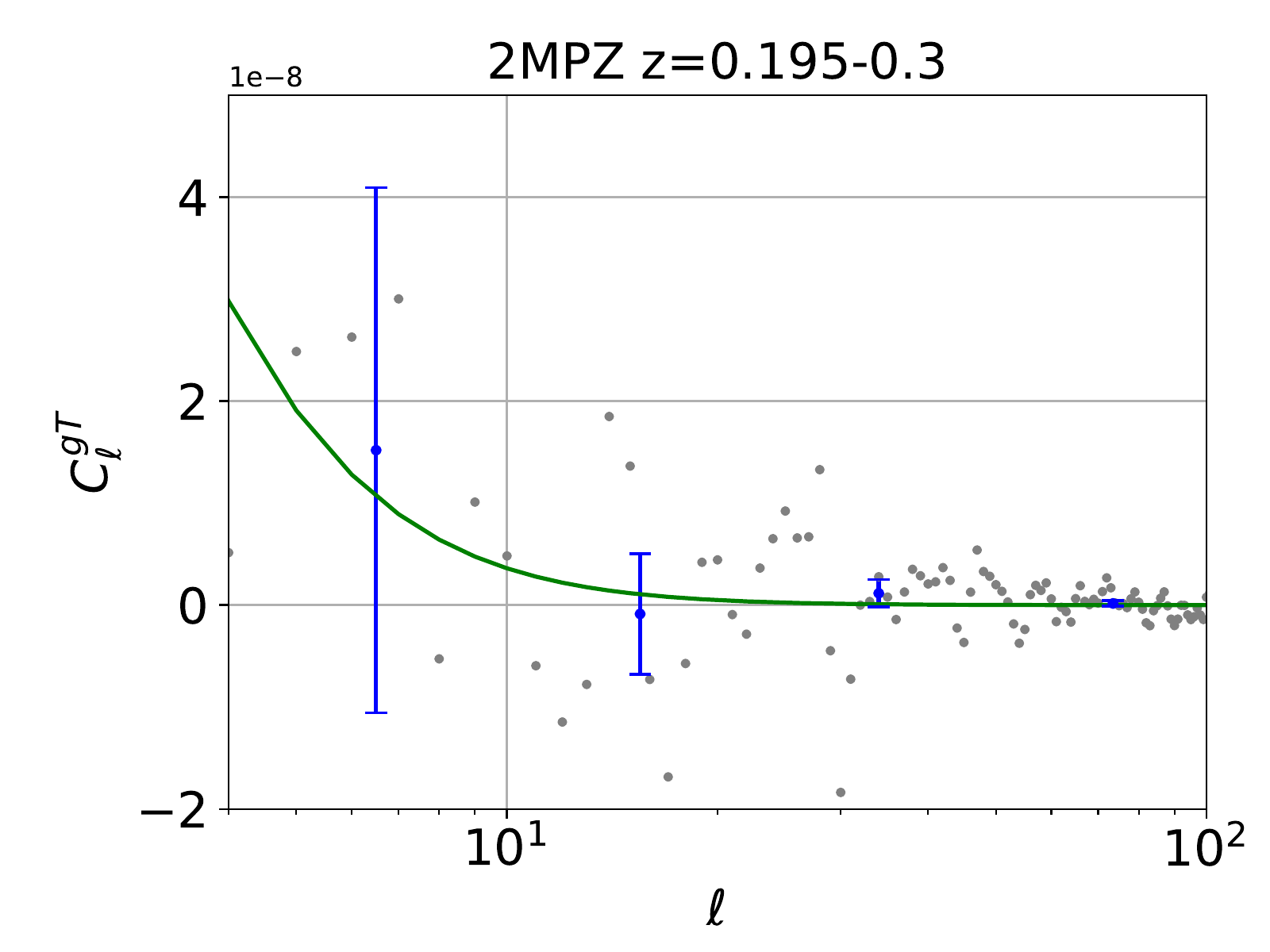}
\includegraphics[width = 0.45\textwidth]{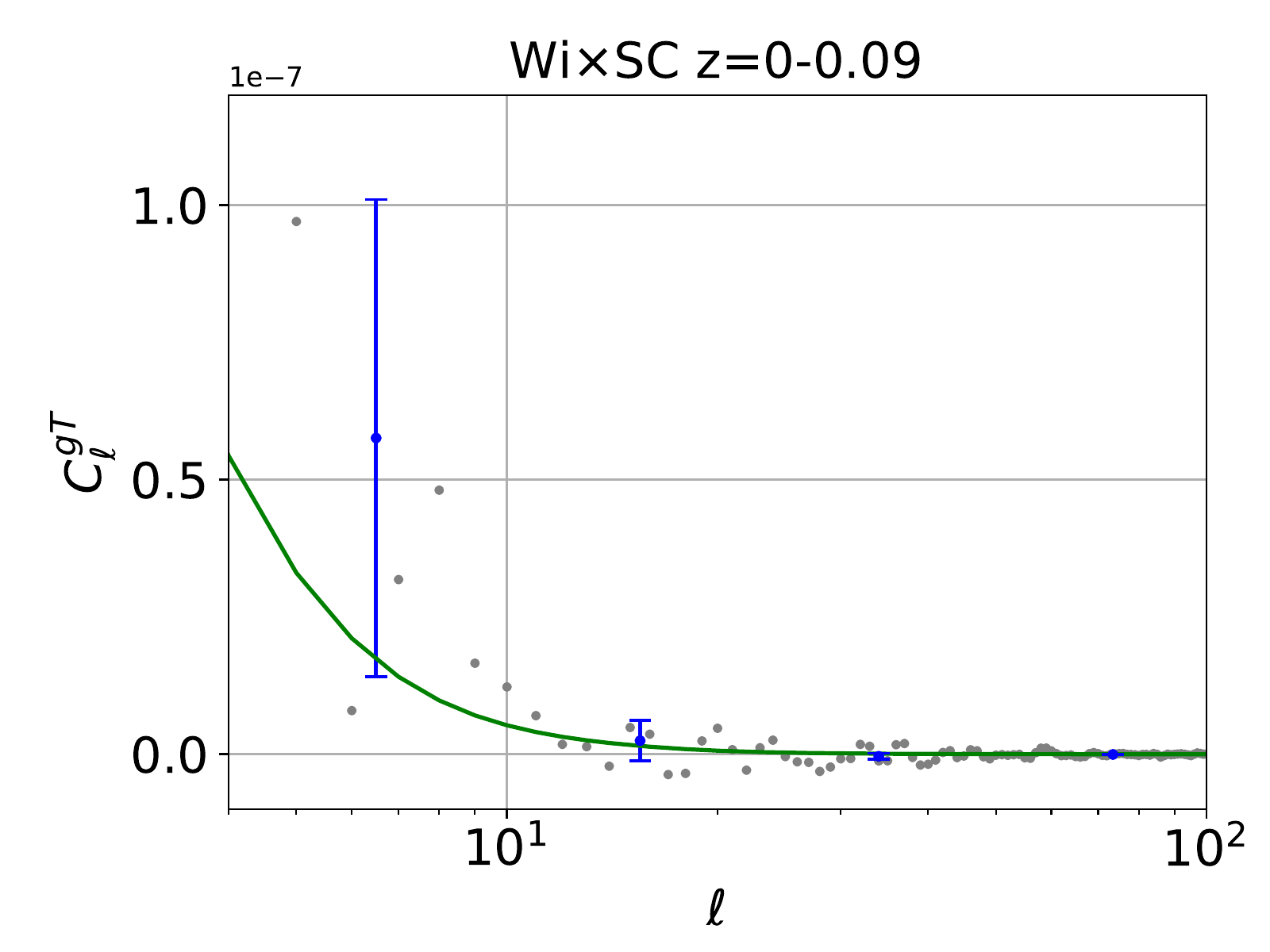}
\includegraphics[width = 0.45\textwidth]{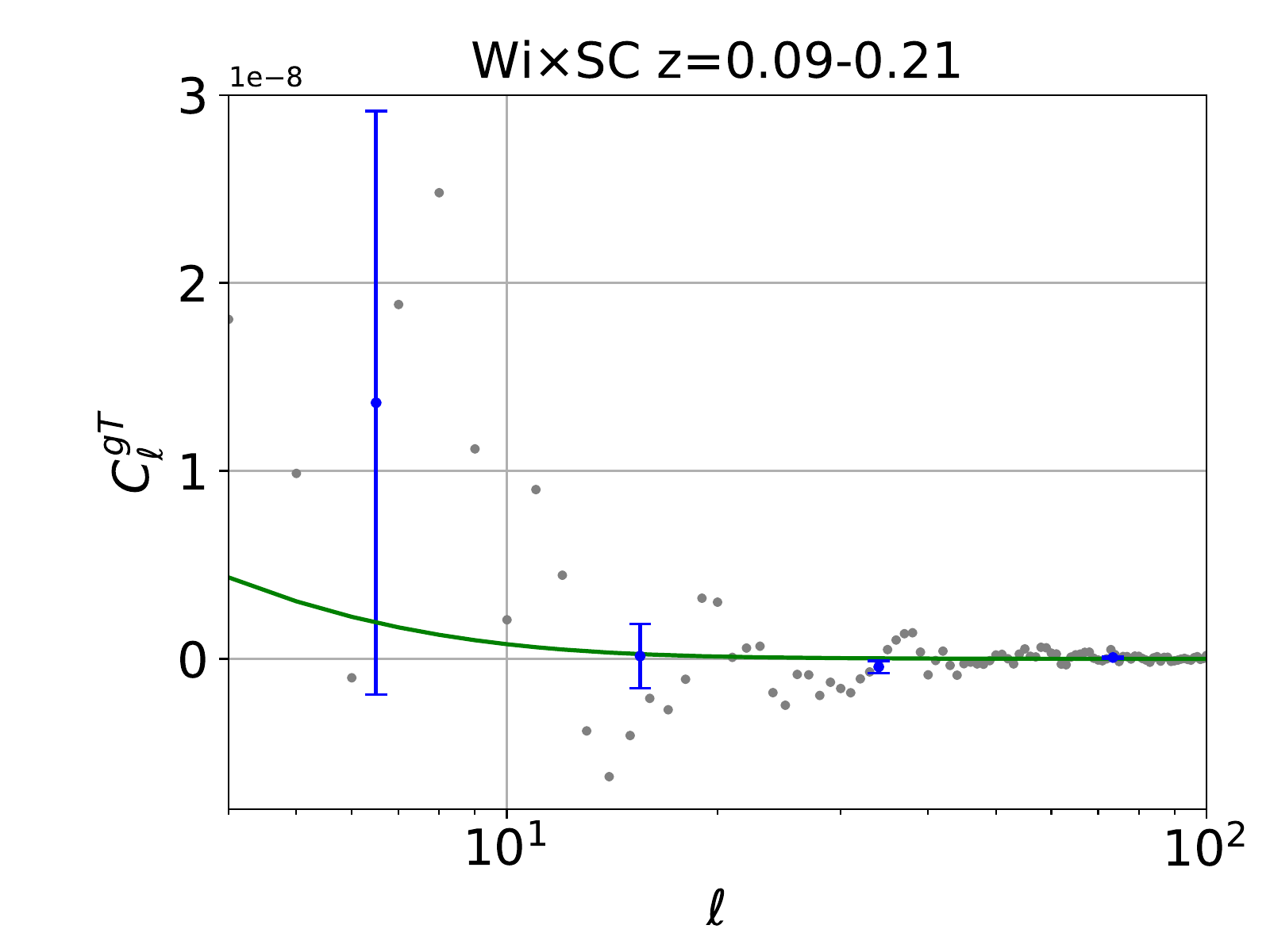}
\includegraphics[width = 0.45\textwidth]{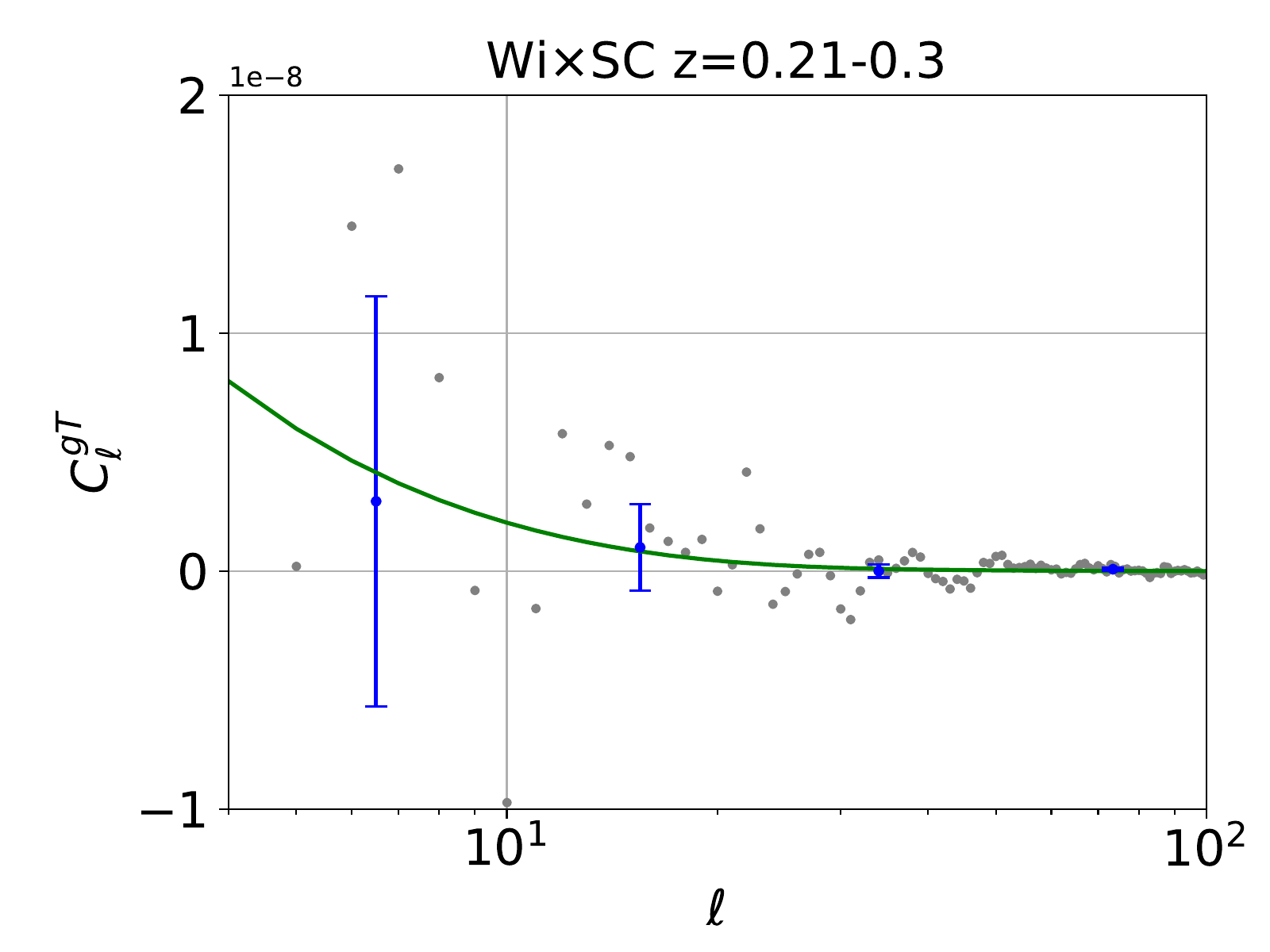}
\caption{Measured cross-correlation with the CMB for different catalogs and redshift bins.}
\end{figure*}

\begin{figure*}
\includegraphics[width = 0.45\textwidth]{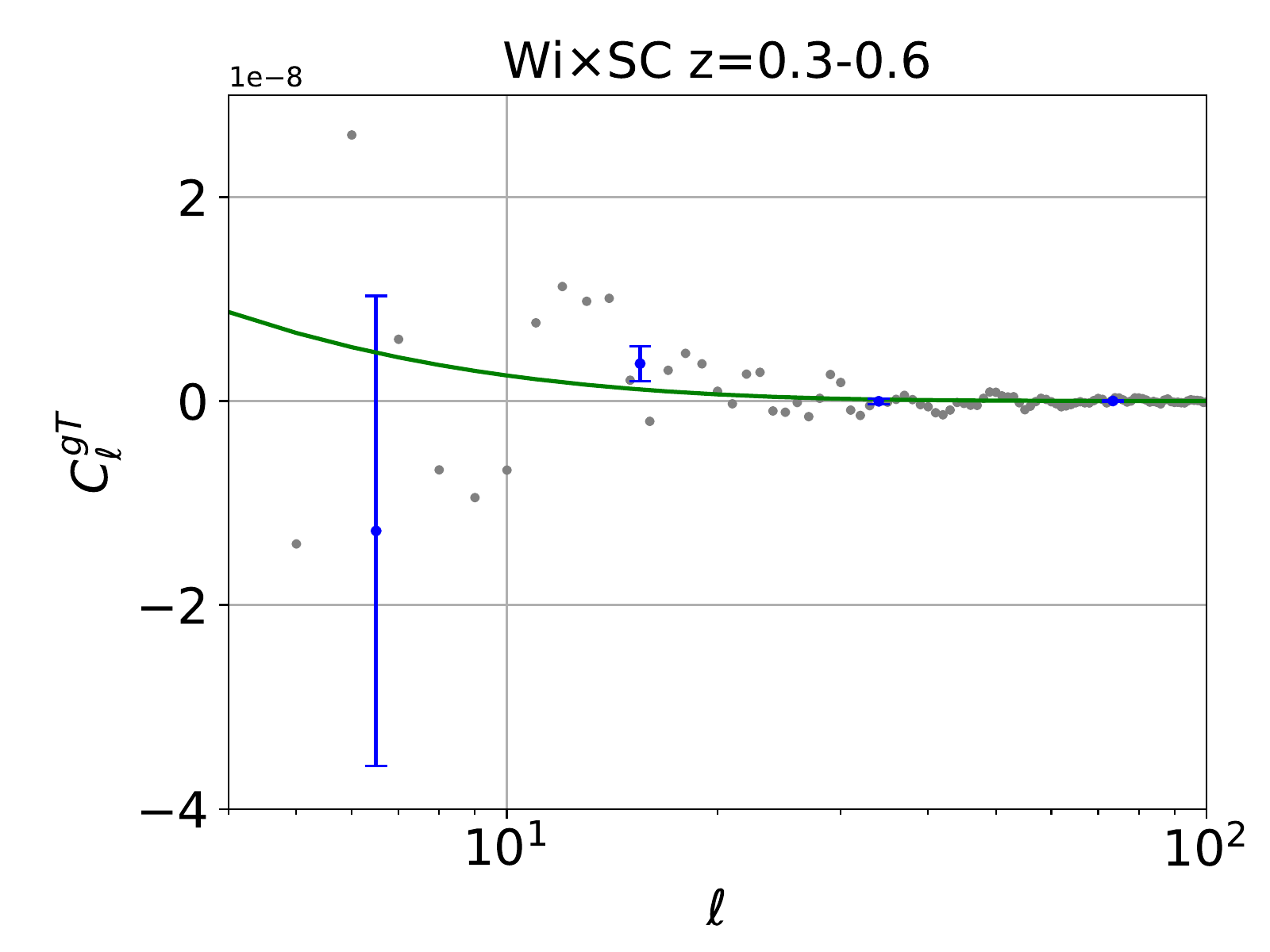}
\includegraphics[width = 0.45\textwidth]{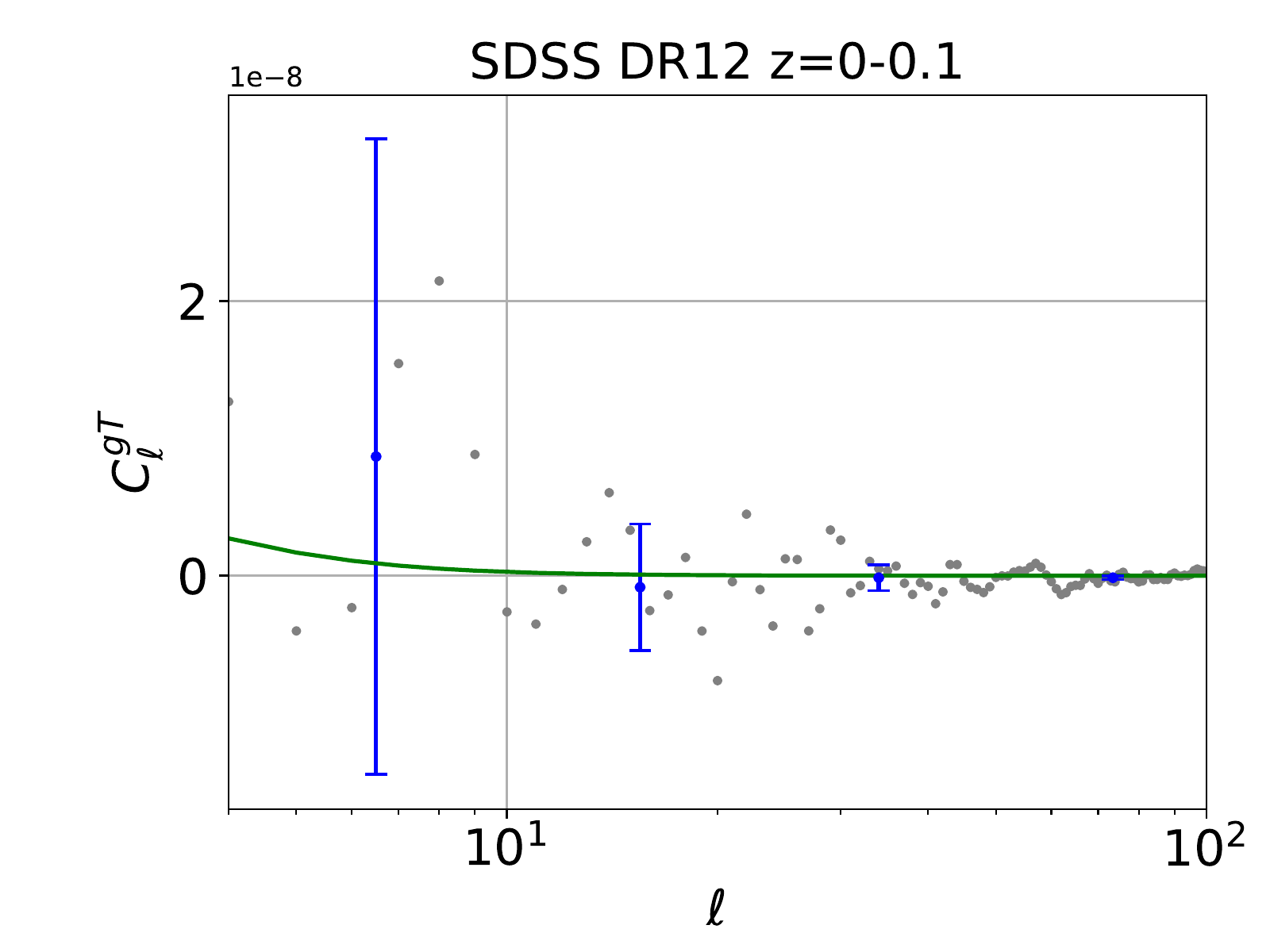}
\includegraphics[width = 0.45\textwidth]{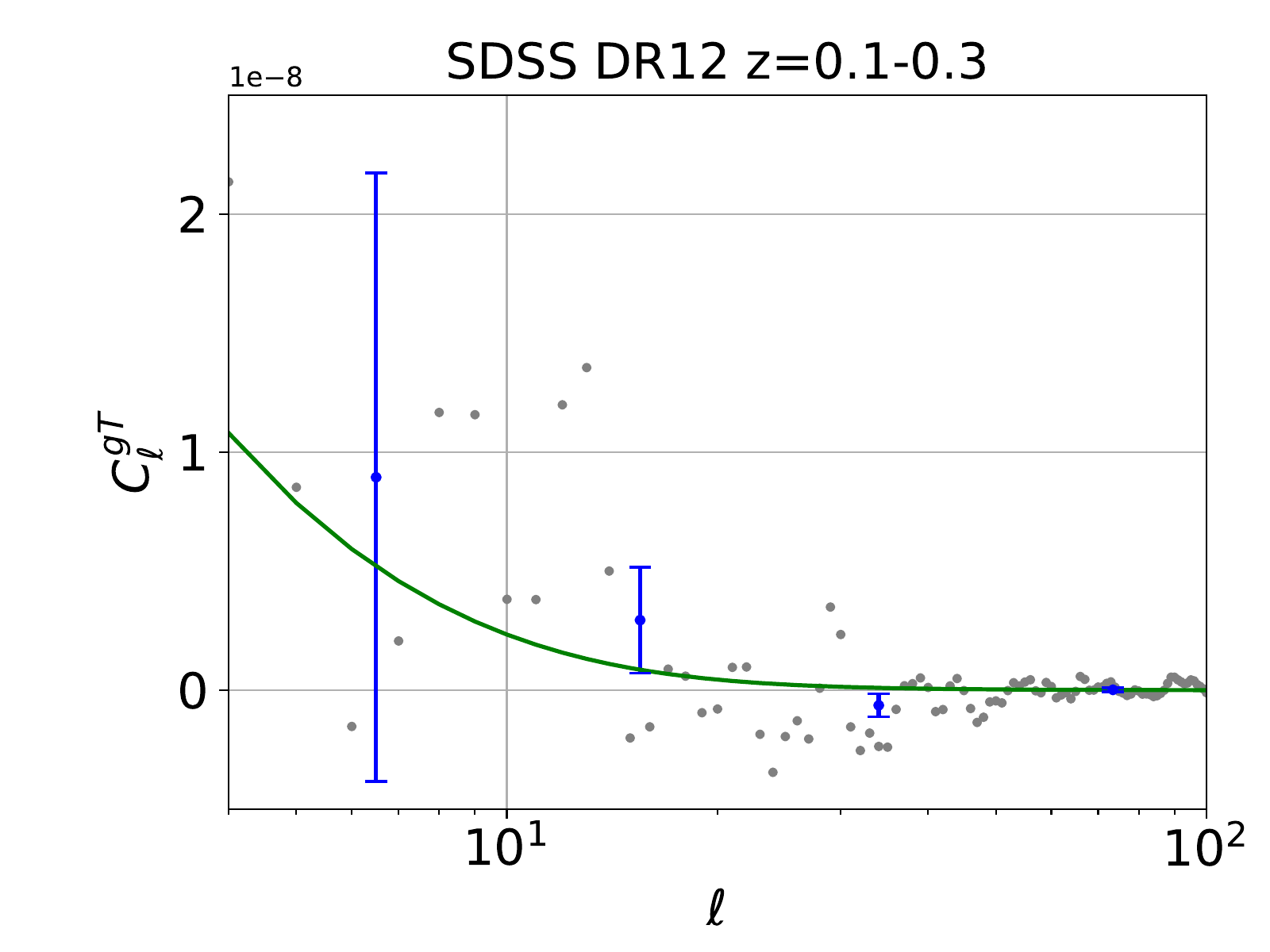}
\includegraphics[width = 0.45\textwidth]{crosscorrelation/sdss_bin_2.pdf}
\includegraphics[width = 0.45\textwidth]{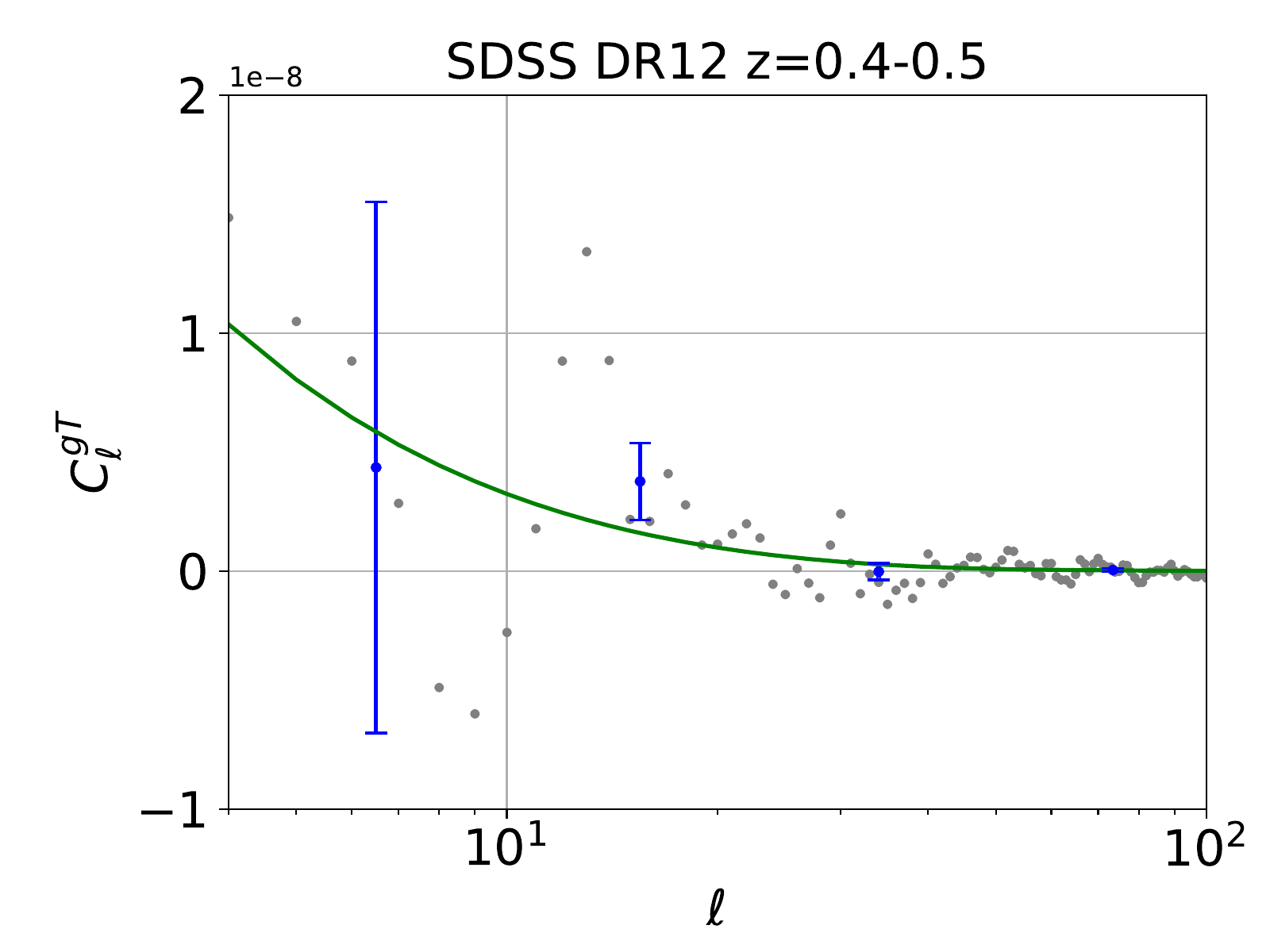}
\includegraphics[width = 0.45\textwidth]{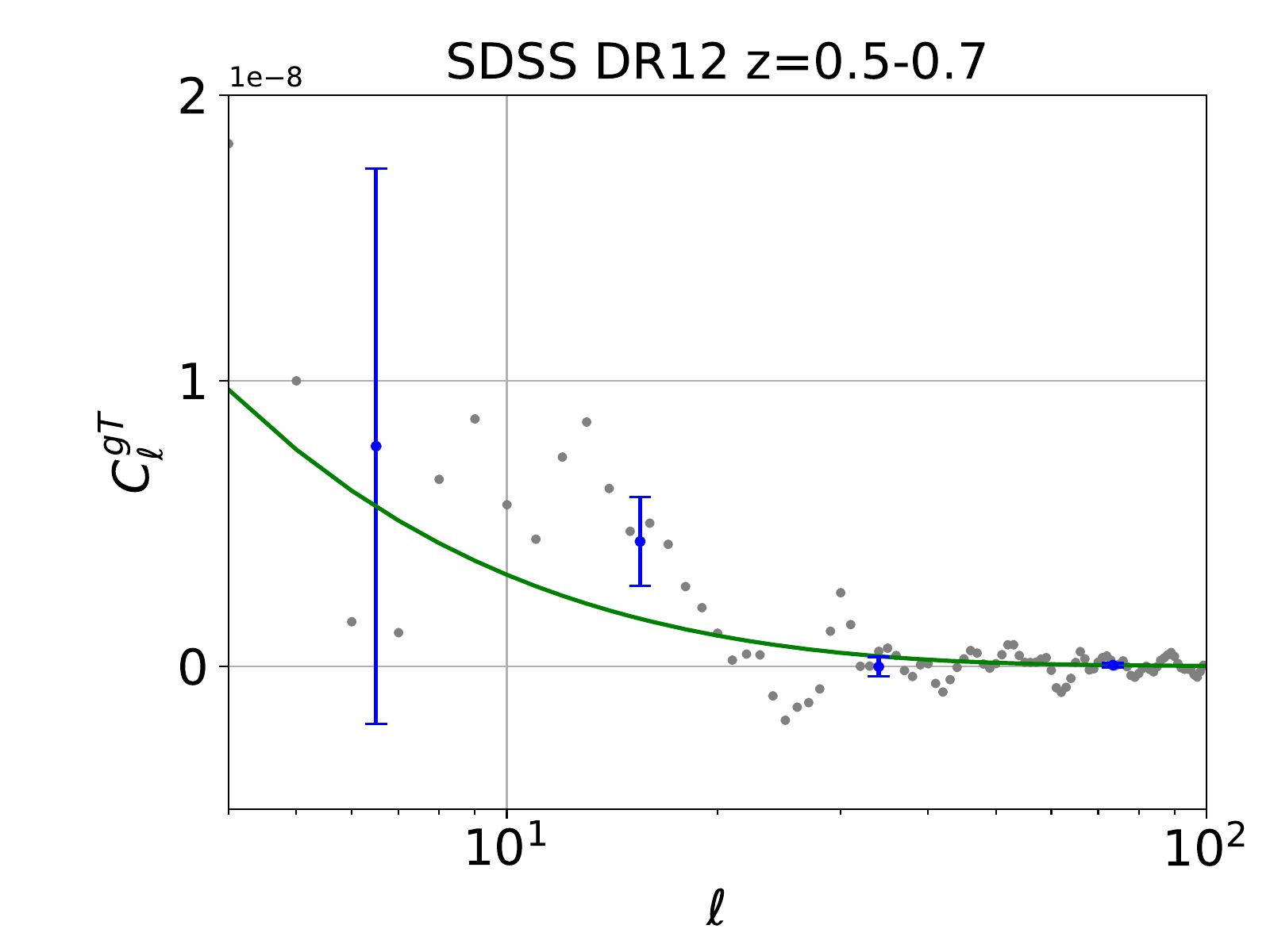}
\caption{Measured cross-correlation with the CMB for different catalogs and redshift bins.}
\end{figure*}

\begin{figure*}
\includegraphics[width = 0.45\textwidth]{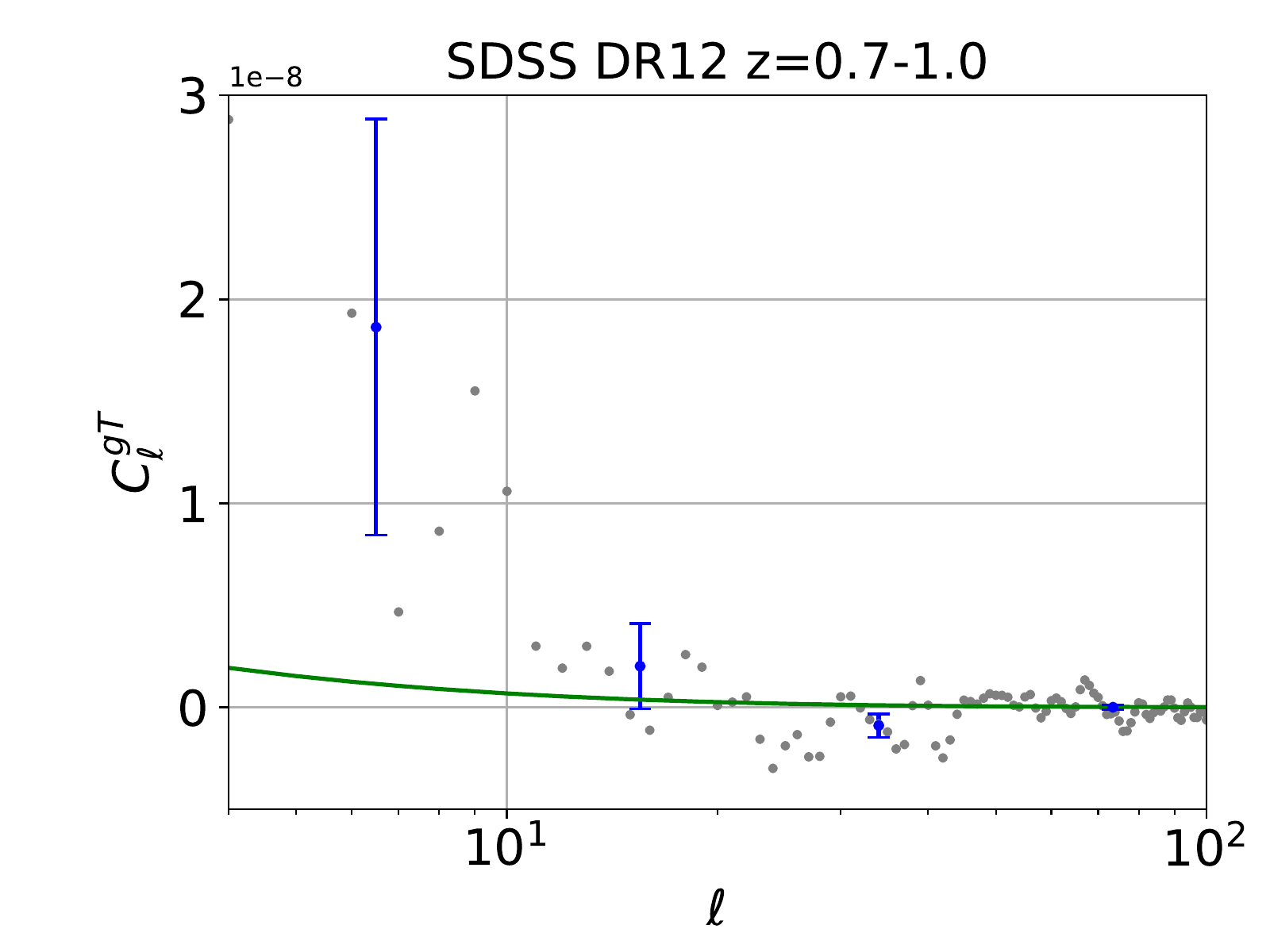}
\includegraphics[width = 0.45\textwidth]{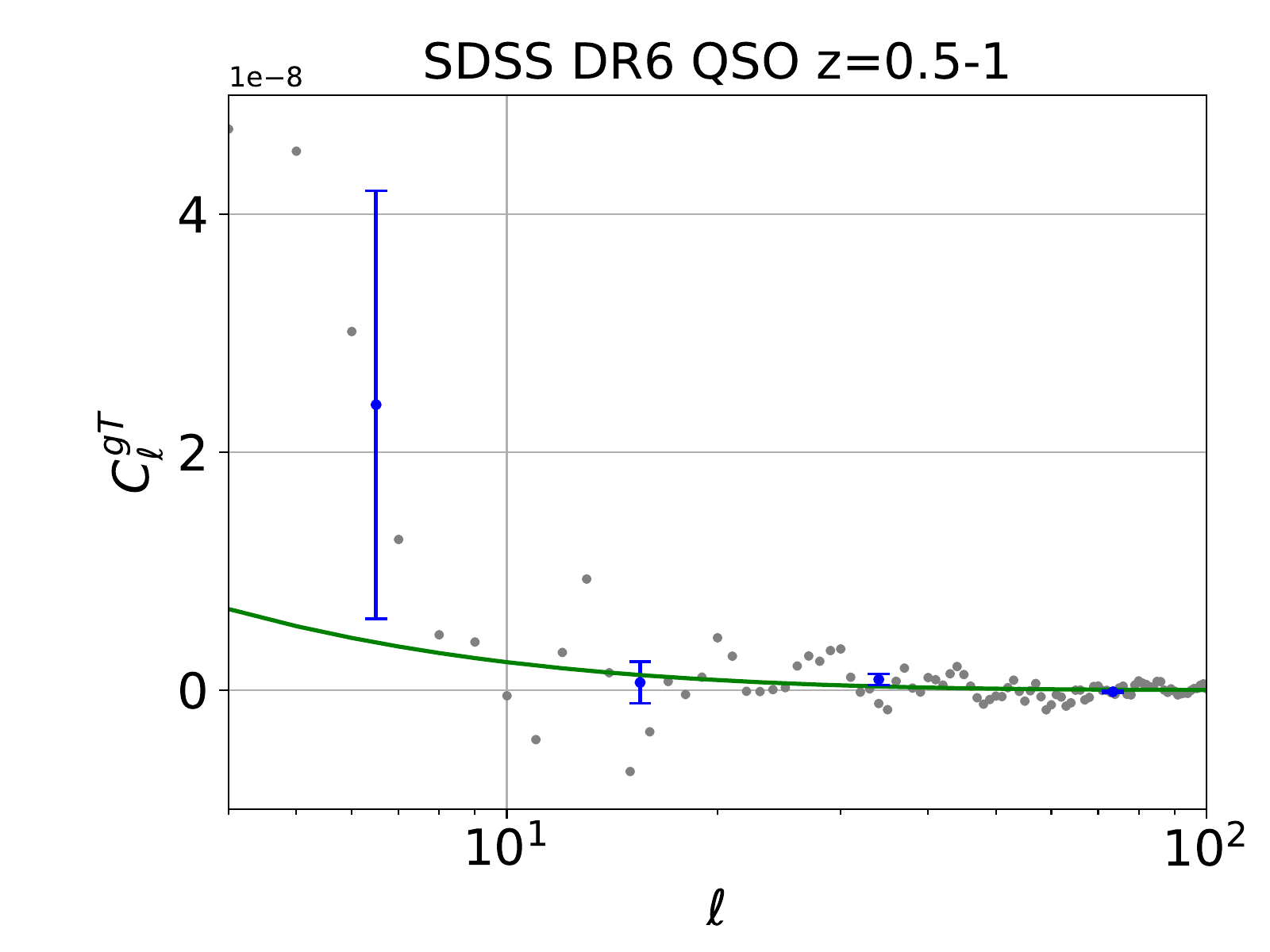}
\includegraphics[width = 0.45\textwidth]{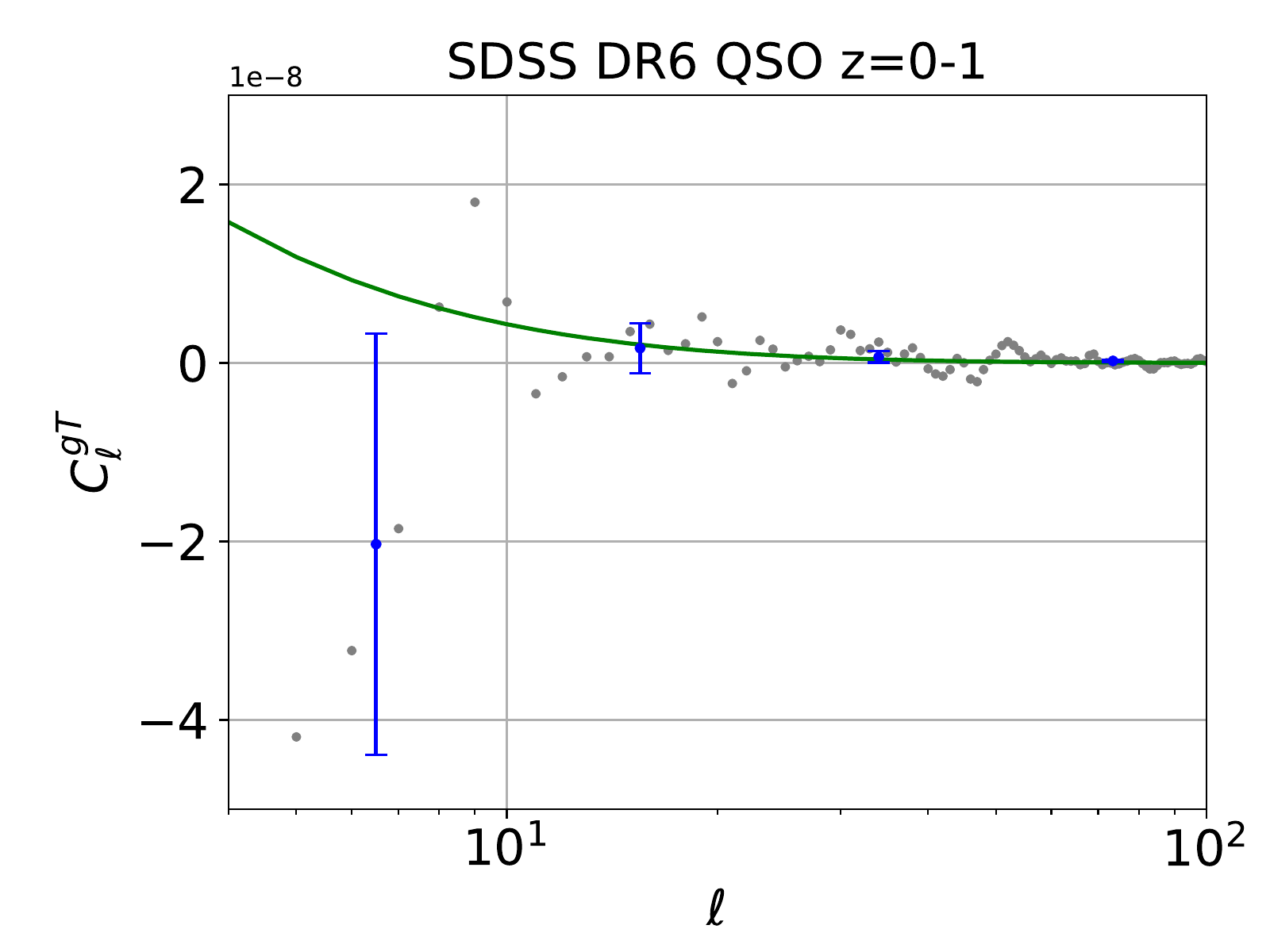}
\includegraphics[width = 0.45\textwidth]{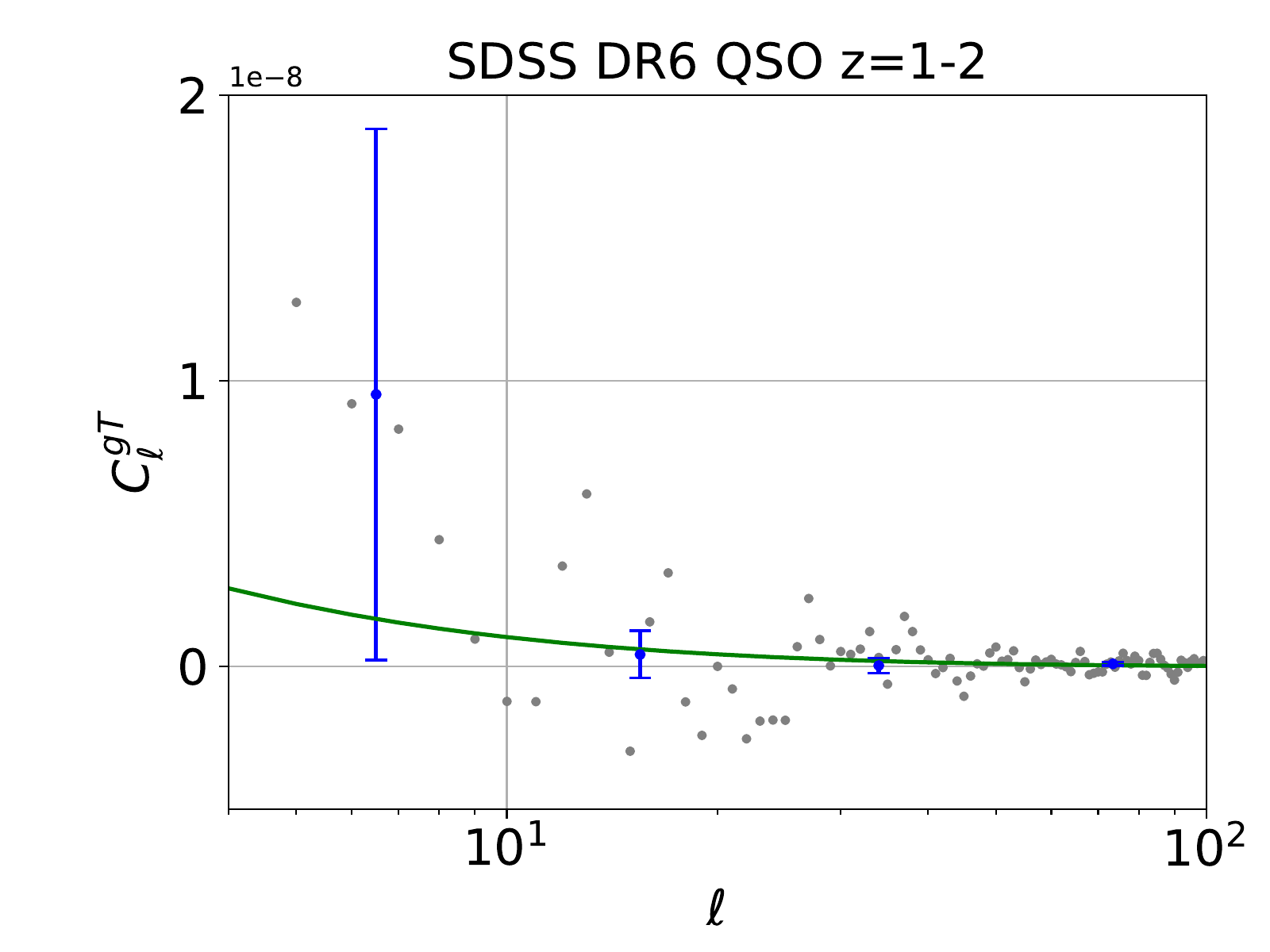}
\includegraphics[width = 0.45\textwidth]{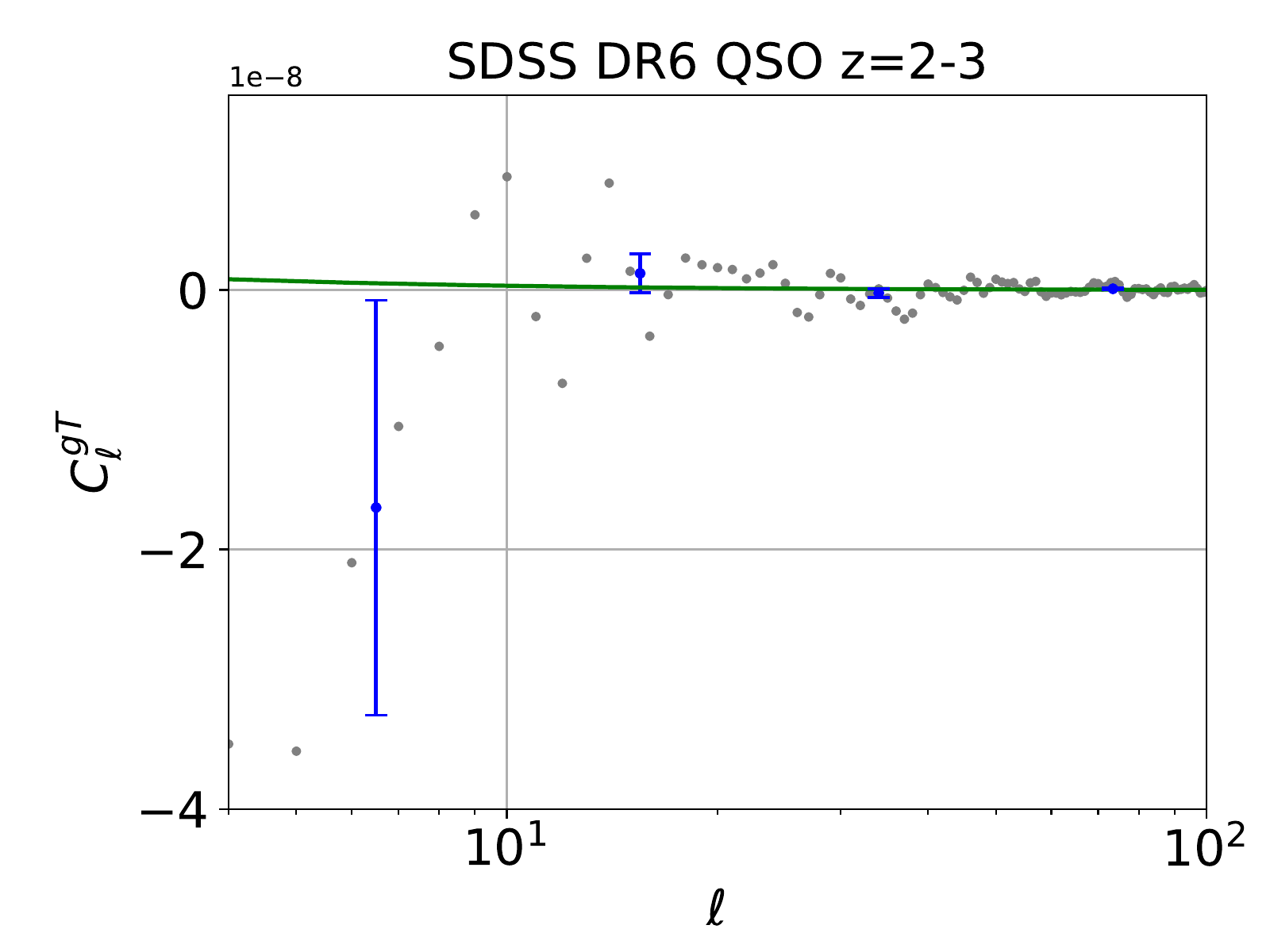}
\includegraphics[width = 0.45\textwidth]{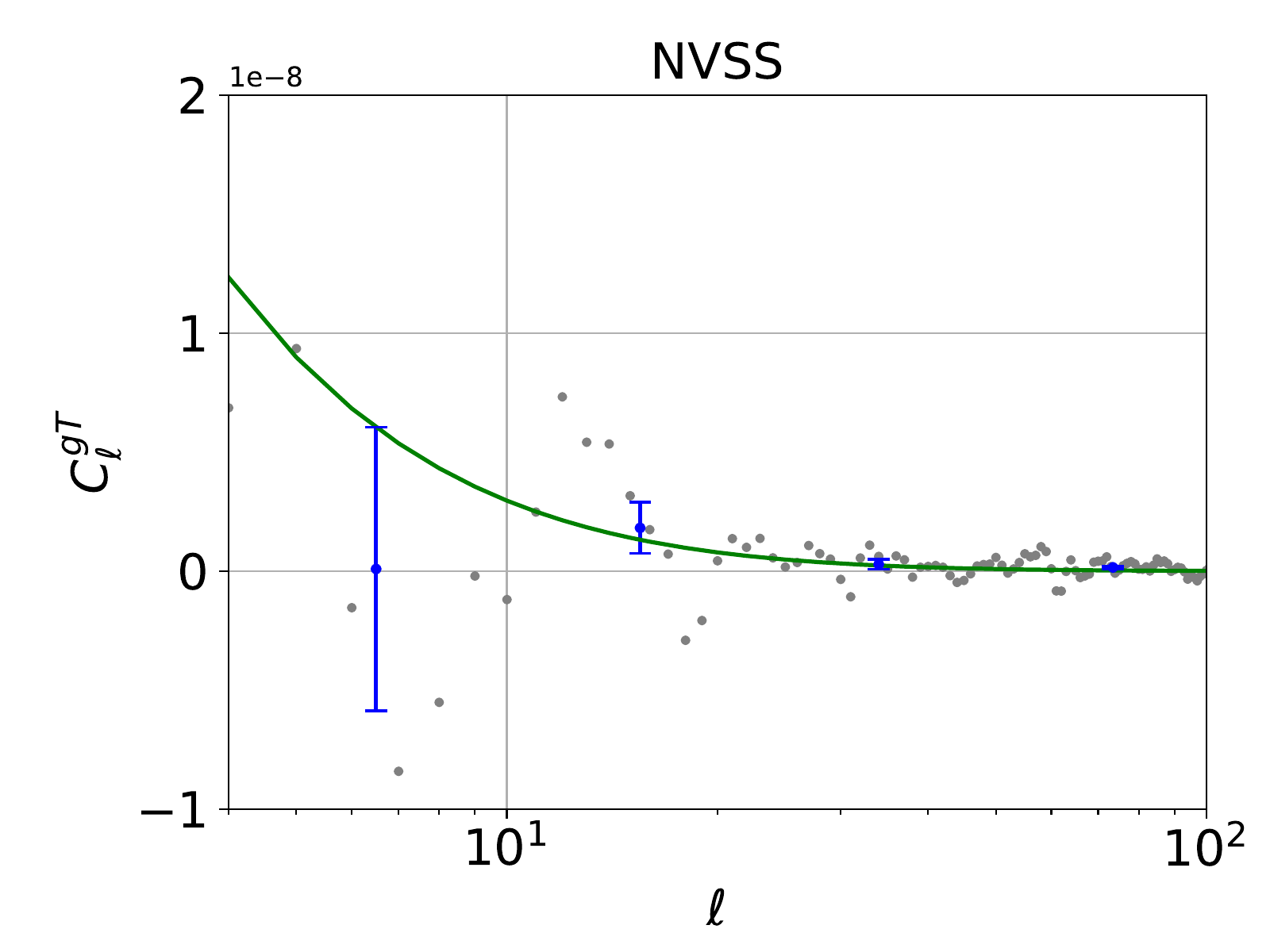}
\caption{Measured cross-correlation with the CMB for different catalogs and redshift bins.}
\end{figure*}

\end{document}